\documentstyle[12pt]{article}
\input{epsf}
\def\roughly#1{\raise.3ex\hbox{$#1$\kern-.75em\lower1ex\hbox{$\sim$}}}
\newcommand{\bra}[1]{\mbox{$\langle #1 $}}
\newcommand{\ket}[1]{\mbox{$| #1 \rangle$}}
\newcommand{\subs}[1]{\subsection{#1}\setcounter{equation}{0}}

\begin{document}

\thispagestyle{empty}
\rightline{NSF-ITP-94-97}
\rightline{hep-th/9411028}

\vskip 3.5 cm

\begin{center}
{\large\bf WHAT IS STRING THEORY?} \break

\vskip 2.5 cm
{\bf Joseph Polchinski}\footnote{Electronic address:
joep@sbitp.itp.ucsb.edu}

\vskip 0.7 cm
\sl
Institute for Theoretical Physics  \break
University of California  \break
Santa Barbara, CA 93106-4030  \break

\end{center}
\vskip 1.6 cm
\rm

\begin{quote}
\begin{center}
{\bf ABSTRACT}
\end{center}
Lectures presented at
the 1994 Les Houches Summer School ``Fluctuating Geometries in
Statistical Mechanics and Field Theory.''  The first part is an
introduction to conformal field theory and string perturbation theory.
The second part deals with the search for a deeper answer to the
question posed in the title.
\end{quote}
\vskip1cm

\normalsize

\newpage
\tableofcontents
\newpage

While I was planning these lectures I happened to reread
Ken Wilson's account of his early work\cite{wilson}, and
was struck by the parallel between string theory today
and quantum field theory thirty years ago.  Then, as now, one had a
good technical control over the perturbation theory but little else.
Wilson saw himself as asking the question ``What is quantum field
theory?''  I found it enjoyable and inspiring to read about the
various models he studied and approximations he tried (he refers
to ``clutching at straws'') before he found the simple and powerful
answer, that the theory is to be organized scale-by-scale rather
than graph-by-graph.  That understanding made it possible to answer
both problems of principle, such as how quantum field theory is to
be defined beyond perturbation theory, and practical problems,
such as how to
determine the ground states and phases of quantum field theories.

In string theory today we have these same kinds of problems, and I
think there is good reason to expect that an equally powerful
organizing principle remains to be found.  There are many reasons,
as I will touch upon later, to believe that string theory is the
correct unification of gravity, quantum mechanics, and particle
physics.  It is implicit, then, that the theory actually exists,
and `exists' does not mean just perturbation theory.  The nature
of the organizing principle is at this point quite open, and may
be very different from what we are used to in quantum field theory.

One can ask whether the situation today in string theory is really
as favorable as it was for field theory in the early 60's.
It is difficult to know.  Then,
of course, we had many more experiments to tell us how
quantum field theories actually behave.  To offset that, we
have today more experience and greater mathematical
sophistication.  As an optimist, I make an encouraging
interpretation of the history, that many of the key
advances in field theory---Wilson's renormalization group, the
discovery of spontaneously broken gauge symmetry as the theory of the
electroweak interaction, the discovery of general relativity
itself---were carried out largely by study of simple model
systems and limiting behaviors, and by considerations of internal
consistency.  These same tools are available in string theory
today.

My lectures divide into two parts---an introduction to
string theory as we now understand it, and a look at attempts to go
further.  For the introduction, I obviously cannot in five
lectures cover the whole of superstring theory.  Given time
limitations, and given the broad range of interests among the
students, I will try to focus on general
principles.  I will begin with conformal field theory (2.5
lectures), which of course has condensed matter applications as well
as being the central tool in string theory.
Section~2 (2.5 lectures)
introduces string theory itself.  Section~3 (1 lecture), on dualities
and equivalences, covers the steadily increasing evidence that what
appear to be different string theories are in many cases different
ground states of a single theory.  Section~4 (1 lecture) addresses the
question of whether `string field theory' is the organizing principle
we seek.  In section~5 (2 lectures) I discuss matrix models, exactly
solvable string theories in low spacetime dimensions.

I should emphasize that this is a survey of many subjects rather
than a review of any single subject (for example $R$-duality, on
which I spend half a lecture, was the subject of a recent
review \cite{GPR} with nearly 300 references).
I made an effort to choose references which will be useful to the
student---a combination
of reviews, some original references, and some interesting recent
papers.

\section{Conformal Field Theory}
\setcounter{footnote}{0}

Much of the material in this lecture, especially the first part,
is standard and can be found in many reviews.  The 1988 Les Houches
lectures by Ginsparg \cite{G_LH} and Cardy \cite{C_LH} focus on
conformal field theory, the latter with emphasis on applications in
statistical mechanics.  Introductions to string theory with
emphasis on conformal field theory can be found in
refs.~\cite{F_LH}-\cite{D'H_TASI}.  There are a number of
recent books on string theory, though often with less emphasis on
conformal techniques~\cite{GSW}-\cite{Hat}
as well as a book~\cite{DI} and reprint collection~\cite{ISZ}
on conformal field theory and statistical mechanics.
Those who are in no
great hurry will eventually find an expanded version of these
lectures in ref.~\cite{JBBS}.  Finally I should mention the seminal
papers~\cite{BPZ} and~\cite{FMS}.

\subs{The Operator Product Expansion}

The operator product expansion (OPE) plays a central role in this
subject.  I will introduce it using the example of a free scalar
field in two dimensions, $X(\sigma^1,\sigma^2)$.
I will focus on two dimensions because this is the case that will
be of interest for the string, and I will refer to these two
dimensions as `space' though later they will be the string
world-sheet and space will be something else.  The action is
\begin{equation}
S= \frac{1}{8\pi} \int d^2\sigma\,
\Bigl\{ (\partial_1 X)^2 + (\partial_2
X)^2 \Bigr\}. \label{xact}
\end{equation}
The normalization of the field $X$ (and so the action) is for later
convenience.  To be specific I have taken two Euclidean dimensions,
but almost everything, at least until we get to nontrivial
topologies, can be continued immediately to the Minkowski
case\footnote{Both the Euclidean and Minkowski cases should be
familiar to the condensed matter audience.  The former would be
relevant to classical critical phenomena in two dimensions, and the
latter to quantum critical phenomena in one dimension.}
$\sigma^2 \to -i \sigma^0$.  Expectation values are defined by the
functional integral
\begin{equation}
< {\cal F}[X] > \ =\ \int [dX]\, e^{-S} {\cal F}[X],
\end{equation}
where ${\cal F}[X]$ is any functional of $X$, such as a product of
local operators.\footnote{Notice that this has not been normalized
by dividing by $< 1 >$.}

It is very convenient to adopt complex coordinates
\begin{equation}
z = \sigma^1 + i \sigma^2, \qquad \bar z = \sigma^1 - i \sigma^2.
\end{equation}
Define also
\begin{equation}
\partial_z = \frac{1}{2} (\partial_1 - i \partial_2),
\qquad  \partial_{\bar z} =
\frac{1}{2} (\partial_1 + i \partial_2). \label{comdir}
\end{equation}
These have the properties $\partial_z z = 1$, $\partial_z \bar z =
0$, and so on.  Note also that $d^2 z = 2 d\sigma^1 d\sigma^2$
from the Jacobian, and that $\int d^2 z \, \delta^2(z,\bar z) = 1$.
I will further abbreviate $\partial_z$ to $\partial$
and $\partial_{\bar z}$ to $\bar\partial$ when this will not be
ambiguous.  For a general vector, define as above
\begin{equation}
v^z = v^1 + i v^2, \qquad v^{\bar z} = v^1 - i v^2, \qquad v_z =
\frac{1}{2}(v^1 - i v^2), \qquad
v_{\bar z} = \frac{1}{2}(v^1 + i v^2).
\end{equation}
For the indices 1, 2 the metric is the identity and we do not
distinguish between upper and lower, while the complex indices are
raised and lowered with\footnote
{A comment on notation: being careful to keep the Jacobian, one
has $d^2 z = 2 d\sigma^1 d\sigma^2$ and $d^2z \, \sqrt{|\det g|}
= d\sigma^1 d\sigma^2$.  However, in the literature one very
frequently finds $d^2 z$ used to mean $d\sigma^1 d\sigma^2$.}
\begin{equation}
g_{z\bar z} = g_{\bar z z} = \frac{1}{2}, \qquad g_{zz} = g_{\bar z
\bar z} = 0, \qquad g^{z\bar z} = g^{\bar z z} = 2, \qquad
g^{zz} = g^{\bar z \bar z} = 0
\end{equation}

The action is then
\begin{equation}
S= \frac{1}{4\pi} \int d^2z\,
\partial X \bar\partial X, \label{xact2}
\end{equation}
and the equation of motion is
\begin{equation}
\partial \bar\partial X(z,\bar z) = 0. \label{xeom}
\end{equation}
The notation $X(z,\bar z)$ may seem redundant, since the
value of $z$ determines the value of $\bar z$, but it is useful to
reserve the notation $f(z)$ for fields whose equation of motion
makes them {\it analytic} in $z$.  For example, it follows at once
from the equation of motion~(\ref{xeom}) that $\partial X$
is analytic and that $\bar\partial X$ is antianalytic (analytic in
$\bar z$), hence the notations $\partial X(z)$ and $\bar\partial
X(\bar z)$.  Notice that under the Minkowski continuation, an
analytic field becomes left-moving, a function only of $\sigma^0
+ \sigma^1$, while an antianalytic field becomes right-moving,
a function only of $\sigma^0 - \sigma^1$.

Now, using the property of path integrals that the integral of a
total derivative is zero, we have
\begin{eqnarray}
0 &=& \int [dX]\,\frac{\delta}{\delta X(z,\bar z) }
\Bigl\{ e^{-S} X(z',\bar z') \Bigr\} \nonumber\\
&=& \int [dX]\,
e^{-S} \Bigl\{ \delta^2(z-z',\bar z - \bar z') + \frac{1}{2\pi}
\partial_z \partial_{\bar z} X(z,\bar z)  X(z',\bar z')
\Bigr\}\nonumber\\
&=&
< \delta^2(z-z',\bar z - \bar z') > + \frac{1}{2\pi} \partial_z
\partial_{\bar z} < X(z,\bar z)  X(z',\bar z') >
\end{eqnarray}
That is, the equation of motion holds except at coincident
points.  Now, the same
calculation goes through if we have arbitrary additional insertions
`$\ldots$' in the path integral, as long as no other fields are at
$(z,\bar z)$ or
$(z',\bar z')$:
\begin{equation}
\frac{1}{2\pi} \partial_z
\partial_{\bar z} < X(z,\bar z)  X(z',\bar z') \ldots >
\ =\ -< \delta^2(z-z',\bar z - \bar z') \ldots >.
\end{equation}
A relation which holds in this sense will simply be written
\begin{equation}
\frac{1}{2\pi} \partial_z
\partial_{\bar z} X(z,\bar z)  X(z',\bar z')
= -\delta^2(z-z',\bar z - \bar z') , \label{xxeom}
\end{equation}
and will be called an {\it operator equation}.  One can think of
the additional fields `$\ldots$' as preparing arbitrary initial
and final states, so if one cuts the path integral open to make an
Hamiltonian description, an operator equation is simply one which
holds for arbitrary matrix elements.  Note also that because of the
way the path integral is constructed from iterated time slices,
any product of fields in the path integral goes over to a
time-ordered product in the Hamiltonian form.  In the
Hamiltonian formalism, the delta-function in eq.~(\ref{xxeom})
comes from the differentiation of the time-ordering.

Now we define a very useful combinatorial tool, {\it normal
ordering}:
\begin{equation}
{ :\! X(z,\bar z)  X(z',\bar z') \! : }\ \equiv\
X(z,\bar z)  X(z',\bar z') + \ln|z-z'|^2.  \label{nord}
\end{equation}
The logarithm satisfies the equation of motion~(\ref{xxeom})
with the opposite sign
(the action was normalized such that this log would have
coefficient~1), so that by construction
\begin{equation}
\partial_z
\partial_{\bar z} { :\! X(z,\bar z)  X(z',\bar z') \! : }
= 0 .  \label{neom}
\end{equation}
That is, the normal ordered product satisfies the naive equation
of motion.  This implies that the
normal ordered product is locally the sum of an
analytic and antianalytic function (a standard result from complex
analysis).  Thus it can be Taylor expanded, and so from the
definition~(\ref{nord}) we have (putting one operator
at the origin for convenience)
\begin{equation}
X(z,\bar z)  X(0,0) = - \ln|z|^2
\ + :\! X^2(0,0) \! : +\ z :\! X\partial X(0,0) \!:  +\ \bar z :\!
X \bar\partial X(0,0) \! : +
\ldots\ .
\label{xxop}
\end{equation}
This is an operator equation, in the same sense as the preceding
equations.

Eq.~(\ref{xxop}) is our first example of an operator product
expansion.  For a general expectation value involving
$X(z,\bar z) X(0,0)$ and other fields, it gives the small-$z$
behavior as a sum of terms, each of which is a known
function of $z$ times the expectation values of a single local
operator.
For a general field theory, denote a complete set of local
operators for a field theory by
${\cal A}_i$.  The OPE then takes the general form
\begin{equation}
{\cal A}_i(z,\bar z) {\cal A}_j(0,0)
= \sum_k c^k\!_{ij}(z,\bar z){\cal A}_k(0,0). \label{gope}
\end{equation}
Later in section~1 I will give a simple derivation of the
OPE~(\ref{gope}), and of a rather broad
generalization of it.  OPE's are frequently used in particle and
condensed matter physics as asymptotic expansions, the first few
terms giving the dominant behavior at small~$z$.  However, I
will argue that, at least in conformally invariant theories,
the OPE is actually a convergent series.  The radius of convergence
is given by the distance to the nearest {\it other} operator in
the path integral.  Because of this the coefficient functions
$c^k\!_{ij}(z,\bar z)$, which as we will see must satisfy various
further conditions, will enable us to reconstruct the entire field
theory.\\[3pt]
{\bf Exercise:} The expectation value\footnote
{To be precise, expectation values of $X(z,\bar z)$ generally
suffer from an infrared divergence on the plane.  This is a
distraction which we ignore by some implicit long-distance
regulator.  In practice one is always interested in `good'
operators such as derivatives or exponentials of $X$, which have
well-defined expectation values.}
$<X(z_1,\bar z_1) X(z_2,\bar z_2) X(z_3,\bar z_3) X(z_4,\bar z_4)>$
is given by the sum over all Wick contractions with the propagator
$- \ln |z_i - z_j|^2$.  Compare the asymptotics as $z_1 \to z_2$
from the OPE~(\ref{xxop}) with the asymptotics of the exact
expression.  Verify that the expansion in $z_1 - z_2$ has the
stated radius of convergence.

The various operators on the right-hand side of the
OPE~(\ref{xxop}) involve products of fields at the same point.
Usually in quantum field theory such a product is divergent and must
be appropriately cut off and renormalized, but here the
normal ordering renders it well-defined.  Normal ordering is thus a
convenient way to define composite operators in free field theory.
It is of little use in most interacting field theories, because
these have additional divergences from interaction vertices
approaching the composite operator or one another.  But many
of the conformal field theories that we will be interested in are
free, and many others can be related to free field theories, so it
will be worthwhile to develop normal ordering somewhat further.

For products of more than 2 fields the definition~(\ref{nord})
can be extended iteratively,
\begin{eqnarray}
&&{ :\! X(z,\bar z) X(z_1,\bar z_1)
\ldots X(z_n,\bar z_n) \! : }
\ \equiv\  X(z,\bar z) \,\,
{ :\! X(z_1,\bar z_1) \ldots X(z_n,\bar z_n) \! : } \label{enord}\\
&&\qquad\qquad
+\ \Bigl\{ \ln|z-z_1|^2  :\! X(z_2,\bar z_2)
\ldots X(z_n,\bar z_n) \! :\ +\ (n-1) {\rm\ permutations} \Bigr\},
\nonumber
\end{eqnarray}
contracting each pair (omitting the pair and subtracting $-\ln |z -
z_i|^2$).  This has the same properties as before: the equation of
motion holds inside the normal ordering, and so the normal-ordered
product is smooth.  ({\bf Exercise:} Show this.  The simplest argument
I have found is inductive, and uses the definition twice to pull
both $X(z,\bar z)$ and $X(z_1,\bar z_1)$ out of the normal ordering.)

The definition~(\ref{enord}) can be written more formally as
\begin{equation}
:\! X(z,\bar z) {\cal F}[X] \! :\ =
X(z,\bar z) :\! {\cal F}[X] \! :
+ \int d^2 z' \, \ln|z-z'|^2 \frac{\delta}{\delta
X(z',\bar z')}:\! {\cal F}[X] \! :,
\end{equation}
for an arbitrary functional ${\cal F}[X]$,
the integral over the functional derivative producing all
contractions.  Finally, the definition of normal ordering can be
written in a closed form by the same strategy,
\begin{equation}
:\! {\cal F}[X] \! :\ =
\exp\biggl\{ \frac{1}{2} \int d^2z\,d^2 z' \, \ln|z-z'|^2
\frac{\delta}{\delta X(z,\bar z)}
\frac{\delta}{\delta X(z',\bar z')} \biggr\}
{\cal F}[X].
\end{equation}
The exponential sums over all ways of contracting zero, one, two,
or more pairs.
The operator product of two normal ordered operators can be
represented compactly as
\begin{equation}
:\! {\cal F}[X] \! :\,\, :\! {\cal G}[X] \! :\ =
\exp\biggl\{ -\int d^2z'\,d^2 z'' \, \ln|z'-z''|^2
\frac{\delta_F}{\delta X(z',\bar z')}
\frac{\delta_G}{\delta X(z'',\bar z'')} \biggr\}
:\! {\cal F}[X]\,{\cal G}[X] \! :,
\end{equation}
where $\delta_F$ and $\delta_G$ act only on the fields in
${\cal F}$ and ${\cal G}$ respectively.  The expressions
$:\! {\cal F}[X] \! :\, :\! {\cal G}[X] \! : $ and
$:\! {\cal F}[X]\,{\cal G}[X] \! :$ differ by the contractions
between one field from ${\cal F}$ and one field from ${\cal G}$,
which are then restored by the exponential.  Now, for ${\cal F}$ a
local operator at $z_1$ and ${\cal G}$ a local operator at
$z_2$, we can expand in $z_1 - z_2$ inside the normal ordering on
the right to generate the OPE.  For example, one finds
\begin{eqnarray}
:\! e^{i k_1 X(z,\bar z)} \! :\,\, :\! e^{i k_2 X(0,0)}
\! :&=& |z|^{2k_1 k_2}
:\! e^{i k_1 X(z,\bar z) + i k_2 X(0,0)}
\! : \nonumber\\
&\sim& |z|^{2k_1 k_2}
:\! e^{i (k_1 + k_2) X(0,0)} \! : \ ,
\end{eqnarray}
since each contraction gives $k_1 k_2 \ln|z|^2$ and the contractions
exponentiate.  Exponential operators will be quite useful to us.
Another example is
\begin{equation}
\partial X(z,\bar z)\, :\! e^{i k X(0,0)} \! :
\ \sim\ -\frac{ik}{z}
:\! e^{i k X(0,0)} \! : \ ,
\end{equation}
coming from a single contraction.

\subs{Ward Identities}

The action~(\ref{xact2}) has a number of important symmetries,
in particular conformal invariance.  Let us first derive the
Ward identities for a general symmetry.  Suppose we have fields
$\phi_\alpha(\sigma)$ with some action $S[\phi]$, and a symmetry
\begin{equation}
\phi'_\alpha(\sigma)
= \phi_\alpha(\sigma) + \epsilon\delta\phi_\alpha(\sigma).
\label{sym}
\end{equation}
That is, the product of the path integral measure and the
weight~$e^{-S}$ is invariant.  For a path integral with general
insertion ${\cal F}[\phi]$, make the change of
variables~(\ref{sym}).
The invariance of the integral under change of variables, and the
invariance of the measure times~$e^{-S}$, give
\begin{equation}
0\ =\ \int d^2\sigma\,\sum_\alpha < \delta\phi_\alpha(\sigma)
\frac{\delta}{\delta \phi_\alpha(\sigma)} {\cal F}[\phi] >
\ \equiv\ < \delta {\cal F}[\phi] >.
\end{equation}
This simply states that the general expectation value is invariant
under the symmetry.

We can derive additional information from the
symmetry: the existence of a conserved current (Noether's theorem),
and Ward identities for the expectation values of the current.
Consider the following change of variables,
\begin{equation}
\phi'_\alpha(\sigma)
= \phi_\alpha(\sigma) + \epsilon\rho(\sigma)
\delta\phi_\alpha(\sigma).
\label{nsym}
\end{equation}
This is not a symmetry, the transformation law being altered by the
inclusion of an arbitrary function $\rho(\sigma)$.  The path integral
measure times~$e^{-S}$
would be invariant if $\rho$ were a constant, so its variation
must be proportional to the gradient $\partial_a \rho$.  Making the
change of variables~(\ref{nsym}) in the path integral thus gives
\begin{eqnarray}
0 &=& \int [d\phi']\, e^{-S[\phi']} - \int [d\phi]\, e^{-S[\phi]}
\nonumber\\
&=&
\frac{i\epsilon}{2\pi} \int [d\phi]\, e^{-S[\phi]} \int d^2\sigma\,
j^a(\sigma)
\partial_a \rho(\sigma).  \label{noe1}
\end{eqnarray}
The unknown coefficient $j^a(\sigma)$ comes from the variation of
the measure and the action, both of which are local, and so it must
be a local function of the fields and their derivatives.
Taking the function $\rho$ to be nonzero only in a small region
allows us to integrate by parts; also, the identity~(\ref{noe1})
remains valid if we add arbitrary distant insertions
`$\ldots$'.\footnote {Our convention is that `$\ldots$' refers to
distant insertions used to prepare a general initial and final state
but which otherwise play no role, while
${\cal F}$ is a general insertion in the region of interest.}
We thus derive
\begin{equation}
\partial_a j^a = 0
\end{equation}
as an operator equation.  This is Noether's theorem.\\[3pt]
{\bf Exercise:}  Use this to derive the classical Noether theorem
in the form usually found in textbooks.  That is,
assume that $S[\phi] = \int d^2 \sigma\,
L(\phi(\sigma), \partial_a \phi(\sigma))$ and ignore the variation
of the measure.  Invariance of the action implies that the variation
of the Lagrangian density is a total derivative, $\delta L
= \epsilon \partial_\mu K^\mu$ under a symmetry
transformation~(\ref{sym}).  Then the classical result is
\begin{equation}
j^\mu = 2\pi i \Biggl( \frac{\partial L}{\partial \phi_{\alpha,\mu}}
\delta \phi_\alpha - K^\mu \Biggr).
\end{equation}
The extra factor of $2\pi i$ is
conventional in conformal field theory. The derivation we have given
is the quantum version of Noether's theorem, and assumes that the path
integral can in fact be defined in a way consistent with the symmetry.

Now to derive the Ward identity, take any closed contour $C$, and
let $\rho(\sigma) = 1$ inside $C$ and 0 outside C.  Also, include in
the path integral some general local operator ${\cal A}(z_0,\bar
z_0)$ at a point $z_0$ {\it inside} $C$, and the usual
distant insertions `$\ldots$'.  Proceeding as above we obtain
the operator relation
\begin{equation}
\frac{1}{2\pi}\oint_C (d\sigma^2 j^1 - d\sigma^1 j^2)\,
{\cal A}(z_0,\bar z_0)\ =\ - i
\delta {\cal A}(z_0,\bar z_0).
\end{equation}
This relates the
integral of the current around any operator to the variation of the
operator.  In complex coordinates, the left-hand side is
\begin{equation}
\frac{1}{2\pi i} \oint_C (dz\, j - d\bar z \, \bar j) \,
{\cal A}(z_0,\bar z_0), \label{ward}
\end{equation}
where the contour runs counterclockwise and we abbreviate $j_z$ to
$j$ and $j_{\bar z}$ to $\bar j$.
Finally, in conformal
field theory it is usually the case that $j$ is analytic
and $\bar j$ antianalytic, except for singularities at the
other fields, so that the integral~(\ref{ward}) just picks out the
residues.  Thus,
\begin{equation}
- i \delta {\cal A}(z_0,\bar z_0)\ =\ {\rm Res}_{z \to z_0}\, j(z)
{\cal A}(z_0,\bar z_0) + {\rm \overline{Res}}_{\bar z \to \bar
z_0}\,
\bar j(\bar z) {\cal A}(z_0,\bar z_0). \label{wardres}
\end{equation}
Here `Res' and `$\overline{\rm Res}$' pick out the coefficients
of $(z - z_0)^{-1}$ and $(\bar z - \bar z_0)^{-1}$ respectively.
This form of the Ward identity is particularly convenient in CFT.

It is important to note that Noether's theorem and
the Ward identity are local properties that do not depend on
whatever boundary conditions we might have far away, not even whether
the latter are invariant under the symmetry.  In particular, since
the function $\rho(\sigma)$ is nonzero only in the interior of $C$,
the symmetry transformation need only be defined there.

\subs{Conformal Invariance}

Systems at a critical point are invariant under overall rescalings
of space, $z \to z' = a z$; if the system is also
rotationally invariant, $a$ can be complex.
These transformations rescale the metric,
\begin{equation}
ds^2 \ =\ d\sigma^a d\sigma^a\ =\ dzd\bar z \quad \to\quad dz'd\bar
z'
\ =\ |a|^2 ds^2.
\end{equation}
Under fairly
broad conditions, a scale invariant system will also be invariant
under the larger symmetry of conformal transformations,
which also will play a central role in string
theory.  These are transformations
$z \to z'(z,\bar z)$ which rescale the metric by a {\it
position-dependent} factor:
\begin{equation}
ds^2 \to \Omega^2(z,\bar z) ds^2.
\end{equation}
Such a transformation will leave invariant ratios of lengths
of infinitesimal vectors located at the same point, and so also
angles between them.  In complex coordinates, it is easy to see that
this requires that $z'$ be an analytic function of
$z$,\footnote
{A antianalytic function $z' = f(\bar z)$ also works, but changes
the orientation.  In most string theories,
including the ones of greatest interest, the orientation is fixed.}
\begin{equation}
z' = f(z), \qquad \bar z' = \bar f(\bar z).
\end{equation}
A theory with this invariance is termed a conformal field theory
(CFT).

The free action~(\ref{xact2}) is conformally invariant with
$X$ transforming as a scalar,
\begin{equation}
X'(z',\bar z') = X(z,\bar z),
\end{equation}
the transformation of $d^2 z$ offsetting that of the derivatives.
For an infinitesimal transformation, $z' = z + \epsilon g(z)$,
we have $\delta X = - g(z) \partial X - \bar g(\bar z) \bar\partial
X$, and Noether's theorem gives the current
\begin{equation}
j(z) = i g(z) T(z), \qquad \bar j(\bar z) = i \bar g(\bar z) \tilde
T(\bar z),
\end{equation}
where\footnote
{Many students asked why $T$ automatically came out normal-ordered.
The answer is simply that in this particular case all ways of
defining the product (at least all rotationally invariant
renormalizations) give the same result; they could differ at most by a
constant, but this must be zero because $T$ transforms by a phase
under rotations.  It was also asked how one knows that the {\it
measure} is conformally invariant; this is evident {\it a posteriori}
because the conformal current is indeed conserved.}
\begin{equation}
T(z) = -\frac{1}{2} :\! \partial X \partial
X \! :, \qquad \tilde T(\bar z) = -\frac{1}{2} :\! \bar\partial X
\bar\partial X \! :\ .
\label{xt}
\end{equation}
Because $g(z)$ and $\bar g(\bar z)$ are linearly independent,
both terms in the divergence $\bar\partial j - \partial \bar j$
must vanish independently,
\begin{equation}
\bar\partial j \ =\ \partial \bar j \ =\ 0,
\end{equation}
as is indeed the case.
The Noether current for a rigid translation is
the energy-momentum tensor, $j_a = i T_{ab}
\delta\sigma^b$ (the $i$ from CFT conventions), so we have
\begin{equation}
T_{zz} = T(z), \qquad T_{\bar z \bar z} = \tilde T(\bar z), \qquad
T_{z\bar z} = T_{\bar z z} = 0.
\end{equation}
With the vanishing of $T_{z\bar z} = T_{\bar z z}$, the conservation
law $\partial_a T^a\!_b = 0$ implies that $T_{zz}$ is analytic and
$T_{\bar z \bar z}$ is antianalytic; this is a general result in CFT.
By the way, $T_{zz}$ and $T_{\bar z \bar z}$ are in no sense
conjugate to one another (we will see, for example, that in they act
on completely different sets of oscillator modes),
so I use
a tilde rather than a bar on $\tilde T$.
The transformation of $X$, with the Ward identity~(\ref{wardres}),
implies the operator product
\begin{equation}
T(z) X(0) = \frac{1}{z} \partial X(0) + {\rm analytic}, \qquad
\tilde T(\bar z) X(0) = \frac{1}{\bar z} \bar\partial X(0)
 + {\rm analytic} .
\end{equation}
This is readily verified from the specific form~(\ref{xt}), and one
could have used it to derive the form of $T$.

For a general operator $\cal A$, the variation under rigid
translation is just $-\delta \sigma^a \partial_a {\cal A}$, which
determines the
$1/z$ term in the $T{\cal A}$ OPE.  We usually deal with operators
which are eigenstates of the rigid rescaling plus rotation $z' =
a z$:
\begin{equation}
{\cal A}' (z',\bar z')\ =\ a^{-h} \bar a^{-\tilde h}
{\cal A} (z,\bar z).
\end{equation}
The $(h,\tilde h)$ are the {\it weights} of
${\cal A}$.  The sum $h+\tilde h$ is the dimension of $\cal A$,
determining its behavior under scaling, while $h - \tilde h$ is the
spin, determining its behavior under rotations.  The Ward identity
then gives part of the OPE,
\begin{equation}
T(z) {\cal A}(0,0)\ =\ \ldots + \frac{h}{z^2} {\cal A}(0,0)
+ \frac{1}{z} \partial {\cal A}(0,0) + \ldots \label{21op}
\end{equation}
and similarly for $\tilde T$.  A special
case is a {\it tensor} or {\it primary} operator ${\cal O}$, which
transforms under general conformal transformations as
\begin{equation}
{\cal O}' (z',\bar z')\ =\ (\partial_z z')^{-h}
(\partial_{\bar z} \bar z')^{-\tilde h} {\cal O} (z,\bar z).
\end{equation}
This is equivalent to the OPE
\begin{equation}
T(z) {\cal O}(0,0) = \frac{h}{z^2} {\cal O}(0,0)
+ \frac{1}{z} \partial {\cal O}(0,0) + \ldots \ ,  \label{prop}
\end{equation}
the more singular terms in the general OPE~(\ref{21op}) being
absent.  In the free $X$ CFT, one can check that $\partial X$
is a tensor of weight $(1,0)$, $\bar\partial X$ a tensor of weight
$(0,1)$, and $:\! e^{ikX} \! :$ a tensor of weight $\frac{1}{2}
k^2$, while $\partial^2 X$ has weight $(2,0)$ but is not a tensor.

For the energy-momentum tensor with itself one finds for the free
$X$ theory
\begin{equation}
T(z) T(0) = \frac{1}{2 z^4} + \frac{2}{z^2} T(0) + \frac{1}{z}
\partial T(0) + {\rm analytic}, \label{ttop}
\end{equation}
and similarly for $\tilde T$,\footnote
{One easily sees that the $T \tilde T$ OPE is analytic.
By the way, unless otherwise stated OPE's hold only at non-zero
separation, ignoring possible delta functions.  For all of the
applications we will have the latter do not matter.  Occasionally it
is useful to include the delta functions, but in general these depend
partly on definitions so one must be careful.}
so this is {\it not} a tensor.  Rather, the OPE~(\ref{ttop})
implies the transformation law
\begin{equation}
\delta T(z) = \frac{1}{12} \partial^3_z g(z) - 2 \partial_z g(z)
T(z) - g(z) \partial_z T(z). \label{ttrans}
\end{equation}
More generally, the $TT$ OPE in any CFT is of the form
\begin{equation}
T(z) T(0) = \frac{c}{2 z^4} + \frac{2}{z^2} T(0) + \frac{1}{z}
\partial T(0) + {\rm analytic}, \label{ttop2}
\end{equation}
with $c$ a constant known as the {\it central charge}.
The central charge of a free boson is~1; for $D$ free
bosons it is $D$.  The finite form of the transformation
law~(\ref{ttrans}) is
\begin{equation}
(\partial_z z')^2 T'(z') = T(z) + \frac{c}{12} \{ z',z\} ,
\label{ttra}
\end{equation}
where $\{ f, z\} $ denotes the {\it Schwarzian derivative},
\begin{equation}
\{ f,z \} = \frac{2 \partial_z^3 f \partial_z f
- 3 \partial_z^2 f \partial_z^2 f}{2\partial_z f \partial_z f}.
\label{schw}
\end{equation}
The corresponding form holds for $\tilde T$,
possibly with a different central charge $\tilde c$.

\subs{Mode Expansions}

For an analytic or antianalytic operator we can make a Laurent
expansion,
\begin{equation}
T(z) = \sum_{m=-\infty}^{\infty} \frac{ L_{m} }{z^{m+2}}, \qquad
\tilde T(\bar z) = \sum_{m=-\infty}^{\infty} \frac{ \tilde{L}_{m}
}{\bar z^{m+2}}.
\label{soro}
\end{equation}
The Laurent coefficients, known as the {\it Virasoro} generators,
are given by the contour integrals
\begin{equation}
L_m = \oint_C \frac{dz}{2\pi i }\, z^{m+1} T(z), \qquad
\tilde{L}_m = -\oint_C \frac{d\bar z}{2\pi i }\, \bar z^{m+1}
\tilde T(\bar z), \label{cont}
\end{equation}
where $C$ is any contour encircling the origin.  This expansion
has a simple and important interpretation \cite{FHZ}.
Defining any monotonic
time variable, one can slice open a path integral along the
constant-time curves to recover a Hamiltonian description.  In
particular, let `time' be $\ln |z|$, running radially outward from
$z=0$.  This may seem odd, but is quite natural in CFT---in terms
of the conformally equivalent coordinate $w$ defined $z = e^{-iw}$,
an annular region around $z=0$ becomes a cylinder, with Im$(w)$
being the time and Re$(w)$ being a spatial coordinate with
periodicity 2$\pi$.  Thus, the radial time slicing is equivalent
to quantizing the CFT on a finite periodic space; this is what
will eventually be interpreted as the quantization of a closed
string.  The `$+2$'s in the exponents~(\ref{soro}) come from the
conformal transformation of $T$, so that in the $w$ frame $m$ just
denotes the Fourier mode; for an analytic field of weight $h$ this
becomes `$+h$'.

In the Hamiltonian form, the Virasoro generators become
operators in the ordinary sense.  Since by analyticity the
integrals~(\ref{cont}) are independent of $C$, they are actually
conserved charges, the charges associated with the conformal
transformations.  It is an important fact that the OPE of currents
determines the algebra of the corresponding charges.  Consider
charges $Q_i$, $i = 1,2$:
\begin{equation}
Q_i\{C\} = \oint_C \frac{dz}{2\pi i }j_{i}.
\end{equation}
Then we have
\begin{equation}
Q_1\{C_1\} Q_2\{C_2\} - Q_1\{C_3\} Q_2\{C_2\} =
[Q_1, Q_2]\{C_2\}
\end{equation}
The charges on the left are defined by the contours shown in
fig.~1a; when we slice open the path integral, operators are
time-ordered, so the difference of contours generates the
commutator.
\begin{figure}
\begin{center}
\leavevmode
\epsfbox{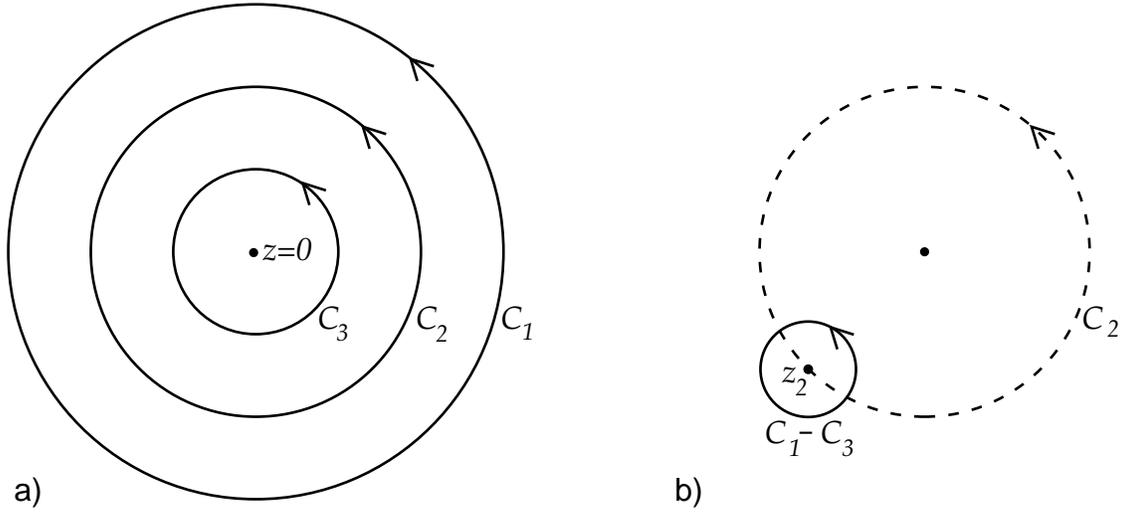}
\end{center}
\caption[]{a) Contours centered on $z=0$.  b) For given
$z_2$ on contour $C_2$, contour $C_1 - C_3$ is contracted.}
\end{figure}
Now, for a given point $z_2$ on the contour $C_2$,
we can deform the difference of the $C_1$ and $C_3$ contours as
shown in fig.~1b, with the result
\begin{equation}
[Q_1, Q_2]\{C_2\} = \oint_{C_2} \frac{dz_2}{2\pi i }\,
{\rm Res}_{z \to z_2}\,j_1(z)j_2(z_2)
\end{equation}

Applying this to the Virasoro generators gives the {\it Virasoro
algebra,}
\begin{equation}
[L_m, L_n] = (m-n) L_{m+n} + \frac{c}{12} (m^3 - m) \delta_{m+n,0}.
\label{vira}
\end{equation}
The $\tilde L_m$ satisfy the same algebra with central charge
$\tilde c$.  For the Laurent coefficients of an analytic tensor
field
${\cal O}$ of weight
$(h,0)$, one finds from the OPE~(\ref{prop}) the commutator
\begin{equation}
[L_m, {\cal O}_n] = ([h-1]m-n) {\cal O}_{m+n}.
\end{equation}
Note that commutation with $L_0$ is diagonal and proportional to
$-n$.  Modes ${\cal O}_n$ for $n > 0$ reduce $L_0$ and are termed
lowering operators, while modes ${\cal O}_n$ for $n < 0$ increase
$L_0$ and are termed raising operators.  From the OPE~(\ref{prop})
and the definitions, we see that a tensor operator is annihilated
by all the lowering operators,\footnote
{Often one deals with different copies of the Virasoro algebra
defined by Laurent expansions in different coordinates $z$, so
I like to put a `$\cdot$' between the generator and the operator
as a reminder that the generators are defined in the coordinate
centered on the operator.}
\begin{equation}
L_n \cdot {\cal O} = 0,\quad n > 0.
\end{equation}
For an arbitrary operator, it follows from the OPE~(\ref{21op})
that
\begin{equation}
L_0 \cdot {\cal A} = h {\cal A}, \qquad
\tilde L_0 \cdot {\cal A} = \tilde h {\cal A}, \qquad
L_{-1} \cdot {\cal A} = \partial {\cal A}, \qquad
\tilde L_{-1} \cdot {\cal A} = \bar\partial {\cal A}. \label{l01}
\end{equation}
Note that $L_0 + \tilde L_0$ is the generator of scale
transformations, or in other words of radial
time translations.  It differs from the Hamiltonian $H$ of the
cylindrical $w$ coordinate system by an additive constant from the
non-tensor behavior of $T$,
\begin{equation}
H = L_0 + \tilde L_0 -\frac{c + \tilde c}{24}. \label{canon}
\end{equation}
Similarly, $L_0 - \tilde L_0$ measures the spin, and is equal to
the spatial translation generator in the $w$ frame, up to an
additive constant.

For the free $X$ CFT, the Noether current of translations is
$(i\partial X(z), i\bar \partial X(\bar z))$.  Again, the
components are separately analytic and antianalytic, which
signifies the existence of an enlarged symmetry $X \to X + y(z) +
\bar y(\bar z)$.
Define the modes
\begin{equation}
i\partial X(z) =
\sum_{m=-\infty}^{\infty} \frac{\alpha_m}{z^{m+1}}, \qquad
i\bar\partial X(\bar z) =\sum_{m = -\infty  }^\infty
\frac{\tilde\alpha_m}{\bar z^{m+1}}. \label{lex1}
\end{equation}
 From the OPE
\begin{equation}
i\partial X(z)\, i\partial X(0) = \frac{1}{z^2} + {\rm analytic},
\end{equation}
we have the algebra
\begin{equation}
[\alpha_m, \alpha_n] = m \delta_{m+n,0},
\end{equation}
and the same for $\tilde \alpha_m$.  As expected for a free field,
this is a harmonic oscillator algebra for each mode; in terms of the
usual raising and lowering operators $\alpha_m \sim \sqrt m a$,
$\alpha_{-m} \sim \sqrt m a^\dagger$.
To generate the whole spectrum we
start from a state $|0,k\rangle$ which is annihilated by the $m > 0$
operators and is an eigenvector of the $m=0$ operators,
\begin{equation}
\alpha_m |0,k\rangle = \tilde \alpha_m |0,k\rangle = 0,
\ \ m>0, \qquad \alpha_0 |0,k\rangle = \tilde\alpha_0
|0,k\rangle = k |0,k\rangle\ .
\end{equation}
The rest of the spectrum is generated by the raising operators
$\alpha_m$ and $\tilde \alpha_m$ for $m < 0$.  Note that the
eigenvalues of $\alpha_0$ and $\tilde\alpha_0$ must be equal because
$X$ is single valued, $\oint (dz\, \partial X + d\bar z\,\bar\partial
X) = 0$; later we will relax this.

Inserting the expansion~(\ref{lex1}) into $T(z)$ and comparing
with the Laurent expansion gives
\begin{equation}
L_m \sim \frac{1}{2} \sum_{n=-\infty}^\infty \alpha_n \alpha_{m-n}.
\end{equation}
However, we must be careful about operator ordering.  The Virasoro
generators were defined in terms of the normal
ordering~(\ref{nord}), while for the mode expansion it is most
convenient to use a different ordering, in which all raising
operators are to the left of the lowering operators.  Both of
these procedures are generally referred to as normal ordering, but
they are in general different, so we might refer to the first as
`conformal normal ordering' and the latter as
`creation-annihilation normal ordering.'  Since conformal normal
order is our usual method, we will simply refer to it as
normal ordering.
We could develop a dictionary between these, but there are
several ways to take a short-cut.  Only for $m = 0$ do
non-commuting operators appear together, so we must have
\begin{eqnarray}
L_0 &=& \frac{1}{2} \alpha_0^2 + \sum_{n = 1}^\infty
\alpha_{-n} \alpha_{n} + A \nonumber\\
L_m &=& \frac{1}{2} \sum_{n=-\infty}^\infty \alpha_n \alpha_{m-n},
\qquad m \neq 0.
\end{eqnarray}
for some constant $A$.  Now use the Virasoro algebra as follows
\begin{equation}
(L_1 L_{-1} - L_{-1} L_1) |0,0\rangle\ =\ 2 L_0 |0,0\rangle
\ =\ 2A|0,0\rangle. \label{nocon}
\end{equation}
All terms on the left have $\alpha_m$ with $m \geq 0$ acting on
$|0,0\rangle$ and so must vanish; thus,
\begin{equation}
A = 0.
\end{equation}
Thus, a general state
\begin{equation}
\alpha_{-m_1} \ldots \alpha_{-m_p} \tilde\alpha_{-m'_1} \ldots
\tilde\alpha_{-m'_q} |0,k\rangle
\end{equation}
has
\begin{equation}
L_0 = \frac{1}{2} k^2 + {\sf L}, \qquad \tilde L_0 = \frac{1}{2}
k^2 + {\sf \tilde L}
\end{equation}
where the levels ${\sf L}$, ${\sf \tilde L}$ are the total oscillator
excitation numbers,
\begin{equation}
{\sf L} = m_1 + \ldots + m_p, \qquad
{\sf \tilde L} = m'_1 + \ldots + m'_q. \label{level}
\end{equation}

One needs to calculate the normal ordering constant $A$ often, so the
following heuristic-but-correct rules are useful:\\[3pt]
\label{rulesec}
1. Add the zero point energies, $\frac{1}{2} \omega$
for each bosonic mode and $- \frac{1}{2} \omega$ for each
fermionic.\\[3pt]
2. One encounters divergent sums of the form
$\sum_{n=1}^\infty (n - \theta)$, the $\theta$ arising when one
considers nontrivial periodicity conditions.  Define this to be
\begin{equation}
\sum_{n=1}^\infty (n - \theta) = \frac{1}{24} -
\frac{1}{8} (2 \theta - 1)^2. \label{zeta}
\end{equation}
I will not try to justify this, but it is the value given by
any conformally invariant renormalization.\\[3pt]
3. The above is correct in the cylindrical $w$
coordinate, but for $L_0$ we must add the non-tensor correction
$c/24$.\\[3pt] For the free boson, the modes are integer so we get
one-half of the sum~(\ref{zeta}) for $\theta = 0$, that is
$-\frac{1}{24}$, after step~2.  This is just offset by the correction
in step~3.  The zero-point sum in step~2 is a Casimir energy, from the
finite spatial size.  For a system of physical size $l$ we must
scale $H$ by $2\pi/l$, giving (including the left-movers) the
correct Casimir energy $-\pi/6l$.  For antiperiodic scalars one
gets the sum with $\theta = \frac{1}{2}$ and Casimir energy
$\pi/12 l$.

To get the mode expansion for $X$, integrate
the Laurent expansions~(\ref{lex1}).  Define first
\begin{equation}
X_L(z) = x_L - i \alpha_0 \ln z
+ i \sum_{m \neq 0} \frac{\alpha_m}{mz^{m}}, \qquad
X_R(\bar z) = x_R - i\tilde\alpha_0 \ln\bar z +
i\sum_{m \neq 0}
\frac{\tilde\alpha_m}{m\bar z^{m}}, \label{lex2}
\end{equation}
with
\begin{equation}
[ x_L, \alpha_0 ] = [x_R, \tilde\alpha_0] = i.
\end{equation}
These give
\begin{equation}
X_L(z) X_L(z') = -\ln (z-z') + {\rm analytic},
\qquad
X_R(\bar z) X_R(\bar z') = -\ln (\bar z - \bar z') + {\rm analytic}.
\end{equation}
Actually, these only hold modulo $i\pi$ as one can check, but we
will not dwell on this.\footnote{But it means that one sometimes
need to introduce `cocycles' to fix the phases of exponential
operators.}  In any case we are for the present only interested in
the sum,
\begin{equation}
X(z,\bar z) = X_L(z) + X_R(\bar z),
\end{equation}
for which the OPE $X(z,\bar z) X(0,0) \sim - \ln|z|^2$ is
unambiguous.

\subs{States and Operators}

Radial quantization gives rise to a natural
isomorphism between the state space of the CFT, in a periodic
spatial dimension, and the space of local operators.  Consider the
path integral with a local operator ${\cal A}$ at the origin,
no other operators inside the unit circle $|z| = 1$, and unspecified
operators and boundary conditions outside.  Cutting open the path
integral on the unit circle represents the path integral as an
inner product $\langle \psi_{\rm out} | \psi_{\rm in} \rangle$,
where $| \psi_{\rm in} \rangle$ is the incoming state produced by
the path integral at $|z| < 1$ and $| \psi_{\rm out} \rangle$
is the outgoing state produced by the path integral at $|z| > 1$.
More explicitly, separate the path integral over fields $\phi$ into
an integral over the fields outside the circle, inside the circle, and
on the circle itself; call these last $\phi_B$.  The outside integral
produces a result $\psi_{\rm out}(\phi_B)$, and the inside
integral a result $\psi_{\rm in}(\phi_B)$, leaving
\begin{equation}
\int [d\phi_B]\, \psi_{\rm out}(\phi_B) \psi_{\rm in}(\phi_B)\ .
\end{equation}
The incoming state depends on $\cal A$, so we denote it
more explicitly as $| \psi_{\cal A}\rangle$.  This is the
mapping from operators to states.  That is, integrating over the
fields on the unit disk, with fixed boundary values $\phi_B$ and with
an operator $\cal A$ at the origin, produces a result $\psi_{\cal A}
(\phi_B)$, which is a state in the Schrodinger representation.
{\it The mapping from operators to states is given by the path
integral on the unit disk.}  To see the inverse, take a state $| \psi
\rangle$ to be an eigenstate of $L_0$ and $\tilde L_0$.  Since $L_0 +
\tilde L_0$ is the radial Hamiltonian, inserting $| \psi \rangle$ on
the unit circle is equivalent to inserting $r^{-L_0 - \tilde L_0}
| \psi \rangle$ on
a circle of radius $r$.  Taking $r$ to be infinitesimal defines a
local operator which is equivalent to $| \psi \rangle$ on the unit
circle.\\[3pt]
{\bf Exercise:}  This all sounds a bit abstract, so here is a
calculation one can do explicitly.  The ground state of the free scalar
is $e^{-\sum_{m=1}^\infty m X_m X_{-m} / 2}$, where $X_m$ are the
Fourier modes of $X$ on the circle.  Derive this by canonical
quantization of the modes, writing them in terms of $X_m$ and
$\partial/\partial X_m$.  Obtain
it also by evaluating the path integral on the unit disk with $X$ fixed
on the boundary and no operator insertions.  Thus the ground state
corresponds to the unit operator.

Usually one does not actually evaluate a path integral as above,
but uses indirect arguments.  Note that if $Q$ is any conserved
charge, the state $Q | \psi_{\cal A} \rangle$ corresponds to
the operator $Q \cdot {\cal A}$, as shown in fig.~2.
\begin{figure}
\begin{center}
\leavevmode
\epsfbox{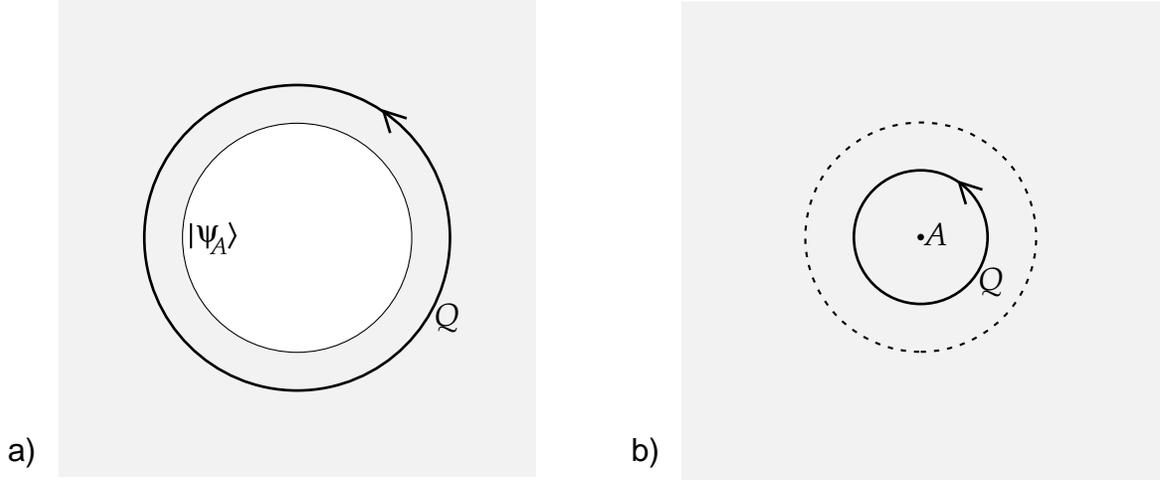}
\end{center}
\caption[]{a) World-sheet (shaded) with
state $| \psi_{\cal A} \rangle$ on the boundary circle, acted upon by
$Q$.  b) Equivalent picture: the unit disk with operator ${\cal A}$
has been sewn in along the dotted line, and $Q$ contracted around the
operator.}
\end{figure}
Now, in
the free theory consider the case that ${\cal A}$ is the unit
operator and let
\begin{equation}
Q = \alpha_m = \oint_C \frac{dz}{2\pi} z^m \partial X,\qquad m
\geq 0.  \label{laurcont}
\end{equation}
With no operators inside the disk, $\partial X$ is analytic and the
integral vanishes for $m \geq 0$.  Thus, $\alpha_m
|\psi_1 \rangle = 0$, $m\geq 0$, which establishes
\begin{equation}
1 \ \leftrightarrow\ |0,0\rangle
\end{equation}
as found directly in the exercise.  Proceeding as above one finds
\begin{equation}
:\! e^{i k X} \! : \ \leftrightarrow\ |0,k\rangle,
\label{eikx}
\end{equation}
and for the raising operators, evaluating the contour
integral~(\ref{laurcont}) for $m < 0$ gives
\begin{equation}
i \frac{1}{(k-1)!} \partial^k X
\ \leftrightarrow\ \alpha_{-k} ,\quad k\geq 1 \ , \label{stoo}
\end{equation}
and in parallel for the tilded modes.
That is, the state obtained by acting with raising operators on
(\ref{eikx}) is given by the product of the exponential with the
corresponding derivatives of $X$; the product automatically comes
out normal ordered.

The state corresponding to a tensor field ${\cal O}$ satisfies
\begin{equation}
L_m | \psi_{\cal O} \rangle = 0, \qquad m > 0.
\end{equation}
This is known as a {\it highest weight} or {\it primary} state.
For almost all purposes one is interested in highest-weight
representations of the Virasoro algebra, built by acting on a given
highest weight state with the $L_m$, $m < 0$.

The state-operator mapping gives a simple derivation of the OPE,
shown in fig.~3.
\begin{figure}
\begin{center}
\leavevmode
\epsfbox{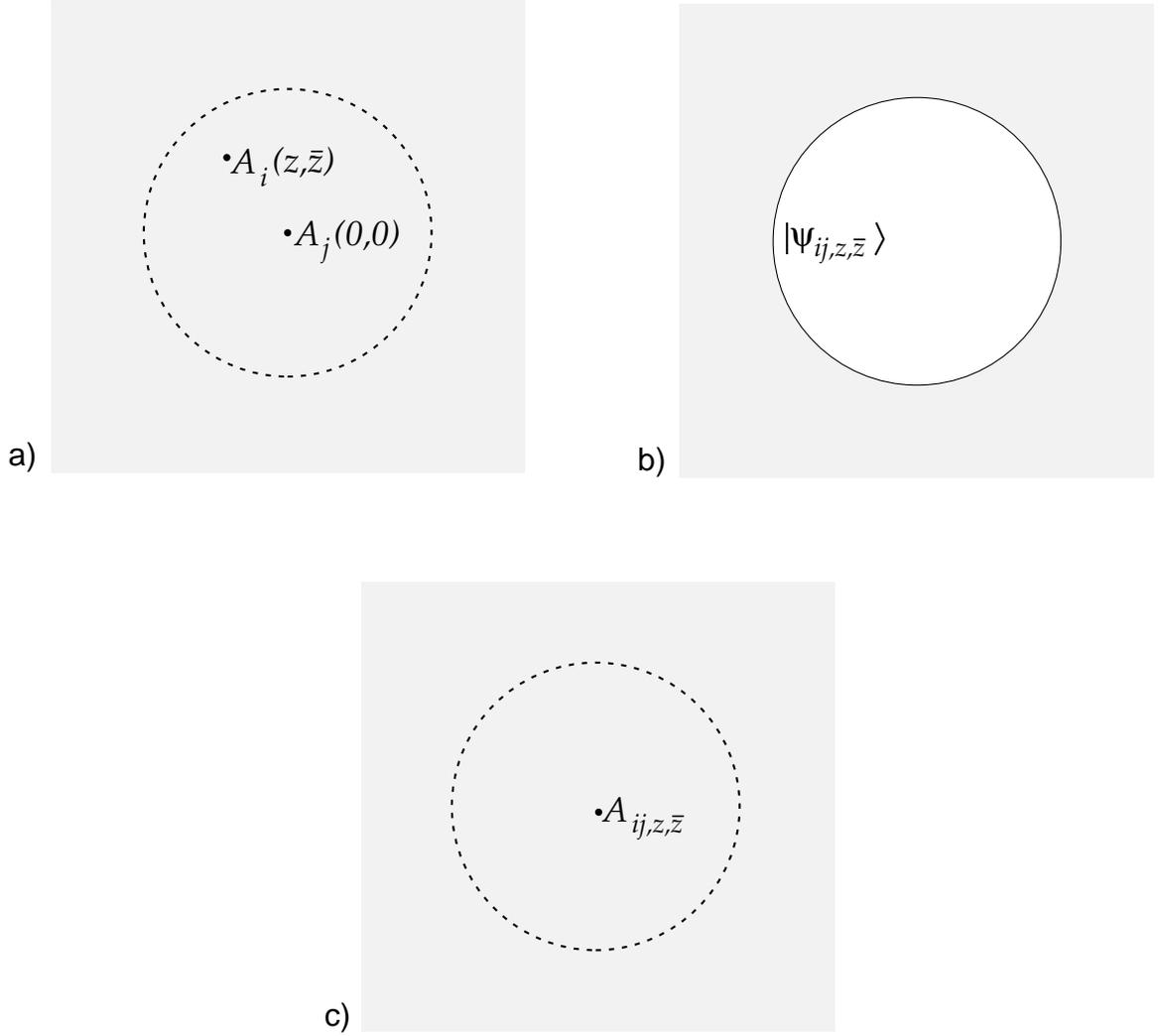}
\end{center}
\caption[]{a) World-sheet with two local operators. b)
Integration over fields on the interior of the disk produces
boundary state $| \psi_{ij,z,\bar z} \rangle$. c) Sewing in a disk
with the corresponding local operator.  Expanding in operators of
definite weight gives the OPE.}
\end{figure}
Consider the product ${\cal A}_i(z,\bar z)
{\cal A}_j(0,0)$, $|z|<1$.  Integrating the fields inside the unit
circle generates a state on the unit circle, which we might call
$| \psi_{ij,z,\bar z} \rangle$.  Expand in a complete set,
\begin{equation}
| \psi_{ij,z,\bar z} \rangle = \sum_k c^k \!_{ij}(z,\bar z)
| \psi_k \rangle.
\end{equation}
Finally use the mapping to replace $| \psi_k \rangle$
on the unit circle with ${\cal A}_k$ at the origin, giving the
general OPE~(\ref{gope}).  The claimed convergence is just the
usual convergence of a complete set in quantum mechanics.  The
construction is possible as long as there are no other operators
with $|z'| \leq |z|$, so that we can cut on a circle of radius
$|z| + \epsilon$.

Incidentally, applying a rigid rotation and scaling to both sides
of the general OPE determines the $z$-dependence of the
coefficient functions,
\begin{equation}
{\cal A}_i(z,\bar z) {\cal A}_j(0,0)
= \sum_k z^{h_k - h_i - h_j} {\bar z}^{\tilde h_k - \tilde h_i -
\tilde h_j} c^k\!_{ij}{\cal A}_k(0,0). \label{gope2}
\end{equation}
 From the full conformal symmetry one learns much more: all the
$c^k\!_{ij}$ are determined in terms of those of the primary
fields.

For three operators, ${\cal A}_i(0) {\cal A}_j(1) {\cal A}_k(z)$,
the regions of convergence of the $z \to 0$ and $z \to 1$ OPE's
($|z| < 1$ and $|1-z| < 1$) overlap.  The coefficient of ${\cal
A}_m$ in the triple product can then be written as a sum involving
$c^l\!_{ik}c^m\!_{lj}$ or as a sum involving $c^l\!_{jk}c^m\!_{li}$.
Associativity requires these sums to be equal; this is represented
schematically in fig.~4.
\begin{figure}
\begin{center}
\leavevmode
\epsfbox{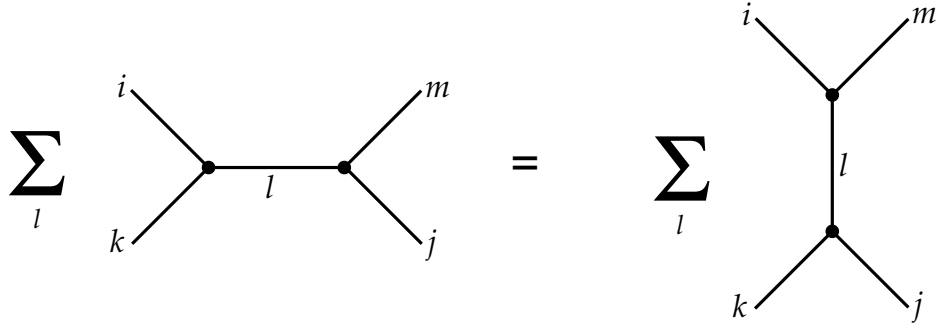}
\end{center}
\caption[]{Schematic picture of OPE associativity.}
\end{figure}

A {\it unitary} CFT is one that has a positive inner product
$\langle
\!\langle \ |\ \rangle$; the double bracket is to distinguish it
from a different inner product to be defined later.  Also, it is
required that $L_m^\dagger = L_{-m}$, $\tilde L_m^\dagger =
\tilde L_{-m}$.   The
$X$ CFT is unitary with
\begin{equation}
\langle\!\langle 0,k|0,k'\rangle = 2\pi \delta(k - k')
\end{equation}
and $\alpha_m^\dagger = \alpha_{-m}$, $\tilde \alpha_m^\dagger =
\tilde \alpha_{-m}$; this implicitly defines the inner product of
all higher states.  Unitary CFT's are highly constrained; I will
derive here a few of the basic results, and mention others later.

The first constraint is that any state in a unitary highest weight
representation must have $h, \tilde h \geq 0$.  Consider first the
highest weight state itself, $|{\cal O}\rangle$.
The Virasoro algebra gives
\begin{equation}
2 h_{\cal O} \langle\!\langle {\cal O}|{\cal O}\rangle
= 2 \langle\!\langle {\cal O}|L_0|{\cal O}\rangle
= \langle\!\langle {\cal O}| [ L_1, L_{-1} ]|{\cal O}\rangle
= \| L_{-1} |{\cal O}\rangle \|^2 \geq 0,
\end{equation}
so $h_{\cal O} \geq 0$.  All other states in the representation,
obtained by acting with the raising generators,
have higher weight so the result follows.
It also follows that if $h_{\cal O} = 0$ then
$L_{-1} \cdot {\cal O} = \tilde L_{-1} \cdot {\cal O}$.  The
relation~(\ref{l01}) thus implies that $\cal O$ is independent of
position; general principle of quantum field theory then
require $\cal O$ to be a $c$-number.  That is, the
 unit operator is the only
(0,0) operator.  In a similar way, one finds that an operator in
a unitary CFT is analytic if and only if $\tilde h = 0$,
and antianalytic if and only if $h = 0$.\\[3pt]
{\bf Exercise:} Using the above argument with the commutator
$[ L_n, L_{-n} ]$, show that $c, \tilde c \geq 0$ in a unitary
CFT.  In fact, the only CFT with
$c = 0$ is the trivial one, $L_n = 0$.

\subs{Other CFT's}

Now we describe briefly several other CFT's of interest.  The
first is given by the same action~(\ref{xact2}) as the earlier $X$
theory, but with energy-momentum tensor~\cite{CT}
\begin{equation}
T(z) = -\frac{1}{2} :\! \partial X \partial X \! : +\,
\frac{Q}{2} \partial^2 X, \qquad
\tilde T(\bar z) = -\frac{1}{2} :\! \bar\partial X
\bar\partial X \! : +\, \frac{Q}{2} \bar\partial^2 X. \label{ldcft}
\end{equation}
The $TT$ operator product is still of the general
form~(\ref{ttop2}),
but now has central charge $c = 1 + 3 Q^2$. The change
in $T$ means that $X$ is no longer a scalar,
\begin{equation}
\delta X = -( g \partial X + \bar g \bar\partial X)
 - \frac{Q}{2} (\partial g + \bar\partial \bar g). \label{LDdelx}
\end{equation}
Exponentials $:\! e^{ikX} \! :$ are still tensors,
but with weight $\frac{1}{2} (k^2 + i k Q)$.  One notable change is
in the state-operator mapping.  The translation current
$j = i \partial X$ is no longer a tensor, $\delta j = - g \partial j
- j \partial g - iQ \partial^2 g /2$.  The finite form is\footnote
{To derive this, and the finite transformation~(\ref{ttra}) of $T$,
you can first write the most general form which has the correct
infinitesimal limit and is appropriately homogeneous in $z$ and $z'$
indices, and fix the few resulting constants by requiring proper
composition under $z \to z' \to z''$.}
\begin{equation}
(\partial_z z') j_{z'}(z') = j_z(z) - \frac{iQ}{2}
\frac{\partial^2_z z'}{\partial_z z'}.
\end{equation}
Applied to the cylinder frame $z' = w = i\ln z$ this gives
\begin{equation}
\frac{1}{2\pi i} \int dw\, j_w = \alpha_0 + \frac{iQ}{2}. \label{lmom}
\end{equation}
Thus a state $|0,k \rangle$ which whose canonical momentum (defined
on the left) is $k$ corresponds to the operator
\begin{equation}
:\! e^{ikX + QX/2} \! :\ . \label{lindvo}
\end{equation}
Note that $i\alpha_0$ just picks out the exponent of the
operator, so $\alpha_0 = k - i Q/2$.

The mode expansion of the $m \neq 0$ generators is
\begin{equation}
L_m = \frac{1}{2} \sum_{n=-\infty}^\infty \alpha_n \alpha_{m-n}
+ \frac{i Q}{2} (m+1) \alpha_m ,
\qquad m \neq 0,
\end{equation}
the last term coming from the $\partial^2 X$ term in $T$.
For $m=0$ the result is
\begin{eqnarray}
L_0 &=&
\frac{1}{8} (2\alpha_0 + iQ)^2
+ \frac{Q^2}{8} + \sum_{n=1}^\infty \alpha^\mu_{-n} \alpha_{\mu,n}
\nonumber\\
&=& \frac{1}{2} k^2
+ \frac{Q^2}{8} + \sum_{n=1}^\infty \alpha^\mu_{-n} \alpha_{\mu,n} .
\end{eqnarray}
The constant in the first line can be obtained from the
$z^{-2}$ term in the OPE of $T$
with the vertex operator~(\ref{lindvo}); this is a quick way to
derive or to check normal-ordering constants.  In the second line,
expressed in terms of the `canonical' momentum $k$ it agrees
with our heuristic rules.

This CFT has a number of applications in string theory, some
of which we will encounter.  Let me also mention a slight variation,
\begin{equation}
T(z) = -\frac{1}{2} :\! \partial X \partial X \! : +\,
i\frac{\kappa}{2} \partial^2 X, \qquad
\tilde T(\bar z) = -\frac{1}{2} :\! \bar\partial X
\bar\partial X \! : -\, i\frac{\kappa}{2} \bar\partial^2 X,
\label{hex}
\end{equation}
with central charge $c = 1 - 3 \kappa^2$.  With the earlier
transformation~(\ref{LDdelx}), the variation of $X$
contains a constant piece under rigid scale transformations ($g$ a
real constant).  In other words, one can regard $X$ as the
Goldstone boson of spontaneously broken scale invariance.  For the
theory~(\ref{hex}), the variation of $X$
contains a constant piece under rigid {\it rotations} ($g$ an
imaginary constant), and $X$ is the
Goldstone boson of spontaneously broken rotational invariance.
This is not directly relevant to string theory (the $i$ in the
energy-momentum tensor makes the theory non-unitary) but occurs for
real membranes (where the unitarity condition is not relevant
because both dimensions are spatial).  In particular the
CFT~(\ref{hex}) describes hexatic membranes,\footnote
{I would like to thank Mark Bowick and Phil Nelson for educating me
on this subject.} in which the rotational
symmetry is broken to ${\bf Z}_6$.  The unbroken discrete symmetry
plays an indirect role in forbidding certain nonlinear couplings
between the Goldstone boson $X$ and the membrane
coordinates.

Another simple variation on the free boson is to make it {\it
periodic}, but we leave this until section~3 where we will discuss
some interesting features.

Another family of free CFT's involves two anticommuting fields
with action
\begin{equation}
S = \frac{1}{2\pi} \int d^2z \,
\{ b \bar\partial c + \tilde{b} \partial \tilde{c} \}. \label{gact}
\end{equation}
The equations of motion are
\begin{equation}
\bar\partial c(z) = \bar\partial b(z) =
\partial \tilde{c} (\bar z) = \partial \tilde{b} (\bar z) = 0,
\end{equation}
so the fields are respectively analytic and antianalytic.
The operator products are readily found as before, with appropriate
attention to the order of anticommuting variables,
\begin{equation}
b(z)c(0) \sim \frac{1}{z}, \qquad
c(z)b(0) \sim \frac{1}{z}, \qquad
\tilde b(\bar z)\tilde c(0) \sim \frac{1}{\bar z},
\qquad
\tilde c(\bar z)\tilde b(0) \sim \frac{1}{\bar z}.
\end{equation}

We focus again on the analytic part; in fact the
action~(\ref{gact}) is a sum, and can be regarded as two
independent CFT's.  The action is conformally invariant if
$b$ is a $(\lambda, 0)$ tensor, and $c$ a $(1-\lambda, 0)$
tensor; by interchange of $b$ and $c$ we can assume $\lambda$
positive.  The corresponding energy-momentum tensor is
\begin{equation}
T(z) = :\! (\partial b) c  \! : - \lambda :\!  \partial( bc ) \! :\ .
\end{equation}
One finds that the $TT$ OPE has the usual form with
\begin{equation}
c = - 3(2 \lambda - 1)^2 + 1.  \label{bccc}
\end{equation}

The fields have the usual Laurent expansions
\begin{equation}
b(z) = \sum_{m=-\infty}^{\infty} \frac{b_m}{z^{m+\lambda}}, \qquad
c(z) = \sum_{m=-\infty}^{\infty}
\frac{c_m}{z^{m+1-\lambda}},
\label{bclaur}
\end{equation}
giving rise to the anticommutator
\begin{equation}
\{ b_m, c_n \} = \delta_{m+n,0}.
\end{equation}
Also, $\{ c_m, c_n \} = \{ b_m, b_n \} = 0$.
Because of the $m=0$ modes there are two natural ground states,
$|\!\uparrow\rangle$ and
$| \!\downarrow \rangle$.  Both are annihilated by $b_m$ and
$c_m$ for $m > 0$, while
\begin{equation}
b_0 | \!\downarrow \rangle = 0, \qquad c_0 |\!\uparrow\rangle = 0.
\label{ghvac}
\end{equation}
These are related $ |\!\uparrow\rangle = c_0  |\! \downarrow
\rangle$,
$ |\! \downarrow \rangle = b_0 |\!\uparrow\rangle$.  With the
antianalytic theory included, there are also the zero modes
$\tilde b_0$ and $\tilde c_0$ and so four ground
states---$|\!\downarrow\downarrow\rangle$, etc.

The Virasoro generators in terms of the modes are
\begin{eqnarray}
L_0 &=& \sum_{n=1}^{\infty}
n (b_{-n} c_n + c_{-n} b_n)
\ -\ \frac{\lambda(\lambda - 1)}{2} \nonumber\\
L_m &=& \sum_{n=-\infty}^{\infty}
\{\lambda m - n \} b_n c_{m-n} , \quad m \neq 0 \ . \label{ghL}
\end{eqnarray}
The ordering constant is found as before.  Two sets
($b$ and $c$) of integer anticommuting modes give $\frac{1}{12}$
at step~2, and the central charge correction then gives the result
above.

The state-operator mapping is a little tricky.  Let
$\lambda$ be an integer, so that the Laurent
expansion~(\ref{bclaur}) has no branch cut.  For the unit operator
the fields are analytic at the origin, so
\begin{equation}
b_m |\psi_1 \rangle = 0, \quad m \geq 1-\lambda,
\qquad
c_m |\psi_1 \rangle = 0, \quad m \geq \lambda.
\end{equation}
Thus, the unit state is in general not one of the ground states,
but rather
\begin{equation}
|\psi_1 \rangle = b_{-1} b_{-2}
\ldots b_{1 - \lambda}|\! \downarrow \rangle,
\end{equation}
up to normalization.  Also, we have the dictionary
\begin{equation}
b_{-m}\ \leftrightarrow\ \frac{1}{(m-\lambda)!}
\partial^{m-\lambda}b, \qquad
c_{-m} \ \leftrightarrow\  \frac{1}{(m+\lambda-1)!}
\partial^{m+\lambda-1} c.
\end{equation}
Thus we have, taking the value $\lambda =
2$ which will be relevant later,
\begin{equation}
|\! \downarrow \rangle = c_1 |\psi_1 \rangle
\ \leftrightarrow\  c, \qquad |\! \uparrow \rangle = c_0 c_1
|\psi_1 \rangle \ \leftrightarrow\  \partial c\,c.
\label{gvop}
\end{equation}
The $bc$ theory has a conserved current $j =\ :\! cb \! :$, called
ghost number, which counts the number of $c$'s minus the number of
$b$'s.  In the cylindrical $w$ frame the vacua have average ghost
number zero, so $-\frac{1}{2}$ for $|\! \downarrow \rangle$
and $+\frac{1}{2}$ for $|\! \uparrow \rangle$.  The ghost numbers
of the corresponding operators are $\lambda - 1$ and $\lambda$,
as we see from the example~(\ref{gvop}).
As in the case of the momentum~(\ref{lmom}), the difference arises
because the current is not a tensor.

For the special case $\lambda = \frac{1}{2}$, $b$ and $c$ have the
same weight and
the $bc$ system can be split in two in a conformally invariant way,
$b = (\psi_1 + i \psi_2)/\sqrt{2}$,
$c = (\psi_1 - i \psi_2)/\sqrt{2}$, and
\begin{equation}
S = \frac{1}{2\pi} \int d^2z \, b \bar\partial c
= \frac{1}{4\pi} \int d^2z \, \Bigl\{ \psi_1 \bar\partial \psi_1 +
\psi_2 \bar\partial \psi_2 \Bigr\}.
\end{equation}
Each $\psi$ theory has central charge $\frac{1}{2}$.
The antianalytic theory
separates in the same way.  We will refer to these as Majorana
(real)
fermions, because it is a unitary CFT with $\psi_m^\dagger =
\psi_{-m}$.

Another family of CFT's differs from the $bc$ system
only in that the fields commute.
The action is
\begin{equation}
S = \frac{1}{2\pi} \int d^2z \,\beta \bar\partial \gamma \ .
\end{equation}
The fields $\beta$ and $\gamma$ are analytic by the equations of
motion; as usual there is a corresponding antianalytic theory.
Because the statistics are changed, some signs in operator products
are different,
\begin{equation}
\beta(z)\gamma(0) \sim -\frac{1}{z}, \qquad
\gamma(z)\beta(0) \sim \frac{1}{z}.
\end{equation}
The action is conformally invariant with $\beta$ a weight $(\lambda,
0)$ tensor and
$\gamma$ a $(1-\lambda, 0)$ tensor.  The energy-momentum tensor is
\begin{equation}
T(z) =\ :\! (\partial \beta) \gamma \! :
- \lambda :\! \partial( \beta \gamma ) \! :\ .
\end{equation}
The central charge has the opposite sign relative to the $bc$
system because of the changed statistics,
\begin{equation}
c =  3(2 \lambda - 1)^2 - 1.
\end{equation}

All of the above are free field theories.  A simple interacting
theory is the non-linear sigma model~\cite{Fnlsm}-\cite{CFMP},
consisting of $D$ scalars $X^\mu$ with a field-dependent
kinetic term,
\begin{equation}
S= \frac{1}{4\pi} \int d^2z\,\Bigr\{ G_{\mu\nu}(X) + i B_{\mu\nu}(X)
\Bigl\}
\partial X^\mu \bar\partial X^\nu, \label{nlsm}
\end{equation}
with $G_{\mu\nu}= G_{\nu\mu}$ and $B_{\mu\nu} = - B_{\nu\mu}$.
Effectively the scalars define a curved field space, with
$G_{\mu\nu}$ the metric on the space.
The path integral is no longer gaussian, but
when $G_{\mu\nu}(X)$ and $B_{\mu\nu}(X)$ are slowly varying
the interactions are weak and there is a small parameter.
The action is naively conformally invariant, but a one-loop
calculation reveals an anomaly (obviously this is closely related
to the $\beta$-function for rigid scale transformations),
\begin{equation}
T_{z\bar z} = \biggr\{- 2 {\bf R}_{\mu\nu} + \frac{1}{2}
H_{\mu\sigma\rho}H_\nu\!^{\sigma\rho} + \nabla^\sigma
H_{\sigma\mu\nu} \biggr\}
\partial X^\mu \bar\partial X^\nu \ . \label{tzzb}
\end{equation}
Here ${\bf R}_{\mu\nu}$ is the Ricci curvature built from
$G_{\mu\nu}$ (I am using boldface to distinguish it from the
two-dimensional curvature to appear later),
$\nabla^\sigma$ denotes the covariant derivative in this metric, and
\begin{equation}
H_{\sigma\mu\nu} =
\partial_\sigma B_{\mu\nu} +
\partial_\mu B_{\nu\sigma} + \partial_\nu B_{\sigma\mu}.
\label{hsmn}
\end{equation}

To this order, any Ricci-flat space with $B_{\mu\nu} = 0$
gives a CFT.  At higher order these conditions receive corrections.
Other solutions involve cancellations between terms in~(\ref{tzzb}).
A three dimensional example is the 3-sphere with a
round metric of radius $r$, and with
\begin{equation}
H_{\sigma\mu\nu} = \frac{4q}{r^3} \epsilon_{\sigma\mu\nu}
\label{heps}
\end{equation}
proportional to the antisymmetric three-tensor.
By symmetry, the first two terms in $T_{z\bar z}$ are proportional
to $G_{\mu\nu}\partial X^\mu \bar\partial X^\nu$ and the third
vanishes.  Thus $T_{z\bar z}$ vanishes for an
appropriate relation between the constants, $r^4 = 4 q^2$.
There is one subtlety.  Locally the form~(\ref{heps}) is compatible
with the definition~(\ref{hsmn}) but not globally.  This
configuration is the analog of a magnetic monopole, with the gauge
potential $B_{\mu\nu}$ now having two indices and the field strength
$H_{\sigma\mu\nu}$.  Then $B_{\mu\nu}$ must have a `Dirac string'
singularity, which is invisible to the string if the field strength
is appropriately quantized; I have normalized $q$ just such that it
must be an integer.  So this defines a discrete
series of models.  The one-loop correction to the central charge is $c
= 3 - 6/|q| + O(1/q^2)$.  The 3-sphere is the $SU(2)$ group space, and
the theory just described is the $SU(2)$ Wess-Zumino-Witten (WZW)
model~\cite{Wwzw}-\cite{GepW} at level $q$.  It can be generalized to
any Lie group. Although this discussion
is based on the one-loop approximation, which is accurate for
large $r$ (small gradients) and so for large $q$, these
models can also be constructed exactly, as will be discussed
further shortly.

For $c < 1$, it can be shown that unitary CFT's can exist only
at the special values~\cite{BPZ},~\cite{FQS}
\begin{equation}
c = 1 - \frac{6}{m(m+1)}, \qquad m = 2,3, \ldots.
\end{equation}
These are the unitary minimal models, and can be solved using
conformal symmetry alone.  The point is that for $c<1$ the
representations are all degenerate, certain linear combinations of
raising operators annihilating the highest weight state, which gives
rise to differential equations for the expectation value of the
corresponding tensor operator.  These CFT's have a $Z_2$ symmetry
and $m-2$ relevant operators, and correspond to interesting critical
systems: $m = 3$ to the Ising model (note that $c = \frac{1}{2}$
corresponds to the free fermion), $m = 4$ to the tricritical Ising
model, $m=5$ to a multicritical $Z_2$ Ising model but also to the
three-state Potts model, and so on.

This gives a survey of the main categories of conformal field
theory, including some CFT's that will be of specific interest to us
later on.  It is familiar that there are many equivalences between
different two dimensional field theories.  For example, the ordinary
free boson is equivalent to a free fermion, which is the same as
the $bc$ system at $\lambda = \frac{1}{2}$.  The
bosonization dictionary is
\begin{equation}
b \equiv\ :\! e^{iX} \! :\ ,\qquad c \equiv\ :\! e^{-iX} \! :\
\end{equation}
In fact this extends to the general $bc$ and $X$ CFT's, with
complex $Q = i(1-2\lambda)$.  The reader can check that the weights
of $b$ and $c$, and the central charge, then match.  There is also
a rewriting of $\beta$ and $\gamma$ in terms of exponentials,
which is more complicated but useful in the superstring.
All of these subjects are covered in ref.~\cite{FMS}.
As a further example,
the level~1 $SU(2)$ WZW model, which we have described in terms of
three bosons, can also be written in terms of a single free boson;
the level~2 $SU(2)$ WZW model can be written in terms of three
Majorana fermions; these will be explained further in the next
section. The minimal models are related to the free $X$ theory with
$Q$ such as to give the appropriate central charge~\cite{DotFat}, but
this is somewhat indirect.

\subs{Other Algebras}

The Virasoro algebra is just one of several important infinite
dimensional algebras.  Another is obtained from $T(z)$ plus any
number of analytic $(1,0)$ tensors $j^a(z)$.  The constraints
obtained at the end of section~1.5 imply that if the algebra is to
have unitary representations the $jj$ OPE can only take the form
\begin{equation}
j^a(z) j^b(0) \sim \frac{k^{ab}}{z^2} + i \frac{f^{ab}\!_c}{z}
j^c(0).
\end{equation}
The corresponding Laurent expansion is
\begin{equation}
j^a(z) =
\sum_{m=-\infty}^{\infty} \frac{j^a_m}{z^{m+1}},
\end{equation}
and the corresponding algebra
\begin{equation}
[j^a_m, j^b_n] = m k^{ab} \delta_{m+n,0} + i f^{ab}\!_c j^c_{m+n}.
\end{equation}
This is known variously as a {\it current algebra,}
an {\it affine Lie algebra}, or
sometimes as a {\it Kac-Moody algebra;}  for general references see
\cite{GO},~\cite{KZ} and~\cite{GepW}.
The $m=n=0$ modes form an ordinary Lie
algebra $g$ with structure constants $f^{ab}\!_c$.  The latter must
therefore satisfy the Jacobi identity; another Jacobi identity
implies that $k^{ab}$ is $g$-invariant.  The energy-momentum tensor
can be shown to separate into a piece built from the current
(the Sugawara construction) and a piece commuting with the current.
The CFT is thus a product of a part determined by the symmetry and
a part independent of the symmetry.

For a single Abelian current we already have the example of the
free $X$ theory.  The next simplest case is $SU(2)$,
\begin{equation}
[j^a_m, j^b_n] = m k \delta^{ab} \delta_{m+n,0} + i
\sqrt 2\epsilon^{abc} j^c_{m+n}.
\end{equation}
The value of $k$ must be an integer, and non-negative in a unitary
theory.\\[3pt]
{\bf Exercise:} Construct an $SU(2)$ algebra containing
$(j^1 + i j^2)_{1}$ and $(j^1 - i j^2)_{-1}$, and use it to show
that $k$ is an integer.\\[3pt]
The Sugawara central charge is $c = 3k/(k+2)$.
The $SU(2)$ WZW model just discussed has $k = |q|$.  The case $k
= 1$ can also be realized in terms of a single free scalar as
\begin{equation}
j^1 = \sqrt{2} :\! \cos\sqrt 2 X \! :\ , \qquad
j^2 = -\sqrt{2} :\! \sin\sqrt 2 X \! :\ , \qquad
j^3 = i \partial X\ .
\end{equation}
The case $k=2$ can also be realized in terms of three Majorana
fermions, $j^a = i \epsilon^{abc} \psi^b \psi^c / \sqrt{2}$.

The energy momentum tensor together with a weight $(\frac{3}{2},0)$
tensor current (supercurrent)
$T_F$ form the $N=1$ superconformal algebra~\cite{Rns},~\cite{rNS}.
The
$T_F T_F$ OPE is
\begin{equation}
T_F(z) T_F(0) \sim \frac{2c}{3z^3} + \frac{2}{z} T(0);
\end{equation}
$TT_F$ has the usual tensor form~(\ref{prop}).
A simple realization is in terms of a free scalar $X$ and a Majorana
fermion $\psi$,
\begin{equation}
T_F = i X \partial \psi, \qquad T = -\frac{1}{2}
( :\! \partial X \partial X \! : + :\! \psi \partial \psi
\!:\ ).
\label{n1free}
\end{equation}
With the Laurent expansions
\begin{equation}
T_F(z) = \sum_{r=-\infty}^{\infty} \frac{G_r}{z^{r+3/2}}, \qquad
\psi(z) = \sum_{r=-\infty}^{\infty} \frac{\psi_r}{z^{r+1/2}},
\label{tfpsl}
\end{equation}
the algebra is
\begin{equation}
\{ G_r, G_s \} = 2 L_{r+s} + \frac{c}{12}(4 r^2 - 1)
\delta_{r+s,0}\ .
\end{equation}
The central charge must be the same is in the $TT$ OPE, by the
Jacobi identity.
Note that for $r$ running over integers, the fields~(\ref{tfpsl})
have branch cuts at the origin, but the corresponding fields in the
cylindrical $w$ frame are periodic due to the tensor transformation
$(\partial w/\partial z)^{h}$.  This is the {\it Ramond sector}.
Antiperiodic boundary conditions on $\psi$ and $T_F$ in the $w$
frame are also possible; this is the Neveu-Schwarz sector.  All of
the above goes through with $r$ running over
integers-plus-$\frac{1}{2}$, and the fields in the $z$ frame are
single valued in this sector.\\[3pt]
{\bf Exercise:} Work out the
expansions of $L_m$ and $G_r$ in terms of the modes of $X$ and
$\psi$.  [Answer: the normal-ordering constant is $0$ in the
Neveu-Schwarz sector and $\frac{1}{16}$ in the Ramond
sector.]\\[3pt]
The operators corresponding to Ramond-sector states
thus produce branch cuts in the fermionic fields, and are known as
{\it spin fields}.  They are most easily described using
bosonization.  With two copies of the free
representation~(\ref{n1free}), the bosonization is $(\psi^1 \pm
i\psi^2)/\sqrt{2} = \ :\! e^{\pm i X} \!:\ $.  There are two Ramond
ground states, which correspond to the operators $:\! e^{\pm i X/2}
\!:$.  Observe that this has the necessary branch cut with $\psi^1$,
$\psi^2$, and also that its weight, $\frac{1}{8} = 2\frac{1}{16}$,
agrees with the exercise.

The energy-momentum tensor with two $(\frac{3}{2},0)$ tensors
$T_F^\pm$ plus a $(1,0)$ current $j$ form the $N=2$ superconformal
algebra~\cite{Aetal}
\begin{eqnarray}
T^+_F(z) T^-_F(0) &\sim& \frac{2c}{3z^3} +
\frac{1}{z^2} j(0) + \frac{2}{z} T(0) + \frac{1}{2z} \partial j(0),
\nonumber\\
j(z) T^{\pm}_F(0) &\sim& \pm \frac{1}{z} T^{\pm}_F(0), \label{n2sc}
\end{eqnarray}
with $T_F^+ T_F^+$ and $T_F^- T_F^-$ analytic.  This can be
generalized to $N$ supercurrents, leading to an algebra with weights
$2, \frac{3}{2}, \ldots, 2 - \frac{1}{2}N$.  From the earlier
discussion we see that there are no unitary representations for
$N > 4$.  There is one $N=3$ algebra and two distinct $N=4$ algebras
\cite{Aetal},\cite{EguT}.

These are the algebras which play a central role in string theory,
but many others arise in various CFT's.  I will briefly discuss
some higher-spin algebras, which have a number of interesting
applications (for reviews of the various linear and nonlinear
higher spin algebras see refs.~\cite{PRS},\cite{BS}).
The free scalar action~(\ref{xact2})
actually has an enormous amount of symmetry, but let us in
particular pick out
\begin{equation}
\delta X(z,\bar z) = - \sum_{l=0}^\infty g_l(z) (\partial
X(z))^{l+1},
\qquad l = 0,1,\ldots
\end{equation}
The Noether currents
\begin{equation}
V^l(z) = -\frac{1}{l+2} :\! (\partial X(z))^{l+2} \!:
\end{equation}
have spins $l+2$.  Making the usual Laurent expansion, one finds
the $w_\infty$ algebra
\begin{equation}
[V^i_m, V^j_n] = \Bigl( [j+1]m - [i+1]n \Bigr) V^{i+j}_{m+n}.
\label{winf}
\end{equation}
The $l = 0$ generators are just the usual Virasoro algebra.
There are a number of related algebras.  Adding in the $l = -1$
generators (which are just the modes of the translation current)
defines the $w_{1+\infty}$ algebra.  Another algebra with the same
spin content as $w_\infty$ but a more complicated commutator is
$W_\infty$; $w_\infty$ can be obtained as a limit (contraction) of
$W_\infty$.
These algebras have a simple and useful
realization in terms of the classical and quantum mechanics of a
particle in one dimension:
\begin{equation}
V^i_m \equiv \frac{1}{4} (p+x)^{i+m+1} (p-x)^{i-m+1}.
\end{equation}
The Poisson bracket algebra of these is the wedge
subalgebra ($m \leq i+1$) of $w_{\infty}$; the commutator algebra is
the wedge subalgebra of $W_{\infty}$.  In the literature one must
beware of differing notations and conventions.

All of these algebras have supersymmetric extensions,
with generators of half-integral spins.  There are also various
algebras with a finite number of higher weights.  One family is
$W_N$, closely related to $W_\infty$, with weights up to $N$.
The commutator of two weight-3 currents in $W_\infty$ contains
the weight-4 current.  In $W_3$ this is not an independent current
but the square of $T(z)$, so the algebra is nonlinear.

\subs{Riemann Surfaces}

Thus far we have focussed on local properties, without regard to
the global structure or boundary conditions.  For string theory we
will be interested in conformal field theories on closed manifolds.
The appropriate manifold for a two-dimensional CFT to live on is a
two-dimensional complex manifold, a {\it Riemann surface}.  One can
imagine this as being built up from patches, patch $i$ having a
coordinate $z^{(i)}$ which runs over some portion of the complex
plane.  If patches $i$ and $j$ overlap,
there is a relation between the
coordinates,
\begin{equation}
z^{(j)} = f_{ij}(z^{(i)})  \label{trfu}
\end{equation}
with $f_{ij}$ an {\it analytic} function.  Two Riemann surfaces
are equivalent if there is a mapping between them such that
the coordinates on one are analytic functions of the coordinates on
the other.  This is entirely parallel to the definition of a
differentiable manifold, but it has more structure---the manifold
comes with a local notion of analyticity.  Since in a CFT each field
has a specific transformation law under analytic changes of
coordinates, the transition function~(\ref{trfu}) is just the
information needed to extend the field from patch to patch.

A simple example is the sphere, which we can imagine as built from
two copies of the complex plane, with coordinates $z$ and $u$, with
the mapping
\begin{equation}
u = -\frac{1}{z}.
\end{equation}
The $z$ coordinate cannot quite cover the sphere, the point at
infinity being missing.
All Riemann surfaces with the topology of the sphere are equivalent.
For future reference let us note that the sphere has a group of
globally defined conformal transformations
(conformal Killing transformations), which in the $z$ patch
take
\begin{equation}
z' = \frac{\alpha z + \beta}{\gamma z + \delta}, \label{mob}
\end{equation}
where $\alpha$, $\beta$, $\gamma$ and $\delta$ are complex
parameters which can be chosen such that
$\alpha \delta - \beta \gamma = 1$.  This is the {\it M\"obius
group}.

The next Riemann surface is the torus.  Rather than build it from
patches it is most convenient to describe it as in fig.~5 by
taking a single copy $z$ of the complex plane and identifying points
\begin{equation}
z \ \cong\ z+2\pi \ \cong\ z+2\pi \tau,
\end{equation}
producing a parallelogram-shaped region with opposite edges
identified.
\begin{figure}
\begin{center}
\leavevmode
\epsfbox{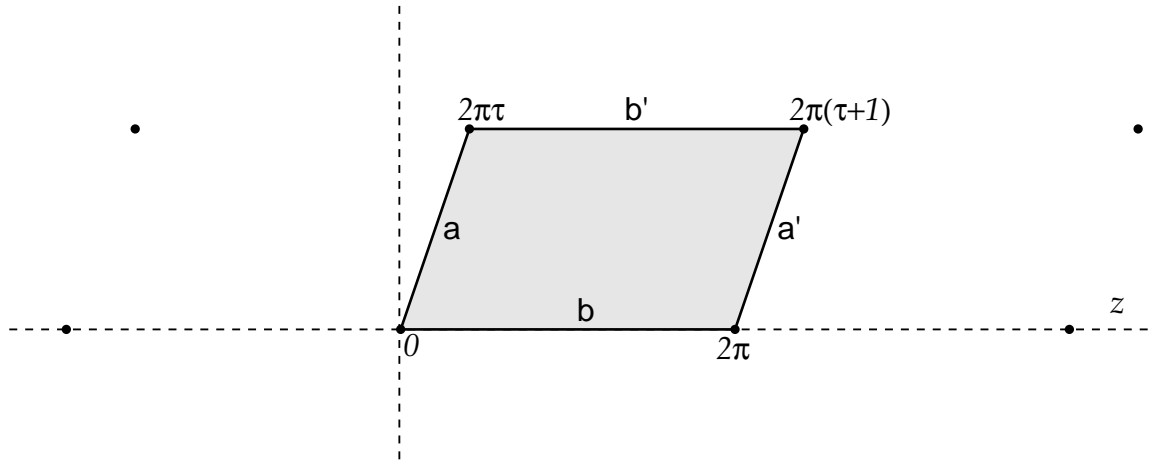}
\end{center}
\caption[]{The torus by periodic identification of the complex
plane.  Points identified with the origin are indicated.
Edges $a$ and $a'$ are identified, as are edges $b$ and
$b'$.}
\end{figure}
Different values of $\tau$ in general define
inequivalent Riemann surfaces; $\tau$ is known as a {\it modulus}
for the complex structure on the torus.  However, there are some
equivalences: $\tau$, $-\tau$, $\tau+1$, and $-1/\tau$ generate the
same group of transformations of the complex plane and so the same
surface (to see the last of these, let $z = - z'\tau$).
So we may restrict to Im$(\tau) > 0$ and moreover identify
\begin{equation}
\tau \sim \tau+1 \sim -\frac{1}{\tau}. \label{tauid}
\end{equation}
These generate the {\it modular group}
\begin{equation}
\tau' = \frac{a \tau + b}{c \tau + d}, \label{sl2z}
\end{equation}
where now $a$, $b$, $c$, $d$ are integers such that $ad-bc= 1$.
A fundamental region for this is given by $|\tau|\geq 1$,
Re$(\tau) \leq \frac{1}{2}$, shown in fig.~6.
\begin{figure}
\begin{center}
\leavevmode
\epsfbox{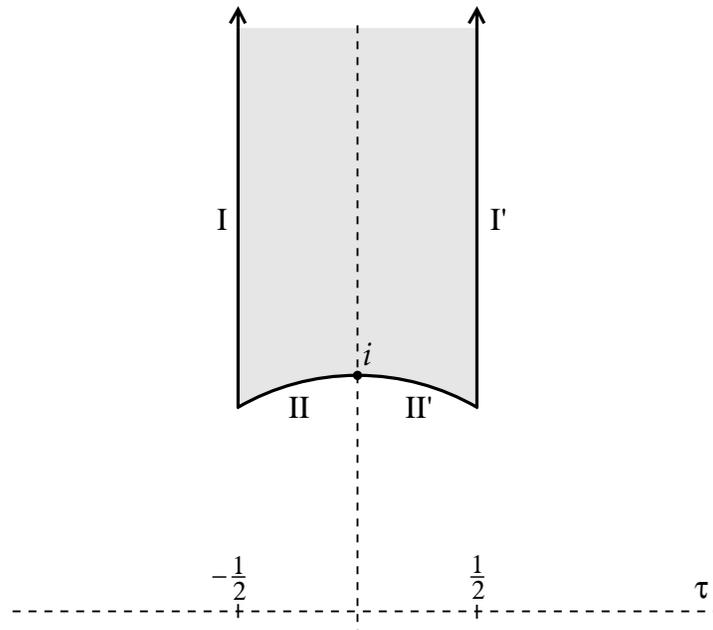}
\end{center}
\caption[]{The standard fundamental region for the modulus
$\tau$ of the torus.  Identifying boundaries I and I$'$, and II and
II$'$, produces the moduli space for the torus.  Note that this is a
closed space except for the limit Im$(\tau) \to \infty$.}
\end{figure}
This is the
{\it moduli space}
for the torus: every Riemann surface with this topology is equivalent
to one with $\tau$ in this region.  The torus also has
a conformal Killing transformation, $z \to z+\alpha$.

Notice the similarity between the transformations~(\ref{mob})
and~(\ref{sl2z}), differing only in whether the parameters are
complex numbers or integers.  You can check that successive
transformations compose like matrix multiplication, so these
are the groups $SL(2,C)$ and $SL(2,Z)$ respectively ($2\times 2$
matrices of determinant one).\footnote
{To be entirely precise, flipping the signs of
$\alpha$, $\beta$, $\gamma$, $\delta$ or
$a$, $b$, $c$, $d$ gives the same transformation, so we have
$SL(2,C)/Z_2$ and $SL(2,Z)/Z_2$ respectively.}
We will meet $SL(2,Z)$ again in a
different physical context.

Any closed oriented oriented two-dimensional surface can be
obtained by adding $h$ handles to the sphere; $h$ is the {\it
genus}.  It is often useful to think of higher genus surfaces built
up from lower via the {\it plumbing fixture construction}.
This essential idea is developed in many places, but my
lectures have been most influenced by the approach in
refs~\cite{V1}-\cite{Son}.  Let
$z^{(1)}$ and $z^{(2)}$ be coordinates in two patches, which may be
on the same Riemann surface or on different Riemann surfaces.  For
complex
$q$, cut out the circles $|z^{(1)}|, |z^{(2)}|
< (1-\epsilon) |q|^{1/2}$
and identify points on the cut surfaces such that
\begin{equation}
z^{(1)} z^{(2)} = q,
\end{equation}
as shown in fig.~7.
\begin{figure}
\begin{center}
\leavevmode
\epsfbox{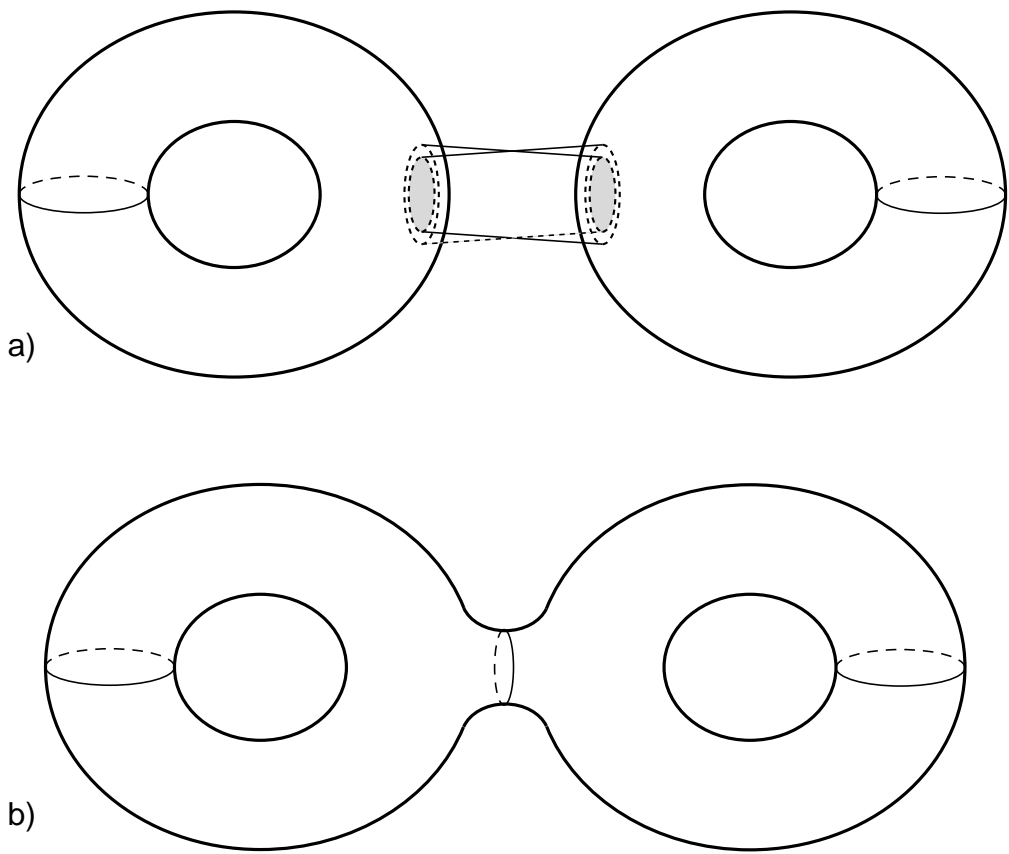}
\end{center}
\caption[]{Plumbing fixture construction.  Identifying annular
regions as in (a) produces the sewn surface (b).}
\end{figure}
If $z^{(1)}$ and $z^{(2)}$
are on the same surface, this
adds a handle.  The genus-$h$ surface can be constructed from the
sphere by applying this $h$ times.  The number of complex parameters
in the construction is $3h$, being $q$ and the position of each end
for each handle, minus 3 from an overcounting due to the
M\"obius group, leaving $3h-3$ which is the correct number of
complex moduli.  An index theorem states that the number of complex
moduli minus the number of conformal Killing transformations is
$3h - 3$, as we indeed have in each case.

Note that for $q < 1$ the region between the circles $|z^{(1)}| = 1$
and
$|z^{(2)}| = 1$ is conformal to the cylindrical region
\begin{equation}
0 < {\rm Im}(w) < 2\pi \ln(1/|q|), \qquad w\ \cong\  w+2\pi,
\end{equation}
which becomes long in the limit $q\to 0$.

Just as conformal transformations can be described as the most
general coordinate transformation which leave $dz$ invariant up to
local multiplication, there is a geometric interpretation for the
superconformal transformations.  The $N=1$ algebra, for example,
can be described in terms of a space with two ordinary and two
anticommuting coordinates $(z,\bar z
,\theta,\bar\theta)$ as the
space of transformations which leave $dz + i\theta
d\theta$ invariant up to local multiplication.  Super-Riemann
surfaces can be defined as above by patching.  The genus-$h$
Riemann surface for $h \geq 2$ has $3h-3$ commuting and
$2h - 2$ anticommuting complex moduli.

\subs{CFT on Riemann Surfaces}

On this large subject I will give here only a few examples and
remarks that will be useful later.  A tensor ${\cal O}$ of weight
$(h,\tilde h)$ transforms as ${\cal O}^{(u)} = {\cal O}^{(z)} z^{2h}
\bar z^{2\tilde h}$ from the $z$ to $u$ patch on the sphere.  It
must be smooth at
$u=0$, so in the $z$ frame we have
\begin{equation}
< {\cal O}^{z}(z,\bar z) \ldots >_{S_2}\ \sim\ z^{-2h} \bar
z^{-2\tilde h}, \qquad z \to \infty
\end{equation}
An expectation value which illustrates this, and will be useful
later, is (all operators implicitly in the $z$-frame unless noted)
\begin{equation}
<:\! e^{i k_1 X(z_1,\bar z_1)} \! :\,
:\! e^{i k_2 X(z_2,\bar z_2)} \! :\,\ldots\,
:\! e^{i k_n X(z_n,\bar z_n)} \! : >_{S_2}\ = 2\pi
\delta(k_1 + \ldots + k_n) \prod_{i < j}
|z_i - z_j|^{2 k_i k_j},
\end{equation}
obtained but summing over all graphs with the propagator
$-\ln |z_i - z_j|^2$.  Using momentum conservation one finds the
appropriate behavior $|z_i|^{-2 k_i^2}$ as $z_i \to \infty$.

Another example involves the $bc$ system, specializing to
the most important case $\lambda = 2$.  Consider an expectation
value with some product of local operators, surrounded by a line
integral
\begin{equation}
\frac{1}{2\pi i} \oint dz j_z
\end{equation}
where $j = :\!cb\!:$.  Contracting $C$ down around the operators
and using the OPE gives $N_c - N_b$ for the contour integral,
counting the total net ghost number of the operators.  We have
previously noted that $j$ is not a tensor, and one finds that
the contour integral above is equal to
\begin{equation}
\frac{1}{2\pi i} \oint du j_u \ +\ 3.  \label{nbnc}
\end{equation}
Now, we can contract contour $C$ to zero in the $u$ patch, so
it must be that $N_c - N_b = 3$ for a nonvanishing
correlator.\\[3pt]
{\bf Exercise:} Using the construction of the genus $h$ surface
from the sphere vis $h$ plumbing fixtures, show that
$N_c - N_b$ must be $3h - 3$.\\[3pt]
The simplest one is
\begin{equation}
<c(z_1) c(z_2) c(z_3)
\tilde c(z_4) \tilde c(z_5) \tilde c(z_6) >_{S_2}
= z_{12}z_{13}z_{23} \bar z_{45} \bar z_{46} \bar z_{56},
\end{equation}
which is completely determined, except for normalization that we
fix by hand, by the requirement that it be analytic, that it be odd
under exchange of anticommuting fields, and that it go as $z_i^2$ or
$\bar z_i^2$ at infinity, $c$ being weight
$(-1,0)$ and $\tilde c$ being $(0,-1)$.

Now something more abstract: consider the general two-point function
\begin{equation}
< {\cal A}'_i(\infty,\infty) {\cal A}_j(0,0) >_{S_2} =
\langle \psi_i | \psi_j\rangle = {\cal G}_{ij} \label{2pt}
\end{equation}
where I use a slightly wrong notation: $(\infty,\infty)$ denotes
the point $z = \infty$ ($u=0$) but the prime denote the $u$ frame
for the operator.
Recall the state-operator mapping.  The operator ${\cal A}_j(0,0)$
is equivalent to removing the disk $|z|<1$ and inserting the
state $| \psi_j\rangle$; the operator ${\cal A}'_i(\infty,\infty)$
is equivalent to removing the disk $|u|<1$ ($|z|>1$) and inserting the
state $| \psi_j\rangle$.  All that is left of the sphere is the overlap
of the two states.  It is also useful to regard this as a metric
${\cal G}_{ij}$ on
the space of operators, the {\it Zamolodchikov metric.}
Note that the path integral~(\ref{2pt}) does not include conjugation,
so if there is a Hermitean inner product $\langle
\!\langle \ |\ \rangle$ these must be related
\begin{equation}
\langle\!\langle \psi_i |\psi_j \rangle =
\langle \psi_i^* |\psi_j \rangle\ ,
\end{equation}
where $^*$ is some operation of conjugation.  For the free
scalar theory, whose Hermitean inner product has already been given,
$^*$ just takes $k \to -k$ and conjugates explicit complex numbers.

Similarly for the three-point function, the operator product
expansion plus the definition~(\ref{2pt}) give
\begin{eqnarray}
&&< {\cal A}'_i(\infty,\infty)
{\cal A}_k(z,\bar z) {\cal A}_j(0,0) >_{S_2}\, =
\ \langle \psi_i | {\cal A}_k(z,\bar z) |\psi_j\rangle \nonumber\\
&&\qquad\qquad =  z^{h_l - h_k - h_j} {\bar z}^{\tilde h_l - \tilde
h_k -
\tilde h_j} {\cal G}_{il}
c^l\!_{kj} = z^{h_l - h_k - h_j} {\bar z}^{\tilde h_l - \tilde h_k -
\tilde h_j} c_{ikj} .
\label{3pt}
\end{eqnarray}
This relates the three-point expectation value on the sphere to a matrix
element and then to an OPE coefficient.

The torus has a simple canonical interpretation: propagate a state
forward by $2\pi$Im$(\tau)$ and spatially by $2\pi$Re$(\tau)$
and then sum over all states, giving the partition function
\begin{equation}
<1>_{T_2(\tau)}\ =\ {\rm Tr}\Bigl(e^{2\pi i \tau (L_0 - c/24)}
e^{-2\pi i \bar\tau (\tilde L_0 - \tilde c/24)} \Bigr),
\label{partit}
\end{equation}
where the additive constant is as in eq.~(\ref{canon}).  This must be
invariant under the modular group.  Let us point out just one
interesting consequence~\cite{Car1}.  In a unitary theory the
operator of lowest weight is the $(0,0)$ unit operator, so
\begin{equation}
<1>_{T_2(\tau)}
\ \sim\ e^{\pi ( c + \tilde c){\rm Im}(\tau)/12}, \qquad \tau \to
i\infty.
\end{equation}
The modular transformation $\tau \to -1/\tau$ then gives
\begin{equation}
<1>_{T_2(\tau)}
\ \sim\ e^{\pi ( c + \tilde c)/12 {\rm Im}(\tau)}, \qquad \tau
\to i0. \label{cdof}
\end{equation}
The latter partition function is dominated by the states of high
weight and is a measure of the density of these states.  We see that
this is governed by the central charge, generalizing the result that
$c$ counts free scalars.

We have described the general Riemann surface implicitly in terms of
the plumbing fixture, and there is a corresponding construction for
CFT's on the surface (again, I follow
refs.~\cite{BPZ}, \cite{V1}-\cite{Son}).
Taking first $q = 1$, sewing the
path integrals together is equivalent to inserting a complete set of
states.  As shown in fig.~8, each can be replaced with a disk
plus vertex operator.
\begin{figure}
\begin{center}
\leavevmode
\epsfbox{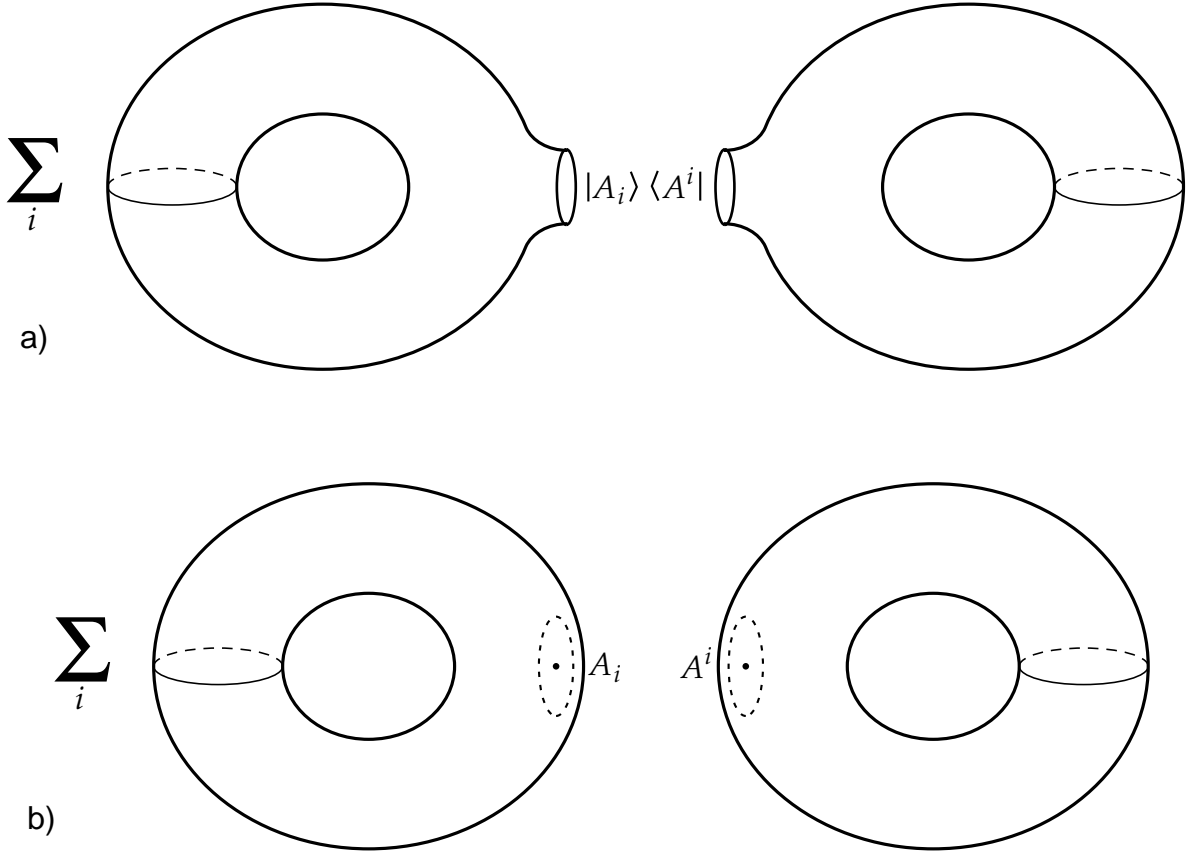}
\end{center}
\caption[]{a) Path integral on sewn surface of fig.~7 written in
terms of a sum over intermediate states. b) Each state replaced by
disk with local operator.}
\end{figure}
Including the radial evolution for general $q$
we have
\begin{equation}
<\ldots_1 \ldots_2 >_{\cal M}\ = \sum_{ij} q^{-h_i} \bar q^{-\tilde
h_i} <\ldots_1 {\cal A}_i>_{{\cal M}_1}\ <\ldots_2 {\cal A}^i>_{{\cal
M}_2},  \label{sew}
\end{equation}
where ${\cal M}$ is sewn from ${\cal M}_{1,2}$, the operators are
inserted at the origins of the $z^{(1,2)}$ frames, and indices are
raised with the inverse of the metric~(\ref{2pt}).\footnote{The one
thing which is not obvious here is the metric to use.  You can check
the result by applying it to sew two spheres together, with
$\ldots_1$ and $\ldots_2$ each being a single local operator, to get
the sphere with two local operators.}

By sewing in this way, an expectation value with any number of
operators on a general genus surface can be related to the
three-point function on the sphere.  For example, fig.~9 shows
three of the many ways to construct the genus-two surface with four
operators.
\begin{figure}
\begin{center}
\leavevmode
\epsfbox{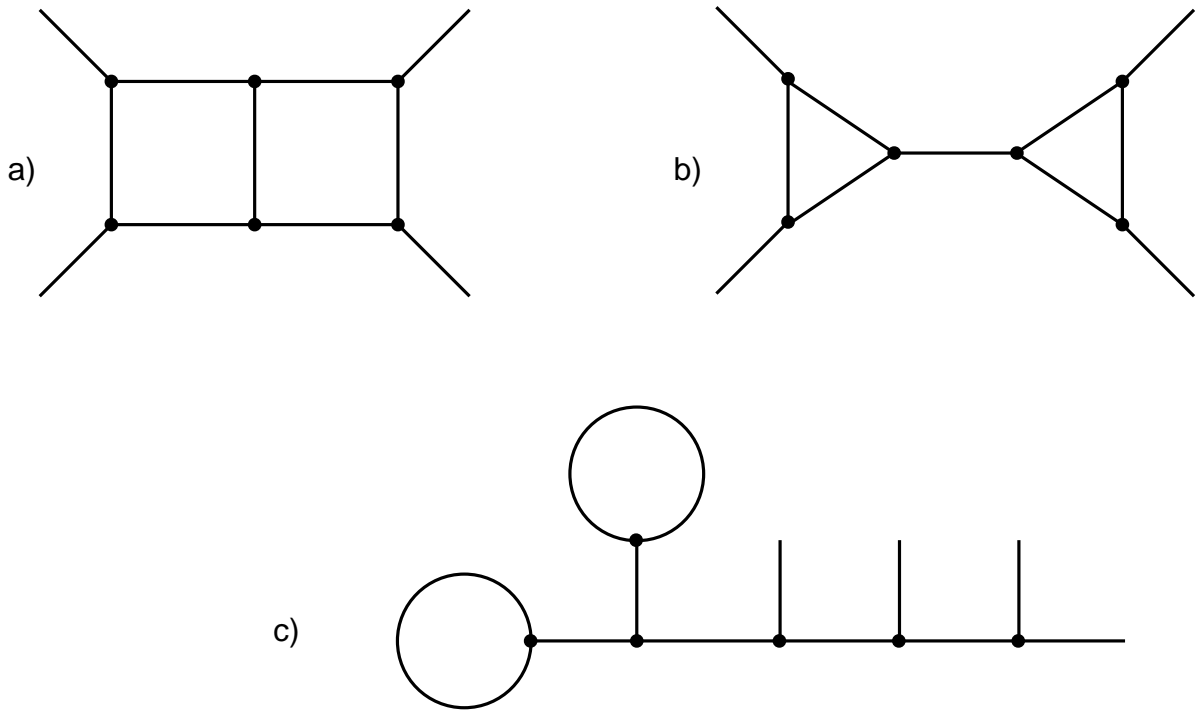}
\end{center}
\caption[]{Some sewing constructions of the genus-two surface with
four operators.  Each vertex is a sphere with three operators, and
each internal line represents the sewing construction.}
\end{figure}
There is one complex modulus $q$ for each
handle, or~7 in all here, corresponding to the $3h - 3 = 3$ moduli for
the surface plus the positions of the four operators.  As a
consequence, the OPE coefficients $c_{ij}\!^k$ implicitly determine all
expectation values, and two CFT's with the same OPE (and same operator
identified as $T(z)$) are the same.  However, the $c_{ij}\!^k$
are not arbitrary because the
various methods of constructing a given surface must agree.  For
example, the constructions of fig.~9a and~9b differ only by a single
move described earlier corresponding to associativity of the OPE,
and by further associativity moves one gets fig.~9c.
The amplitudes must also be modular invariant.  In fig.~9c we see
that the amplitude has been factorized into a tree amplitude times
one-loop one-point amplitudes, so  modular invariance of the latter
is sufficient.
It can be shown generally that all constructions
agree and are modular invariant
given two conditions~\cite{Son}: associativity of the OPE and modular
invariance of the torus with one local operator (which constrains
sums involving $c_{ij}\!^j$).

The classification of all CFT's can
thus be reduced to the algebraic problem of finding all sets
$c_{ij}\!^k$ satisfying the constraints of conformal invariance plus
these two conditions.  This program, the conformal
bootstrap~\cite{BPZ}, has been carried out only for cases where
conformal invariance (or some extension thereof) is sufficient to
reduce the number of independent $c_{ij}\!^k$ to a finite
number---these are known as {\it rational conformal field theories}.

My description of higher-genus surfaces and the CFT's on them has
been rather implicit, using the sewing construction.  This is
well-suited for my purpose, which is to understand the general
properties of amplitudes.  For treatments from a more explicit point
of view see refs.~\cite{Oalv},~\cite{DFrmp}.

Unoriented surfaces, and surfaces with boundary, are also of interest.
In particular, CFT's with boundary have many interesting condensed
matter applications.  I do not have time for a detailed discussion,
but
will make a few comments about boundaries.  Taking coordinates such
that the boundary is Im$(z) = 0$ and the interior is Im$(z) > 0$, the
condition that the energy-momentum be conserved at the boundary is
\begin{equation}
T(z) = \tilde T(z), \qquad {\rm Im}(z)=0. \label{tbound}
\end{equation}
It is convenient to use the doubling trick, extending $T$ into the
lower half-plane by defining
\begin{equation}
T(z) = \tilde T(\bar z), \qquad {\rm Im}(z) < 0.
\end{equation}
The boundary conditions plus conservation of $T$ and $\tilde T$ are
all implied by the analyticity of the extended $T$.  Then $T$ and
$\tilde T$ together can be expanded in terms of a single Virasoro
algebra, the first of equations~(\ref{soro}).  The doubling trick is
also useful for free fields.  For free scalar with Neumann boundary
conditions, $\partial_n X = 0$, the mode expansion is
\begin{equation}
X(z,\bar z) = X_L(z) + X_L(\bar z), \label{neu}
\end{equation}
with
\begin{equation}
X_L(z) = \frac{x}{2} + i \alpha_0 \ln z
- i \sum_{m \neq 0} \frac{\alpha_m}{mz^{m}}. \label{osmode}
\end{equation}
There is a factor of~2 difference from the earlier~(\ref{lex2});
with this, $x$ is the mean value of $X$ at time $|z|=1$.
The commutator $[x,\alpha_0]$ is found to be $2i$, so $\alpha_0 =
2k$ where $k$ is the conjugate to $x$.  The leading
surfaces with boundary are the disk and annulus.

\section{String Theory}
\setcounter{footnote}{0}

\subs{Why Strings?}

The main clue that leads us to string theory is the short-distance
problem of quantum gravity.  Figure~10 shows some process,
say two particles propagating, and corrections due to
one-graviton exchange and two-graviton exchange.
\begin{figure}
\begin{center}
\leavevmode
\epsfbox{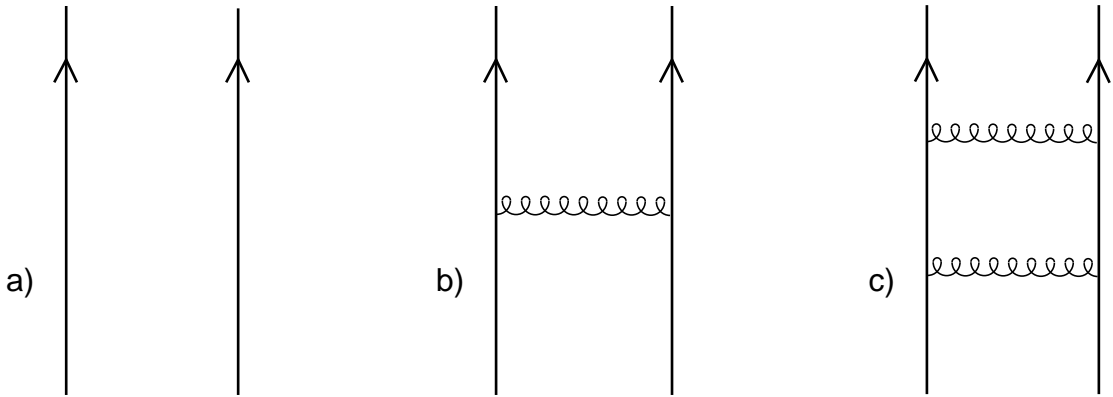}
\end{center}
\caption[]{a) Two particles propagating freely.
b) Correction from one-graviton exchange.
c) Correction from two-graviton exchange.}
\end{figure}
The one graviton exchange is proportional to Newton's constant
$G_{\rm N}$, which with $\hbar = c = 1$ has units of length$^2$ or
mass$^{-2}$: $G_{\rm N} = M_{\rm P}^{-2}$ where the Planck mass
$M_{\rm P} = 1.2 \times 10^{19}$GeV.  The dimensionless ratio of the
one-graviton correction to the original amplitude must then be of
order $E^2 / M_{\rm P}^2$, where $E$ is the characteristic energy
of the process.
This is thus an {\it irrelevant} coupling, growing weaker at long
distance, and in particular is negligible at particle physics
energies of hundreds of GeV.  By the same token, the coupling
grows stronger at high energy and at $E > M_{\rm P}$ perturbation
theory breaks down.  This shows up as the nonrenormalizability of the
theory: the two-graviton correction (c) is of order
\begin{equation}
M_{\rm P}^{-4} \int^\infty dE' E'^3
\end{equation}
where $E'$ is the energy of the virtual intermediate state,
and so diverges if the theory is extrapolated to arbitrarily high
energies.

There are two main possibilities.  The first is that the theory
has a nontrivial ultraviolet fixed point and is fine at high energy,
the divergences being an artifact of naive perturbation theory.
The second is that there is new physics at some energy and the
extrapolation of the low energy theory beyond this point is invalid.

The existence of a nontrivial fixed point is hard to determine.
One of the usual tools, Monte Carlo simulation, is extremely
difficult because of the need to retain coordinate invariance
in the discretized theory.  Expansion
around the critical dimension $d = 2$ indicates a nontrivial UV
fixed point when gravity is coupled to certain kinds of matter,
but it is impossible to say whether this persists to $d=4$.\footnote
{The idea that the divergence problems of
quantum gravity might be solved
by a resummation of perturbation theory has been examined from many
points of view, but let me mention in particular the
approach ref.~\cite{Weps} as one that will be familiar to the
condensed matter audience.}

The more common expectation, based in part on experience (such as
the weak interaction), is that the nonrenormalizability indicates a
breakdown of the theory, and that at short distances we will find a
new theory in which the interaction is spread out in
spacetime in some way that cuts off the divergence.  At this point
the condensed matter half of the audience is thinking, ``OK,
so put the thing on a lattice.''  But it is not so easy.  We know
that Lorentz invariance holds to very good approximation in the low
energy theory, and that means that if we spread the interaction in
space we spread it in time as well, with consequent loss of
causality or unitarity.  Moreover we know that we have local
coordinate invariance in nature---this makes it even harder to
spread the interaction out without producing inconsistencies.

In fact, we know of only one way to spread out the gravitational
interaction and cut off the divergence without spoiling the
consistency of the theory.  That way is string theory, in which the
graviton and all other elementary particles are one-dimensional
objects, strings, rather than points as in quantum field theory.
Why this should work and not anything else is not at all obvious
a priori, but as we develop the theory we will see that if we try
to make a consistent Lorentz-invariant quantum theory of strings
we are led inevitably to include gravity~\cite{Yon},~\cite{SS}, and
that the short distance divergences of field theory are no longer
present.\footnote
{There is an intuitive answer to at least one common question:
why not membranes, two- or higher-dimensional objects?
The answer is that as we spread out particles in more
dimensions we reduce the spacetime divergences, but encounter
new divergences coming from the increased number of {\it
internal} degrees of freedom.  One dimension appears to be the
one case where both the spacetime and internal divergences are
under control.  But this is far from conclusive: just as
pointlike theories of gravity are still under study, so are
membrane theories, as we will mention in section~3.5.}

Perhaps we merely suffer from a lack of imagination, and there are
many other consistent theories of gravity with a short-distance
cutoff.  But experience has shown that the divergence problems of
quantum field theory are not easily resolved, so if we have even one
solution we should take it very seriously.  In the case of the weak
interaction, for example, there is only one known way to spread out
the nonrenormalizable four-fermi theory consistently.\footnote
{During the lecture, Prof.~Zinn-Justin reminded me that there is
evidence from the large-$N$ approximation that some four-fermi
theories have nontrivial fixed points.  So perhaps there is more
than one way to smooth the weak interaction---but perhaps also we
should take this as an indication that, given a choice between new
physics and a nontrivial ultraviolet fixed point, nature will choose
the former.  At any right, given my understanding of renormalizable field
as an effective theory that emerges at long distance, I would find the
fixed point resolution very unappealing.}
That way is
spontaneously broken Yang-Mills theory, which did indeed turn out to
be the correct theory of the weak interaction. Indeed, we are very
fortunate that consistency turns out to be such a restrictive
principle, since the unification of gravity with the other
interactions takes place at such high energy, $M_{\rm P}$, that
experimental tests will be difficult and indirect.

So what else do we find, if we pursue this idea?  We find that
string theory fits very nicely into the pre-existing picture of
what physics beyond the Standard Model might look like.  Besides
gravity, string theory necessarily incorporates a number of
previous unifying ideas (though sometimes in transmuted form):
grand unification, Kaluza-Klein theory (unification via extra
dimensions), supersymmetry and extended supersymmetry.  Moreover it
unifies these ideas in an elegant way, and resolves some of the
problems which previously arose---most notably difficulties of
obtaining chiral (parity-violating) gauge interactions and the
renormalizability problem of Kaluza-Klein theory, which is even more
severe than for four-dimensional gravity.
Further, some of the simplest string theories~\cite{CHSW}
give rise to
precisely the gauge groups and matter representations which previously
arose in grand unification.  Finally, the whole subject has a unity and
structure far nicer than anything I have seen or expect to see in
quantum field theory.  So I am strongly of the opinion, and I think
that almost all of those who have worked in the subject would agree,
that string theory is at least a step toward the unification of
gravity, quantum mechanics, and particle physics.

In this lecture and the next I will try to sketch our current
understanding of the answer to the question posed in the title.
Given limits of time and the nature of the audience,
I will focus on broad dynamical issues, especially the mechanics by
which string theory cuts off gravity in a consistent way.
Most notably,
spacetime supersymmetry and the superstring will be underemphasized.

There is one graph, fig.~11, that I want to show you before I launch
into the introduction to string theory.
\begin{figure}
\begin{center}
\leavevmode
\epsfbox{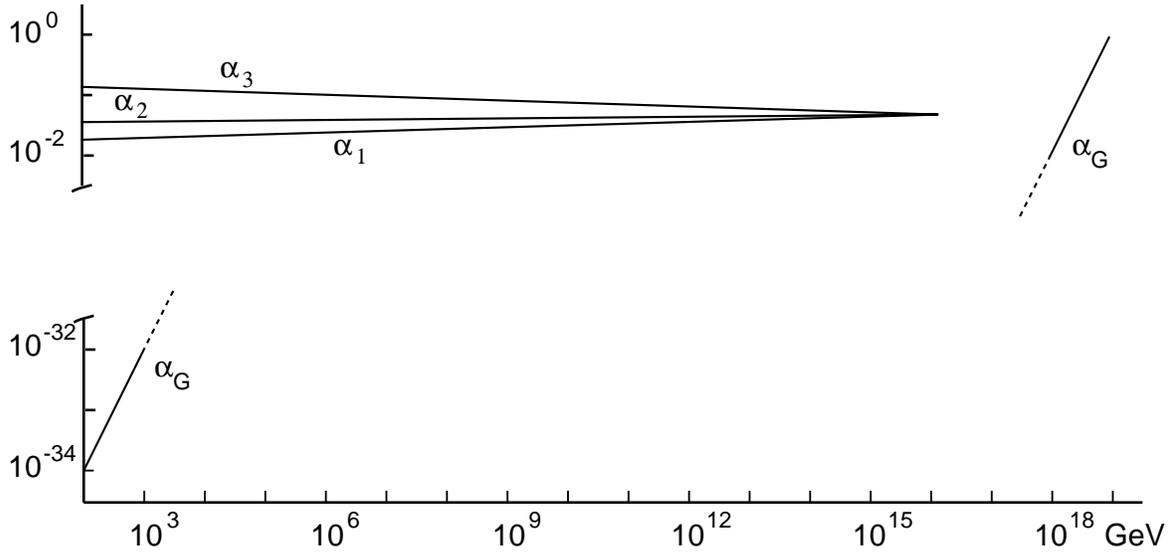}
\end{center}
\caption[]{Energy dependence of dimensionless gauge couplings
$\alpha_{1,2,3}$ (running logarithmically) and $\alpha_{\rm G}
= E^2 / M_{\rm P}^2$.}
\end{figure}
It shows how the three
dimensionless gauge couplings and the dimensionless gravitational
coupling $E^2 / M_{\rm P}^2$ depend on energy.  The gauge couplings
evolve slowly (logarithmically); the big news of recent years is
that in the minimal supersymmetric extension of the Standard Model
they meet to high accuracy at a common scale, of order $10^{16}$GeV,
giving evidence for supersymmetric grand unification~\cite{AdBF}.
The gravitational coupling starts much smaller but grows as a power
and so is just a bit late for its meeting with the others,
missing by two orders of magnitude or perhaps a little less.  These
extrapolations are sensitive to assumptions about the spectrum, so
perhaps all four couplings meet at a single energy, a very
grand unification.  Or perhaps there is a small hierarchy of scales
near the Planck scale.  But the near meeting in this minimal
extrapolation suggests that nature may have been kind and put
little new physics between current energies and the Planck scale.
With the thorough exploration of the weak interaction scale in
coming years, and hopefully the discovery of supersymmetry, we will
have several additional extrapolations of the same sort and so
several handles on physics near the Planck scale.  Also, proton
decay reaches into the same region, and if we are lucky it will
occur at a rate that will one day be seen.

\subs{String Basics}

We want to describe the dynamics of one-dimensional objects.  The
first thing we need is an action, and the simplest that comes to
mind is the Nambu-Goto action,\footnote
{Indices $\mu, \nu = 0, 1, \ldots D-1$ are raised and lowered with the
flat-space metric $\eta_{\mu\nu} = {\rm diag}(-,+,+,\ldots,+)$.}
\begin{eqnarray}
S &=& -\frac{1}{2\pi \alpha'} (\mbox{Area of world-sheet})
\nonumber\\
&=& -\frac{1}{2\pi \alpha'} \int d^2 \sigma \,
\sqrt{ -\det \partial_a X^\mu \partial_b X_\mu }\ . \label{nambu}
\end{eqnarray}
This generalizes the relativistic action for a point particle,
which is minus the mass times the invariant length of the
world-line.  For a static string, this action reduces to minus the
length of the string times the time interval times $1/2\pi \alpha'$,
so the latter is the string tension.  Note that in the second line
we are describing the world-sheet by $X^\mu(\sigma_1,\sigma_2)$,
using a parameterization $\sigma^a$ of the world-sheet, but the
action is independent of the choice of parameterization (world-sheet
coordinate invariant).  This will play an important role soon.

In quantum field theory we are familiar with a variety of
one-dimensional objects---magnetic flux tubes in superconductors
and other spontaneously broken gauge theories, color-electric flux
tubes in QCD.  Also, the classical statistical mechanics of
membranes is given by a sum over two-dimensional surfaces, and so is
closely related to the quantum-mechanical path integral for the
string.  In all of these cases the leading term in the action is
the tension~(\ref{nambu}).  But these are all composite
objects, with a thickness, and so there will be higher-dimension
terms in the action, such as a rigidity term, multiplied by powers
of the thickness.  The strings I am talking about, the
`fundamental' strings which give rise to gravity, are {\it exactly}
one-dimensional objects, of zero thickness.  Composite strings also
have a large contact interaction when they intersect; fundamental
strings do not.  Fundamental strings are thus simpler than the various
composite strings, simpler in particular than the hypothetical string
theory of QCD.\footnote{In the coming
sections I will discuss ideas that strings are in some sense
composite, but not composites of ordinary gauge fields.}  Nevertheless
there has been a great deal of cross-fertilization between the
theories of fundamental and composite strings.

It is useful to rewrite the action~(\ref{nambu}) in a form which
removes the square root from the derivatives.  Add a world-sheet
metric
$g_{ab}(\sigma)$ and let
\begin{equation}
S_{\rm P} = \frac{1}{4\pi \alpha'} \int d^2 \sigma \,\sqrt{g}
g^{ab} \partial_a X^\mu \partial_b X_\mu \ , \label{poly}
\end{equation}
where $g = \det g_{ab}$.  This is commonly
known as the Polyakov action because he emphasized its virtues for
quantization~\cite{Poly1}.  The equation of motion for the metric
determines it up to a position-dependent normalization
\begin{equation}
g_{ab}\ \propto\ \partial_a X^\mu \partial_b X_\mu; \label{polnam}
\end{equation}
inserting this back into the Polyakov action gives the Nambu
action.\footnote{So these are classically equivalent.  How about
quantum-mechanically, say in a path integral?  The glib answer is
that the Nambu action is hard to use in a path integral, so the way
to define it is via the Polyakov path integral.  On the the other
hand, ref.~\cite{PolStr} shows an example of a composite string
where the Nambu description is more natural than the Polyakov.}
Now, the Polyakov
action makes sense for either a Lorentzian metric, signature
$(-,+)$, or a Euclidean metric, signature $(+,+)$.\footnote{Though
the Lorentzian case needs an overall minus sign and one in the
square root.}  Much of the development can be carried out in
either case.  These are presumably related by a contour rotation in
the integration over metrics, since the light-cone quantization
(Lorentzian) gives the same theory as the Euclidean Polyakov
quantization that I will describe.  The relation between path
integrals over Lorentzian and Euclidean metrics is a complicated
and confusing issue in four-dimensional gravity.  It seems to work
out simply in two dimensions, though I don't have a simple
explanation of why---the demonstration of the equivalence is rather
roundabout~\cite{DG}.  Perhaps it is simply that there is enough
gauge symmetry to remove the metric entirely.\footnote{I would like
to thank M. Natsuume for this suggestion.} In any case I will take a
Euclidean metric henceforth as defining the theory.

In addition to the two-dimensional coordinate invariance mentioned
earlier ({\it diff invariance} for short),
\begin{equation}
X'(\sigma') = X(\sigma), \qquad
\frac{\partial \sigma'^a}{\partial \sigma^c}
\frac{\partial \sigma'^b}{\partial \sigma^d} g'_{ab}(\sigma') =
g_{cd}(\sigma)\ ,
\end{equation}
the Polyakov action has another local symmetry, {\it Weyl invariance},
position-dependent rescalings of the metric,
\begin{equation}
g'_{ab}(\sigma) = e^{2\omega(\sigma)} g_{ab}(\sigma).
\end{equation}

To proceed with the quantization we need to remove the redundancy
from the local symmetries, to fix the gauge.  Noting that the
metric has three components and there are three local symmetries
(two coordinates and the scale of the metric), it is natural to
do this by conditions on the metric, setting
\begin{equation}
g_{ab}(\sigma) = \delta_{ab}. \label{flat}
\end{equation}
This is always possible at least locally.
The Polyakov action~(\ref{poly}) then reduces to $D$ copies of
the earlier scalar action~(\ref{xact}),\footnote
{To be precise, because of the Minkowski signature, the action for
$X^0$ has the opposite sign and gives a divergent gaussian path
integral.  This is not problem; the path integral is implicitly
defined by the Euclidean rotation~$X^0 \to -iX^D$.  This is similar
to the treatment of Grassman path integrals---we don't have to take
them seriously as integrals, as long as they have certain key
properties, most notably factorization (so we can cut them open to
get a Hamiltonian formalism) and the integral of a derivative
vanishing (so we can derive equations of motion).} provided that
we choose units such that $\alpha'$, which has units of
length-squared, is equal to~2.

It is not an accident that the gauge-fixed action is conformally
invariant.  Fixing the flat metric does not fully determine the
local coordinate system.  From the discussion of conformal
invariance we know that if two coordinate systems are related by
\begin{equation}
\sigma'^1 + i \sigma'^2 = f(\sigma^1 + i \sigma^2) \label{conf}
\end{equation}
for analytic $f$, the metric changes only by a position-dependent
rescaling, so that a Weyl transformation restores it to its
original form.  In other words, the coordinate
transformation~(\ref{conf}) combined with the appropriate Weyl
transformation leaves the metric in flat gauge and so is a
conformal symmetry of the flat world-sheet action.\footnote
{Of course there will be some global conditions that fix most or all
of this residual invariance, as we will discuss further later.
This is not relevant now:
we noted earlier that to derive
Noether's theorem and the Ward identities we only need a symmetry
transformation to be defined in a region.}

So the two-dimensional spacetime of the
previous section is now the string world-sheet, while
spacetime is the field space where the $X^\mu$ live, the
{\it target space} of the map $X^\mu:$ world-sheet $\to$
spacetime.

\subs{The Spectrum}

For a closed string, where the spatial coordinate $\sigma^1$
is periodic,
we can immediately use the earlier results to write down the
spectrum.  We have $D$ sets of harmonic oscillators,
\begin{equation}
[\alpha^\mu_m, \alpha^\nu_n] = m \delta_{m+n,0} \eta^{\mu\nu} ,
\qquad
[\tilde\alpha^\mu_m, \tilde\alpha^\nu_n] = m \delta_{m+n,0}
\eta^{\mu\nu} , \label{osc}
\end{equation}
the covariant generalization of the earlier commutator, as well as
the $D$ momenta $\alpha^\mu_0 = \tilde \alpha^\mu_0 = k^\mu$.
Starting from the states $|0,k\rangle$ which are annihilated by
the $m < 0$ modes, we build the spectrum by acting any number of
times with the $m > 0$ modes.  By choice of conformal gauge, the
string thus separates into a superposition of harmonic oscillators.

But there is one more point to deal with.  The index $\mu$ on the
oscillators~(\ref{osc}) runs over $D$ values.  A
string stretched out in the
$X^1$-direction should by able to oscillation in the
$D-2$ transverse directions $\mu = 2, \ldots, D-1$, but
oscillation along the $X^1$-direction leaves the world-sheet
unchanged---according to the earlier discussion it is just an
oscillation of the parameterization~$\sigma$.  The same is true
of oscillation in the $X^0$-direction.  It is essential that this
be true, because from the oscillator algebra we have the inner
product
\begin{equation}
\langle\!\langle 0,k|\alpha^\mu_1 \alpha^\nu_{-1}|0,k'\rangle
= (2\pi)^D \delta^D (k-k') \eta^{\mu\nu}.
\end{equation}
The timelike oscillation $\alpha^\nu_{0}|0,k'\rangle$ thus has a
negative norm and so had better not be in the Hilbert space of the
theory.

The point is that when we fix $g_{ab}$ we lose the equations of
motion we get from varying $g_{ab}$, and we have to restore them
as constraints.  Varying the metric gives
\begin{equation}
T_{ab} = 0.  \label{tcon}
\end{equation}
In fact $T_{z\bar z}$ vanishes as a consequence of the
$X^\mu$ equation of motion, but $T_{zz}$ and $T_{\bar z \bar z}$ do
not.  The equation of motion does imply that if they vanish at one
time they vanish for all times; this is a general feature of such
missing equations of motion.  Classically, then, we impose these
equations on
the initial values; quantum mechanically we impose them on the
states.  In either case the constraint is then preserved by the
dynamics.

Later we will discuss a general and powerful way to implement the
constraints, the BRST quantization, but it is useful to proceed
first by a bit of trial and error (so-called {\it old covariant
quantization}~\cite{Brow}).  Going to the Laurent modes, we
could try to impose
$L_n |\psi\rangle = 0$ for all $n$.  But this is inconsistent with
the Virasoro algebra, since it would imply that
$0 = [L_m, L_{-m}]  |\psi\rangle = \frac{c}{12} (m^3 - m)
|\psi\rangle$.  Instead we require physical states to satisfy
\begin{eqnarray}
(L_0 - a) |\psi\rangle \ =\ (\tilde L_0 - a) |\psi\rangle &=& 0
\nonumber\\
L_n |\psi\rangle \ =\ \tilde L_n |\psi\rangle &=& 0, \qquad n>0\ ,
\label{ocq}
\end{eqnarray}
allowing a possible ordering constant in the $L_0$ condition, which
will turn out to be necessary.
This implies that matrix elements
of~(\ref{tcon}) between physical states vanish for {all} $n$,
\begin{equation}
\langle \psi | (L_n - a \delta_{n,0}) | \psi' \rangle =
\langle \psi | (\tilde L_n - a \delta_{n,0}) | \psi' \rangle = 0,
\end{equation}
the $n < 0$ generators annihilating the bra since $L_{-n} =
L^\dagger_n$.

There is one more provision.  A state of the form
\begin{equation}
L_{-n} | \chi \rangle + \tilde L_{-n} | \tilde\chi \rangle,
\qquad n>0
\end{equation}
for any $|\chi \rangle, | \tilde\chi \rangle$
is orthogonal to all physical states and so is called spurious.
A physical state which is also spurious is called null.
All physical amplitudes involving such a state vanish, so it is
physically equivalent to the zero state.  Thus we define an
equivalence relation between physical states
\begin{equation}
|\psi\rangle \cong |\psi'\rangle \ \ {\rm if}
\ \ |\psi'\rangle - |\psi\rangle = L_{-n} | \chi \rangle + \tilde
L_{-n} | \tilde\chi \rangle.
\end{equation}
The `observable' Hilbert space is the set of equivalence classes,
physical states modulo null states.

Let us see how this works for some of the lowest levels.  We focus
on the open string because it has only one set of modes.  At the
first level are the states $|0,k\rangle$, with all internal
oscillators in their ground states.  The physical state
conditions~(\ref{ocq}) for $n>0$ all involve lowering operators
and so hold.  There remains
\begin{equation}
0 = (L_0 - a)|0,k\rangle = (2k^2 - a) |0,k\rangle.
\end{equation}
Thus we obtain a mass shell condition,
\begin{equation}
M^2 = -k^2 = -\frac{a}{2} \ \to\ -\frac{a}{2} \frac{2}{\alpha'}
= -\frac{a}{\alpha'}. \label{tach}
\end{equation}
where we have restored $\alpha'$ by dimensional analysis.

At the next level the states are
\begin{equation}
|e,k\rangle = e_\mu \alpha^\mu_{-1} |0,k\rangle.
\end{equation}
for some polarization vector $e_\mu$.
The nontrivial physical state conditions are
\begin{eqnarray}
0&=& (L_0-a)|e,k\rangle \ =\ (2k^2 + \alpha_{-1}\cdot\alpha_1 - a)
|e,k\rangle\ =\ (2k^2 + 1 - a)|e,k\rangle \nonumber\\
0&=& L_1|e,k\rangle \ =\ 2k \cdot \alpha_1
|e,k\rangle\ =\ 2k \cdot e |0,k\rangle.
\end{eqnarray}
These give a mass shell condition and a transversality condition
on $e$,
\begin{equation}
M^2 = -k^2 = \frac{1-a}{\alpha'},\qquad
k \cdot e = 0. \label{lev1}
\end{equation}
There is a spurious state,
\begin{equation}
L_{-1} |0,k\rangle = 2k \cdot \alpha_{-1} |0,k\rangle.
\end{equation}
Thus $|e,k\rangle$ is spurious for $e^\mu \ \propto k^\mu$.
According to the conditions~(\ref{lev1}), this is physical, and so
null, only if
$k \cdot k = 0$, which is the case only if $a = 1$.

There are now three cases:\\[3pt]
{\it i}) If $a < 1$, the mass-squared is positive.  Going to the
rest frame, $k^\mu = (m,0,0,\ldots,0)$, the physical state
condition $k \cdot e = 0$ removes the negative norm timelike
polarization and leaves the $D-1$ spacelike polarizations.  There are
no null states, and the spectrum consists of the
$D-1$ positive-norm states of a massive vector particle.\\[3pt]
{\it ii}) If $a = 1$, the mass-squared is zero.
Going to a frame in which
$k^\mu = (\omega,\omega,0,0,\ldots,0)$, the physical
states are $e^\mu \propto k^\mu$ plus the $D-2$ transverse
polarizations $\mu = 2, \ldots, D-1$.  The state $e^\mu \propto k^\mu$
is null,
leaving the $D-2$ positive-norm transverse states of a massless vector
particle.\\[3pt]
{\it iii}) If $a > 1$, the mass-squared is
negative and we can go to a frame $k^\mu = (0,k^1,0,\ldots,0)$.
The physical state condition
removes a positive-norm {\it spacelike} polarization.  There is
no null state, so we are left
with $D-2$ positive-norm spacelike polarizations and
one negative-norm timelike polarization.

Case {\it (iii)} is obviously unacceptable, but either {\it (i)}
or {\it (ii)} seem satisfactory so far.  It is case~{\it
(ii)} that agrees with the BRST quantization, and also with the
light-cone quantization, a different gauge in which the number of
oscillators is reduced to $D-24$ from the start.  In fact, there
is no consistent way known to introduce interactions in
case~{\it (i)}.

The result at the next level, states
\begin{equation}
f_\mu \alpha^\mu_{-2} |0,k\rangle + f_{\mu\nu} \alpha^\mu_{-1}
 \alpha^\nu_{-1} |0,k\rangle,
\end{equation}
is quite interesting.  It
depends on the constant
$a$ and also on the spacetime dimension $D$.  If $a > 1$ or $D >
26$, there are negative-norm states. If  $a \leq 1$ and $D \leq
26$ the OCQ spectrum has positive norm.  For the particular case
$a = 1$ and $D = 26$ the observable spectrum is the same as the
BRST and light cone spectra;
otherwise there are extra states.  The derivation is left to the
reader.

This pattern persists at all higher levels as well: the
observable spectrum has only positive norm states provided
$a \leq 1$ and $D \leq 26$, and if $a = 1$ and $D = 26$ it is
the same as the BRST and light cone spectra.
This is the {\it no-ghost theorem} for the spectrum.  This is our
first encounter with the critical dimension; later we will
understand it more deeply.

For the value $a=1$, the second level is a
massless vector particle, a gauge boson.  This implies a
spacetime gauge symmetry.  In amplitudes this symmetry appears as
spacetime (not world-sheet) Ward identities, to the effect that
the unphysical polarization is not produced, and the amplitudes
for equivalent polarizations $e^\mu \cong e^\mu + c k^\mu$ are
equal.  We will see how this works when we discuss interactions.
Since null and unphysical states appear at all levels, this
means that ordinary gauge symmetry is only one piece of
some much large gauge symmetry in string theory.  We will
discuss the form of this later.

Notice also that for $a=1$ the lowest state (\ref{tach}) is a
tachyon, $M^2 = -1/\alpha'$.  Since the potential for a scalar
field is $\frac{1}{2} M^2 \varphi^2$, this means that the ground
state is unstable.  (One way to think about this is as resulting from
the negative Casimir energy of the $X^\mu$'s.)  We are using the
bosonic string only as a toy model, and are indeed expanding around an
unstable state.  The superstring does not have a tachyon, though
unfortunately time will not permit me in these lectures to give a
detailed treatment of this.

The closed string spectrum is just the tensor product of two copies
of the above, one right-moving and one left-moving.  The lowest
state $|0,k\rangle$ satisfies\footnote {I am using the same
notation for open and closed string states. It should always be
clear which is meant.}
\begin{equation}
M^2 = -\frac{4a}{\alpha'}.
\end{equation}
The next level
\begin{equation}
|e,k\rangle = e_{\mu\nu} \alpha^\mu_{-1} \tilde\alpha^\nu_{-1}
|0,k\rangle
\end{equation}
satisfies
\begin{equation}
M^2 = \frac{4(1-a)}{\alpha'}, \qquad k^\mu e_{\mu\nu} = k^\nu
e_{\mu\nu} = 0. \label{ctens}
\end{equation}
The correct values are again $a=1$, $D=26$, so these states are
massless, and there are null states leading to the equivalence
relation
\begin{equation}
e_{\mu\nu} \sim e_{\mu\nu} + k_\mu \zeta_\nu + \zeta'_\mu k_\nu
\end{equation}
with $k \cdot \zeta = k \cdot \zeta' = 0$.  Again letting
$k^\mu = (\omega,\omega,0,0,\ldots,0)$, a complete set of
$(D-2)^2$ observable states is obtained from transverse
$e_{\mu\nu}$, $\mu,\nu \in 2, \ldots, D-1$.  This set can be
decomposed under the $SO(D-2)$ transverse rotation group,
into a traceless symmetric tensor, antisymmetric tensor and
invariant.  These are respectively the graviton, antisymmetric
tensor, and dilaton.  Again there are spacetime gauge invariances
associated with the graviton and antisymmetric tensor.

\subs{The Weyl Anomaly}

The fact that strings can be consistently quantized only in the
critical dimension is due to an anomaly, a quantum violation of
the local world-sheet symmetries.
To see the anomaly, we will work in a more general gauge
in which the gauge symmetry is used to fix the
metric to some general form $ g_{ab}(\sigma)$, not necessarily
flat.  The result should be independent of what
$ g_{ab}(\sigma)$ we choose; let us see if it is. I should warn
you that the next lecture or so will get steadily more
technical, but then things will get better again.  I will try to
highlight the main results that we will need later on.

We start by examining the path integral over $X^\mu$ in the
fixed metric $ g_{ab}$.  The action is diff $\times$
Weyl invariant, but we have to define the path integration.
It is easy to preserve the diff invariance.  For example, expand
$X^\mu(\sigma)$ in a complete set of eigenfunctions of
the invariant Laplacian $\nabla^2$, and put a cutoff on the
eigenvalues.  Any cutoff will refer to
world-sheet lengths, so the Weyl invariance is not automatically
preserved and there may be a Weyl anomaly~\cite{CD,Poly1}.
We must check the Weyl invariance by
explicit calculation.  This is one
virtue of the Polyakov action---any possible anomaly appears in
the Weyl symmetry, which is somewhat easier to work with than an
anomaly in the diff invariance.

Let us expand around the flat metric in
the plane, $ g_{ab}(\sigma) = \delta_{ab} + h_{ab}(\sigma)$.
The variation of the action with respect to the metric is the
energy-momentum tensor, so to second order in $h_{\bar z\bar z}$
the path integral is
\begin{equation}
<1> \stackrel{O(h^2)}{=}
\frac{1}{8\pi^2} \int d^2z\, d^2z'\, h_{\bar z\bar z}(z,\bar z)
h_{\bar z\bar z}(z', \bar z')
< T_{zz}(z)  T_{zz}(z') >\ +\ {\rm local}. \label{Oh2}
\end{equation}
The local term comes from the second order variation of the action;
we will not need its explicit form.
The expectation value is evaluated on the flat world-sheet,
and from the OPE we know it to be
\begin{equation}
< T_{zz}(z)  T_{zz}(z') >\ = \frac{c}{2(z - z')^4} =
-\frac{c}{12} \partial_z^4 \ln |z-z'|^2
= \frac{\pi c}{6}  \partial_z^4 (\partial_z \partial_{\bar
z})^{-1}
\delta^2(z - z', \bar z - \bar z'),
\end{equation}
where $c=D$ for the $X$ theory.
Thus the second order term~(\ref{Oh2}) becomes
\begin{eqnarray}
<1>&\stackrel{O(h^2)}{=}&\frac{c}{48\pi} \int d^2z\,
d^2z'\, \partial^2 h_{\bar z\bar z}(z,\bar z)
\frac{1}{\partial \bar \partial} \partial^2 h_{\bar
z\bar z}(z', \bar z') \nonumber\\
&\to & \frac{c}{96\pi} \int d^2\sigma\, d^2\sigma'\,
\sqrt{ g} R(\sigma) \frac{1}{\nabla^2}(\sigma,\sigma')
\sqrt{ g} R(\sigma')\ .  \label{liou}
\end{eqnarray}
The second line, using the
curvature scalar built from $g_{ab}$,
is the unique coordinate invariant form to order
$h^2$ with the given nonlocal $(z - z')^{-4}$ term.  Now, under a
Weyl transformation $g'_{ab} = e^{2\omega } g_{ab}$, the curvature
changes
\begin{equation}
\sqrt{g'}R' = \sqrt{g}R - 2\nabla^2 \omega. \label{Rweyl}
\end{equation}
Thus
\begin{equation}
\delta_{\rm Weyl} <1>
\ \stackrel{O(h^2)}{=} \frac{c}{24\pi} <1> \int
d^2\sigma\,\sqrt{ g} R\delta\omega.
\label{weylan}
\end{equation}

Although derived to second order in the background, the
result~(\ref{weylan}) is in fact the full answer.  Given that the
Weyl variation must be local (since it comes from the violation
of the symmetry by the cutoff) and that the theory is
conformally invariant on the flat world-sheet,
eq.~(\ref{weylan}) is the only possible form.\footnote{For
example, terms with more derivatives would be suppressed by
powers of the cutoff.  A possible Weyl variation
$\int d^2\sigma\,\sqrt{g}\delta\omega$ can be removed by a
counterterm $\int d^2\sigma\,\sqrt{g}$ in the action.} This
goes through for any CFT, so we have the result that if we put
a CFT on a curved world-sheet the Weyl-dependence is
determined entirely by the central charge of the flat
world-sheet theory,
\begin{equation}
\delta_{\rm Weyl} \ln < \ldots >
\ = \frac{c}{24\pi} \int d^2\sigma\,\sqrt{g}R\delta\omega,
\end{equation}
where as usual the transformation properties of the insertions
`$\ldots$' are a separate issue.
Also, we can integrate the Weyl anomaly to give the full
result\footnote
{But what if we have a CFT with $c \neq \tilde c$?
Then we cannot extend the result as in the second line of
eq.~(\ref{liou}) to a diff invariant: the $h_{zz}^2$ and $h_{\bar z
\bar z}^2$ terms are inconsistent.  That is, there {\it is} in
this case an anomaly in the two-dimensional coordinate
invariance~\cite{A-GW}.  The necessary and sufficient condition for
this to be absent is $c = \tilde c$.}
\begin{equation}
<1>\ = \exp\biggl\{\frac{c}{96\pi} \int\!\int d\sigma\,d\sigma'\,
R \nabla^{-2} R \biggr\}.
\end{equation}

To complete the determination of the Weyl anomaly we need to
carry out the gauge fixing carefully, taking into account the
Fadeev-Popov determinant.  That is, we write the integral over
metrics as in integral over the gauge group times a Jacobian,
and divide by the gauge volume.  Under a small diff $\times$
Weyl transformation, the change in the metric is
\begin{eqnarray}
\delta g_{ab} &=& - (\nabla_a \delta \sigma_b
+ \nabla_b \delta \sigma_a - g_{ab} \nabla \cdot \delta \sigma)
- g_{ab} (2\delta\omega + \nabla \cdot \delta \sigma)
\nonumber\\
&=& -(P_1 \delta \sigma)_{ab} - 2g_{ab} \delta\omega'.
\end{eqnarray}
In the first line we have separated the variation into
a traceless part and a part proportional to the metric.
In the second line we have defined the differential operator
$P_1$ taking vectors to traceless symmetric tensors, and
absorbed the $\nabla \cdot \delta \sigma$ term in a shift of
$\omega$.  Short circuiting some formal steps that are
parallel to the standard treatment of non-Abelian gauge
theories, we get
\begin{equation}
\int [dg] \to \det(P_1) V_{\rm diff \times Weyl}\ .
\end{equation}
The $\delta\omega'$ part being purely local gives a trivial
Jacobian, equivalent to local counterterms.  Thus,
\begin{equation}
\frac{1}{V_{\rm diff \times Weyl}} \int [dX\,dg]\, e^{-S_{\rm
P}[g]} = \det(P_1) \int [dX]\, e^{-S_{\rm P}} . \label{gafix}
\end{equation}

In order to expose the general structure of amplitudes it
is useful to rewrite the determinant as a path integral over
anticommuting Fadeev-Popov ghost fields.  From the definition of
Grassman integration we have
\begin{equation}
\int db\,dc\, e^{-bMc} = \int db\,dc\,(1 - bMc) = M
\end{equation}
for two anticommuting variables $b, c$.  This generalizes to
\begin{equation}
\int {\textstyle \prod_i} (db_i\,dc_i)\,
e^{-b_i M_{ij} c_j}   = \det M\ ,
\end{equation}
as is evident by diagonalizing $M$.  This applies to functional
determinants as well, so
the path integral~(\ref{gafix}) becomes
\begin{equation}
\int [dX\, db\, dc]\, e^{-S_{\rm P} - S_{\rm g}}
\end{equation}
where
\begin{equation}
S_{\rm g} = \frac{1}{2\pi}  \int d^2\sigma\, \sqrt{g}\,
b^{ab} (P_1 c)_{ab}. \label{ghact}
\end{equation}
To see the conformal transformation properties, consider a
conformally flat metric
\begin{equation}
g_{ab}(\sigma) = e^{\phi (\sigma)} \delta_{ab}. \label{conflat}
\end{equation}
The action~(\ref{ghact}) becomes
\begin{equation}
S_{\rm g} = \frac{1}{2\pi}  \int d^2z\,
\Bigl\{ b_{zz} \partial_{\bar z} c^z
+ b_{\bar z \bar z} \partial_{z} c^{\bar z} \Bigr\}.
\label{ghact2}
\end{equation}
We have used a convenient trick here.  The
metric~(\ref{conflat}) is not flat, so the covariant
derivatives are nontrivial.  But for the special case of a $z$
derivative acting on a tensor with only $\bar z$ indices, or
vice versa, the covariant derivatives reduce to the ordinary
ones.  We have raised and lowered indices so as to take
advantage of this.  In the form~(\ref{ghact2}) it is evident
that the action is Weyl-invariant with $c^{z}, c^{\bar z},
b_{zz}$, and $b_{\bar z \bar z}$ neutral under the Weyl
transform.  The conformal transformation of these fields comes
only from the coordinate transformation, so $c^z$ is a $(-1,0)$
tensor,
$c^{\bar z}$ a $(0,-1)$ tensor, $b_{zz}$ a $(2,0)$ tensor, and
$b_{\bar z\bar z}$ a $(0,2)$ tensor.

The ghosts thus are a $\lambda = 2$ $bc$ system.  Referring
back to the earlier result~(\ref{bccc}), the ghost system has
central charge $-26$.  As we have seen, this determines the
Weyl transformation properties, so for the combined
$X$ and ghost system we have
\begin{equation}
\delta_{\rm Weyl} \ln < \ldots >
\ = \frac{D-26}{24\pi} \int d^2\sigma\,\sqrt{g}R\delta\omega.
\end{equation}
So the Weyl invariance is anomalous except in the critical
dimension $D = 26$~\cite{Poly1}!

Is there any possibility for making sense of the theory without
Weyl invariance?  One possibility is to make some
choice for the scale factor of the metric, but there is no
natural way to do this.  For example, the rather natural choice
$\sqrt{g} = 1$ is not diff invariant and just moves the
anomaly into that symmetry.  The other possibility is to integrate
over the scale factor---that is, to treat the theory as one that
has only coordinate invariance, and use this to fix only two of the
three components of the metric~\cite{Poly1}.  This is a rather large
change in the theory, introducing a new degree of freedom.  Such
`non-critical' string theories are of great interest, and we
will return to them later.

For now, let us just point out the
following.  We can use the coordinate invariance to bring the
metric to the form
\begin{equation}
g_{ab}(\sigma) = e^{\varphi(\sigma)} \hat g_{ab}(\sigma),
\label{cnonc}
\end{equation}
where $\varphi(\sigma)$ is to be integrated.  Now, this theory has
a `fake' Weyl invariance, under which $g_{ab}$ is neutral but
\begin{equation}
\hat g_{ab} (\sigma) \to e^{2\omega(\sigma)}
\hat g_{ab} (\sigma), \qquad \varphi(\sigma) \to
\varphi(\sigma) - 2\omega(\sigma).
\end{equation}
But this is indistinguishable from a theory in which $\hat
g_{ab}$ is the `real' metric and $\varphi$ is another field.
So we can regard this as a Weyl-invariant theory with an extra
degree of freedom.  Strominger and Verlinde will be interested
in two-dimensional coordinate invariance without Weyl
invariance, and will make use of this trick to use techniques
from CFT.  It is interesting to apply this to a theory whose
Weyl-variation comes only from the central charge.  Inserting
the metric~(\ref{cnonc}) into the action~(\ref{liou})
and using the relation~(\ref{Rweyl}) gives the
$\varphi$-dependent
\begin{equation}
\frac{c}{96\pi} \int d^2\sigma \sqrt{\hat g}
\Bigl\{ \hat g^{ab} \partial_a \varphi \partial_b \varphi + 2 \hat R
\varphi \Bigl\}\ .  \
\end{equation}
The first term is the same as the action for one of the
coordinates $X$, so this looks like an extra dimension.
The second term is not translationally invariant, so it means
that this theory may have some cosmological interpretation, but
is not relevant to the translationally invariant surroundings
that we find ourselves in.

It may seem odd that a theory without Weyl invariance can be
regarded as having it.  This is the first example of a general
theme which will arise again, both on the world-sheet and in
spacetime.  This is that gauge invariance is in the end just a
redundancy, though a useful one.  We can always add redundant
fields and sometimes it is useful to do so, as here where
it will enable us to apply critical string methods to the
noncritical string.

\subs{BRST Quantization}

Representing the Fadeev-Popov determinant in terms of ghosts gives us an
even larger Hilbert space, and we need to identify the observable states.
This leads us to
BRST quantization, which is a general method for quantizing systems with
gauge symmetries and is indispensable for understanding the general
structure of string amplitudes.\footnote{My treatment is similar
to that in many modern field theory texts.  For other points of
view, see refs.~\cite{KO'},~\cite{HT}.}
One way to motivate it is to
imagine some
arbitrary small change in the gauge-fixing condition.  Thus far we have
taken a gauge in which $g_{ab}(\sigma)$ was fixed.  That is,
$g_{ab}(\sigma) - \overline g_{ab}(\sigma) = 0$ for
some fixed function $\overline g_{ab}(\sigma)$.  In the previous
section we in effect
checked that the result was invariant under a Weyl
transformation of $\overline g_{ab}$.  But we could imagine a much
more general change of gauge
\begin{equation} g_{ab}(\sigma) -
\overline g_{ab}(\sigma) - \delta F_{ab}(g,X) = 0\ . \label{genga}
\end{equation}
For example, the `manifestly unitary' light-cone gauge places conditions on
$X$ as well as $g_{ab}$; to interpolate between the conformal and
light-cone gauges one would have to consider deformations of the
form~(\ref{genga}).

In order to derive the full invariance condition, it is useful to take
a more general and abstract point of view.  Consider a path integral
with a local symmetry.  The path integral fields are denoted
$\phi_i$, which in the present case would be $X^\mu(\sigma)$ and
$g_{ab}(\sigma)$.  Here we use a very condensed notation where $i$ labels the
field, the component, and {\it also} the coordinate $\sigma$.  The gauge
invariance is $\epsilon^\alpha \delta_\alpha$, where again $\alpha$ labels
component and also coordinate.  By assumption the gauge parameters
$\epsilon^\alpha$ are real, since we can alway separate a complex parameter
into its real and imaginary parts.
The gauge transformations satisfy an algebra\footnote{This is not the most
general gauge symmetry possible, but is sufficient for our application
to the string.}
\begin{equation}
[\delta_\alpha, \delta_\beta] = f_{\alpha\beta}\!^\gamma \delta_\gamma.
\label{gaalg}
\end{equation}
Now fix the gauge by conditions
\begin{equation}
F^A(\phi) = 0,
\end{equation}
where again $A$ includes the coordinate.  Following the usual Fadeev-Popov
procedure, the path integral becomes
\begin{equation}
\int \frac{[d\phi_i]}{V_{\rm gauge}}\, e^{-S_1} \to
\int [d\phi_i\, dB_A\, db_A\, dc^\alpha]\, e^{-S_1 - S_2 - S_3},
\end{equation}
where $S_1$ is the original gauge invariant action, $S_2$ is the
gauge-fixing action
\begin{equation}
S_2 = i B_A F^A(\phi),
\end{equation}
and $S_3$ is the Fadeev-Popov action
\begin{equation}
S_3 = b_A c^\alpha \delta_\alpha F^A(\phi).
\end{equation}
We have introduced the field $B_A$ to produce an integral representation of
the gauge-fixing $\delta(F^A)$.

There are two things to notice about this
action.  The first is that it is invariant under the {\it
Becchi-Rouet-Stora-Tyupin (BRST)
transformation}~\cite{BRS},
\begin{eqnarray}
\delta_{\rm B} \phi_i &=& -i\epsilon c^\alpha \delta_\alpha \phi_i,
\qquad
\delta_{\rm B} B_A \ =\ 0,
\nonumber\\
\delta_{\rm B} b_A &=& -\epsilon B^A,
\qquad
\delta_{\rm B} c^\alpha \ =\ -\frac{i}{2}
\epsilon c^\beta c^\gamma f_{\beta\gamma}\!^\alpha. \label{btran}
\end{eqnarray}
Note that the transformation mixes commuting and anticommuting objects,
so that $\epsilon$ must be taken to be anticommuting.
The original action $S_1$ is invariant by itself, because the action
of $\delta_{\rm B}$ on $\phi_i$ is just a gauge transformation with parameter
$c^\alpha$.  The variation of $S_2$ cancels the variation of $b_A$ in
$S_3$, while the variations of $\delta_\alpha F^A$ and $c^\alpha$ in $S_3$
cancel.  The second key property is that
\begin{equation}
\delta_{\rm B} (b_A F^A) = -i\epsilon(S_2 + S_3). \label{s2s3}
\end{equation}

Now consider a small local change $\delta F$ in the gauge-fixing condition.
The change in the gauge-fixing and ghost actions gives
\begin{equation}
\epsilon \delta \langle \psi |\psi' \rangle =
-i \langle \psi | \delta_{\rm B} (b_A \delta F^A) |\psi' \rangle
= \langle \psi | \{ Q_{\rm B}, b_A \delta F^A \}|\psi' \rangle, \label{fqbi}
\end{equation}
where we have written the BRST variation as an anticommutator with the
corresponding conserved charge $Q_{\rm B}$.  For physical states the
amplitude must be independent of the gauge condition.
In order that this hold for arbitrary $\delta F$, it must be that
\begin{equation}
Q_{\rm B} \ket{\psi} = Q_{\rm B} \ket{\psi'} = 0,
\end{equation}
with $Q_{\rm B}^\dagger = Q_{\rm B}$.
This is the essential condition:
{\it physical states must be BRST invariant}.

There is one more key idea.  In order to move around in the space of gauge
choices, the BRST charge must remain conserved.  Thus it must commute with
the change in the Hamiltonian,
\begin{eqnarray}
0 &=& [Q_{\rm B}, \{ Q_{\rm B}, b_A \delta F^A \} ]   \nonumber\\[2pt]
&=& Q_{\rm B}^2 b_A \delta F^A - Q_{\rm B} b_A \delta F^A Q_{\rm B}
+ Q_{\rm B} b_A \delta F^A Q_{\rm B}
- b_A \delta F^A Q_{\rm B}^2\nonumber\\[2pt]
&=& [Q_{\rm B}^2,  b_A \delta F^A ].
\end{eqnarray}
In order for this to vanish for general changes of gauge, we need
\begin{equation}
Q_{\rm B}^2 = 0.
\end{equation}
That is, the BRST charge is {\it nilpotent}.
You can check that acting twice with the BRST
transformation~(\ref{btran}) does indeed leave all fields invariant.

The nilpotence of $Q_{\rm B}$ has an important consequence.  A state of the
form
\begin{equation}
Q_{\rm B} \ket{\chi} \label{nulst}
\end{equation}
will be annihilated by $Q_{\rm B}$ for any $\chi$ and so is physical.  However,
it is orthogonal to all physical states including itself:
\begin{equation}
\bra{\psi}| \Bigl( Q_{\rm B} \ket{\chi} \Bigr) =
\Bigr( \bra{\psi}| Q_{\rm B} \Bigr) \ket{\chi} = 0
\end{equation}
if $Q_{\rm B} \ket{\psi} = 0$.  All physical amplitudes involving such a {\it
null
state} thus vanish.  Two physical states which differ by a null state,
\begin{equation}
\ket{\psi'} = \ket{\psi} + Q_{\rm B} \ket{\chi}
\end{equation}
will have the same inner products with all physical states and are therefore
indistinguishable.  So we again identify the observable Hilbert space as a
set of equivalence classes, the {\it cohomology} of $Q_{\rm B}$.
Following terminology from cohomology, a BRST-invariant state is also
called {\it closed}, and a null state {\it exact}.

Applying this to string theory~\cite{KO} gives the BRST
transformation\footnote {A few details are left to the student: we
have kept only the ghosts associated with the coordinate part,
because the Weyl ghosts are auxiliary fields (no derivatives) and
can be integrated out.  And we have implicitly integrated $B_{ab}$
to give the delta-function on the metric, so the equation of motion
has been used for $B_{ab}$.}
\begin{eqnarray}
&&\delta_{\rm B} X^\mu = i\epsilon (c\partial + \tilde c
\bar\partial) X^\mu
\nonumber\\
&&\delta_{\rm B} c = i\epsilon (c\partial + \tilde c \bar\partial) c
\qquad
\delta_B \tilde c = i\epsilon (c\partial + \tilde c \bar\partial)
\tilde c
\nonumber\\
&&\delta_{\rm B} b =  i\epsilon(T^X + T^{\rm g})\qquad
\delta_{\rm B} \tilde b =  \epsilon(\tilde T^X + \tilde
T^{\rm g}), \label{bzt}
\end{eqnarray}
where the energy-momentum tensor has been divided into matter and
ghost parts.  Noether's theorem gives the BRST current
\begin{equation}
j_{\rm B} = cT^{ X} + \frac{1}{2} :\! cT^{\rm g}\! :
= cT^{ X} + :\! b c \partial c \!: , \label{brsj}
\end{equation}
and correspondingly for $\bar j_{\rm B}$.  (A total derivative can
be added to make this a tensor.)  This form is rather general: the
$c$-ghost times the matter gauge current plus half the ghost gauge
current.

Based on our previous experience, we expect that something will go
wrong outside the critical dimension.  The problem is with $Q_{\rm
B}^2$.  We can calculate this from the $j_{\rm B} j_{\rm B}$ OPE, but
the following  gives a slight shortcut.  The OPE
\begin{equation}
j^{B}(z)b(0) \sim
-\frac{1}{z^2} :\! bc(0) \!:  + \frac{1}{z}
( T^{X}(0) + T^{\rm g}(0) ) \label{jbb}
\end{equation}
gives the commutator for the corresponding charge,
\begin{equation}
\{Q_{\rm B}, b_m \} = L^{X}_m + L^{\rm g}_m.
\end{equation}
{\bf Exercise:}  Using this anticommutator and the Jacobi
identity, show that
$\{ [ Q_{\rm B},
L^{X}_m + L^{\rm g}_m ], b_n \}$ vanishes if and only if the
total central charge $D - 26$ vanishes.  This implies in fact that
$[ Q_{\rm B},
L^{X}_m + L^{\rm g}_m ]=0$, because by ghost number any term in
the commutator would have to contain at least one $c$ mode.  Now
extend this to $[ \{ Q_{\rm B}, Q_{\rm B} \}, b_n ]$: it vanishes,
and so also does
$\{ Q_{\rm B}, Q_{\rm B} \}$, if and only if $D=26$.\\[3pt]
So the Weyl
anomaly shows up as an anomaly in the nilpotence $Q_{\rm B}^2 =
0$~\cite{KO}, which is necessary for the consistency of the
formalism.

To see how all the formalism works, let us look again at the lowest
levels of the open string.  There is one more condition that must be
imposed on the states, namely
\begin{equation}
b_0 | \psi \rangle = 0. \label{b0}
\end{equation}
The way I think about this is that when we work out the interactions later
on we will find that the string propagator always includes a factor $b_0$,
which projects onto states
satisfying~(\ref{b0})
because $b_0 b_0 = 0$.  Recalling the notation~(\ref{ghvac}), this
means that we are interested in states built with raising operators
acting on the ghost vacuum $|\! \downarrow \rangle$. The
condition~(\ref{b0}) implies in turn that physical states satisfy
\begin{equation}
\{Q_{\rm B}, b_0\} | \psi \rangle =
(L^{X}_0 + L^{\rm g}_0) | \psi \rangle = 0, \label{L0}
\end{equation}
where $L^{X}_0 + L^{\rm g}_0
= 2k^2 + {\sf L} - 1$,
with ${\sf L} = {\sf L}^{X } + {\sf L}^{\rm g}$
the total ghost plus $X$ excitation
level, while the total normal ordering constant is taken from the
earlier result~(\ref{ghL}).  The condition~(\ref{L0}) thus relates
the mass of a string state to its level of excitation,
\begin{equation}
M^2 = \frac{ {\sf L} - 1 }{2}
\label{mshell}
\end{equation}
In terms of the modes,
\begin{equation}
Q_{\rm B} =
 \sum_{n= -\infty}^{\infty} c_n L^X_{-n} +
\sum_{m,n=-\infty}^{\infty}\frac{(m-n)}{2}
( c_m c_n b_{-m-n} )_{{}_{\it CA}}
- c_0\ , \label{qex}
\end{equation}
where the subscript {\it CA} denotes creation-annihilation normal
ordering.  The $c_0$ term comes from the normal ordering
constant in $L_0^{\rm g}$, as follows from $\{Q_{\rm B}, b_0
\} = L^{X}_0 + L^{\rm g}_0$.

At the lowest level, ${\sf L}=0$, the states are $|\! \downarrow, 0, k
\rangle$, denoting the ghost vacuum, $X$ vacuum and momentum.
Then
\begin{equation}
0 = Q_{\rm B} |\! \downarrow, 0, k \rangle = (2k^2 - 1)
c^0 |\! \downarrow, 0, k \rangle ,
\end{equation}
giving the same mass shell condition as in the old covariant
quantization.  In this formalism, the constant $a = 1$ is from the
ghosts.  There are
there are no exact states at this level,
so we have the same tachyon states as before.

At the next level, $N=1$, there are 26+2 states,
\begin{equation}
|\psi \rangle = (e \cdot \alpha_{-1} + \beta b_{-1} + \gamma c_{-1} )
|\! \downarrow, 0, k \rangle, \label{psi1}
\end{equation}
depending on a 26-vector $e_\mu$
and two constants $\beta$ and $\gamma$.
The BRST condition is
\begin{eqnarray}
0  =  Q_{\rm B} \ket{\psi} = 2(k^2 c_0 + k \cdot e c_{-1} + \beta k
\cdot \alpha_{-1})  |\! \downarrow, 0, k \rangle ,
\end{eqnarray}
so an invariant state satisfies $k^2 = 0$ and $k \cdot e = \beta = 0$.
There are $26$ linearly independent states left.  A general
$|\chi\rangle$ is of the same form~(\ref{psi1}) with constants
$e'_\mu$, $\beta'$, $\gamma'$, so the general BRST-exact state at this
level, with $k^2 = 0$, is
\begin{equation}
Q_{\rm B} \ket{\chi} = 2(k \cdot e' c_{-1} + \beta' k \cdot
\alpha_{-1})  |\! \downarrow, 0, k \rangle.
\end{equation}
Thus the ghost state $c_{-1}\ket{0;k}$ is BRST-exact,
while the polarization is
transverse with the equivalence relation
$e_\mu \cong e_\mu +2 \beta' k_\mu$.
This leaves the $24$ positive-norm states expected for a massless vector
particle.\\[3pt]
{\bf Exercise:} Do the first massive level.

This pattern is general: with the ghosts there are $26 + 2$
oscillators at each level.  The BRST condition eliminates two
of these, and two others are exact, leaving $26 - 2$ oscillators.
The BRST quantization is equivalent to the old covariant and
light-cone quantizations~\cite{KO},~\cite{FO}.  In fact, the BRST
quantization reduces to the old covariant quantization if we
consider only states where the ghosts are in their ground state.
The no-ghost theorem states that every cohomology class includes a
state of this form. The generalization to the closed string is again
straightforward, with $k^\mu \to k^\mu/2.$

This derivation of the BRST formalism makes it look like a
consequence of gauge fixing, but should be thought of as more
fundamental.  It carries the full information of the original gauge
symmetry; in effect, we increase the redundancy of the theory by the
addition of the ghosts, but the BRST principle then singles out the
physical observables.  When we try to generalize string theory, it is
generally easiest to generalize it within the BRST formalism, rather
than the locally invariant formalism.

It is worth noting that in most familiar
circumstances, Ward identities are important but are only part of
the story---one also has dynamics.  But string theory has coordinate
invariance, both on the world-sheet and in spacetime, so that time
itself is a gauge degree of freedom.  As a consequence the dynamics
becomes part of the constraints.  As an example, the mass shell
condition (which is the Fourier transform of a Klein-Gordon equation
in spacetime) arose from the physical state condition.

\subs{Generalizations}

We have developed the importance of the Weyl invariance,
and its relation to the central charge and the nilpotence of the BRST
operator.  Now we are in a position to
state one answer to the question, ``What is the most general
consistent string theory?''  We will restrict attention to
theories for which the action for the embedding
time takes the Polyakov form
\begin{equation}
- \frac{1}{8\pi} \int d^2 \sigma \,\sqrt{g}
g^{ab} \partial_a X^0 \partial_b X^0\ . \label{timeact}
\end{equation} \
That is, the theory is stationary, and (jumping ahead a bit) the
spacetime metric has $G_{0\mu} = -\delta_{0\mu}$.
Even in field theory, non-stationary situations are more
complicated.  There is no longer a distinguished zero-particle
vacuum state,
and the meaning of a `particle' becomes
ambiguous. This subject is much less developed in string theory, and
I will avoid it here.

Now, the immediate problem is the one we have discussed, that the
quantization of the action~(\ref{timeact}) leads to negative
norm states, and we need the observable spectrum to have a
positive norm.  The general solution is simply stated: it
depends only on the local symmetry, or BRST invariance, being
preserved.  That is, the no-ghost theorem holds if we
{\it replace the 25 spatial $X^\mu$ fields with any $(c, \tilde
c) = (25,25)$ unitary CFT.}  In particular, the calculation of
$Q_{\rm B}^2$ goes precisely as before, with $T^X$ replaced by the
$T$ of the CFT.  The unitary condition is necessary
because we do not want any other negative norm states besides
those from $X^0$.
Needless to say this generalization has something to do with the idea
of compactifying some of the spatial dimensions, but we will discuss
this in section~3.  We have not yet described string
interactions, but this is also the condition under which we
can introduce consistent interactions (the CFT
must satisfy the two conditions discussed earlier, OPE associativity
and one-loop one-point modular invariance).

I should emphasize that all of these theories have tachyons.
Whatever the spatial CFT is, it contains the unit operator.
The state $|0, k^0\rangle$ from the $X^0$ theory times
$|1\rangle$ from the spatial theory is in the physical spectrum
for $1 = L_0 = -\frac{1}{2} (k^0)^2$, which is unstable
(tachyonic).  To eliminate the tachyon we have to generalize in
a different way, enlarging the world-sheet gauge symmetry.
Everything that we have done generalizes readily to the $N=1$
superconformal algebra~\cite{Rns},~\cite{rNS}.  A simple
$N=1$-invariant extension of the bosonic string action uses $D$
copies of the theory~(\ref{n1free}),
\begin{equation}
S= \frac{1}{4\pi} \int d^2z\,
\Bigl\{\partial X^\mu \bar\partial X_\mu + \psi^\mu \bar\partial
\psi_\mu + \tilde\psi^\mu \partial \tilde\psi_\mu\Bigr\}\ .
\label{n1act}
\end{equation}
To be precise this is $(1,1)$ supersymmetric, with one analytic and
one antianalytic supercurrent.  To build a BRST charge we need a $bc$
or $\beta\gamma$ theory for each gauge current, with the $b$ or
$\beta$ ghost having the same weight and opposite statistics from the
current.  So in this case we need a $\lambda = 2$ $bc$ system and a
$\lambda = \frac{3}{2}$ $\beta\gamma$ system.  The BRST current
follows the same pattern as~(\ref{brsj}),
\begin{equation}
j_{\rm B} = c T^X + \gamma T^X_F + \frac{1}{2}:\!  (c T^{\rm g}
+ \gamma T^{\rm g}_F ) \! :\ .
\end{equation}
This is nilpotent in the critical dimension, which is again where
the total central charge vanishes.  The $\beta\gamma$ ghosts
contribute $c = \tilde c = 11$, while each $\psi^\mu$ contributes
$\frac{1}{2}$, for
\begin{equation}
c = \tilde c = \frac{3}{2} D - 26 + 11\ \Rightarrow\ D_{crit} = 10.
\end{equation}
Both the $\psi^0$ and $\alpha^0$ oscillators create negative norm
states, which in the critical dimension are removed by the
constraints.  The spectrum is equivalent to the light-cone gauge, in
which only $D-2$ transverse sets of modes remain.  The same
generalization as above can be made: {\it replace the 9 spatial
$X\psi$ theories with any $(c, \tilde c) =
(\frac{27}{2},\frac{27}{2})$ unitary $(1,1)$ super-CFT.}
\label{gensec}

The details of the spectrum---constraints from modular invariance,
the GSO projection, absence of tachyons, and spacetime
supersymmetry---make a long story that we do not have time for.  Let
me just mention that in the Ramond sector (integer modes) the Ramond
generator $G_0$ is
\begin{equation}
G_0 = k_\mu \psi^\mu + \ldots,
\end{equation}
terms with lowering operators being omitted, and
\begin{equation}
\{ \psi^\mu_0, \psi^\nu_0 \} = \eta^{\mu\nu}. \label{cliff}
\end{equation}
The algebra~(\ref{cliff}) is the same as that of the Dirac matrices,
and its unique representation is the $2^{D/2}$-dimensional Dirac
spinor representation.  The $G_0$ constraint then gives the Dirac
equation, just as the $L_0$ constraint gave the Klein-Gordon
equation.

The analytic and antianalytic constraint algebras need not be the
same~\cite{GHMR}.  The $(1,0)$ heterotic string combines an analytic
$N=1$ superconformal symmetry with antianalytic conformal symmetry.
The ghosts have central charge $(c,\tilde c) = (-15,-26)$.  For
$D=10$ the action~(\ref{n1act}) with $\tilde\psi$ omitted has central
charge $(15,10)$, so an additional $(0,16)$ is needed from another
unitary CFT.  The constraints of modular invariance are strong, and
the only solutions are an $E(8)\times E(8)$ or $SO(32)$ level~1
current algebra~\cite{GHMR},~\cite{GSan}.  The generalization is to
combine the
$(c,\tilde c) = (\frac{3}{2},1)$ CFT from $\mu = 0$ with the ghosts plus
any $(c,\tilde c) = (\frac{27}{2},25)$ theory with $(1,0)$
superconformal symmetry.  This string is the most promising for a
unified theory, the superconformal side eliminating the tachyon and
producing spacetime supersymmetry, and the side with the current
algebra giving rise to the gauge symmetries.

Moving on, the $N=2$ superconformal algebra needs $\lambda =
2$ and $\lambda = 1$ $bc$ systems plus two $\lambda = \frac{3}{2}$
$\beta\gamma$ systems for total central charge $-26-2 + 2 \cdot 11 =
-6$.  The basic free representation consists of one complex $X$ and
one complex $\psi$ (= two real) for $c=3$, so the critical theory
has two such representations.  As usual the constraints remove two
sets of modes, leaving in this case zero transverse dimensions.
The only degree of freedom is a scalar from the center of mass
motion of the string.  Nevertheless, there is some interesting
structure~\cite{OV}.

Higher-$N$ extended superconformal algebras have zero or negative
critical dimension.  These are the only possibilities as long as we
are restricted to half-integer spins $\leq 2$.  Both the fractional
case~\cite{Tye93} and the higher spin case~\cite{Pw93} (particularly
$W_3$) are under investigation.  In the fractional case, the OPE has
branch cuts so a projection is needed to get well-defined
amplitudes.  For both cases the product of two representations is not
a representation, owing to the nonlocality in the fractional case and
the nonlinearity in the $W_3$ case.  This has made it hard to find
theories with a spacetime interpretation, and also makes it hard to
construct the BRST operator (since the ghosts and matter do not
separate).  For $W_3$, a theory with a spacetime interpretation has
been found and the BRST operator constructed, but the result is
disappointing: it turns out to be a special case of the bosonic
string.
So the theories described above are for now the most general
known.\footnote
{I have restricted attention to algebras which have nontrivial unitary
representations, because this is necessary to get a spacetime
interpretation.  This implicitly excludes topological string
theories, about which I have little to say.}

String theories are also distinguished by the world-sheet topologies
allowed: closed only, or with boundaries, and oriented only, or
unoriented also.  Boundaries must be consistent with the constraint
algebra, as in~eq.(\ref{tbound}) for the conformal case.\footnote
{There are also additional conditions generalizing the OPE
associativity and modular invariance~\cite{CLew}.}  In
particular, the constraint algebra must be left-right symmetric, so
that boundaries are not possible in the heterotic case. Unoriented
surfaces are allowed only if the full world-sheet action is
invariant under world-sheet parity (exchange of $z$ and $\bar z$).
Theories without boundaries have closed strings only; theories with
boundaries have {\it both} closed and open strings.  There are no
string theories with open strings alone---a surface with boundary
can be cut open along a curve running from boundary to boundary,
corresponding to open string intermediate states, {\it or} along a
closed curve, corresponding to closed string intermediate states.
Inclusion of unoriented surfaces has the effect of projecting the
spectrum onto states of even parity.

\subs{Interactions}

For composite strings, there are a variety of contact and
long-range interactions.  In the fundamental string, such
interactions cannot be introduced without spoiling the local
symmetries, and are not possible.\footnote
{For example, a contact interaction $\int d^2\sigma\,d^2\sigma'\,
\sqrt{g(\sigma)}\sqrt{g(\sigma')} \delta(X(\sigma) - X(\sigma'))$
is not Weyl invariant (Fourier transforming the delta function,
one obtains exponentials with a continuum of weights.)}
The only interactions that are possible are those that are already
contained in the sum over all surfaces.  For example, fig.~12
shows a world-sheet in which one closed string splits into two,
or the time-reversed process in which two join into one.
\begin{figure}
\begin{center}
\leavevmode
\epsfbox{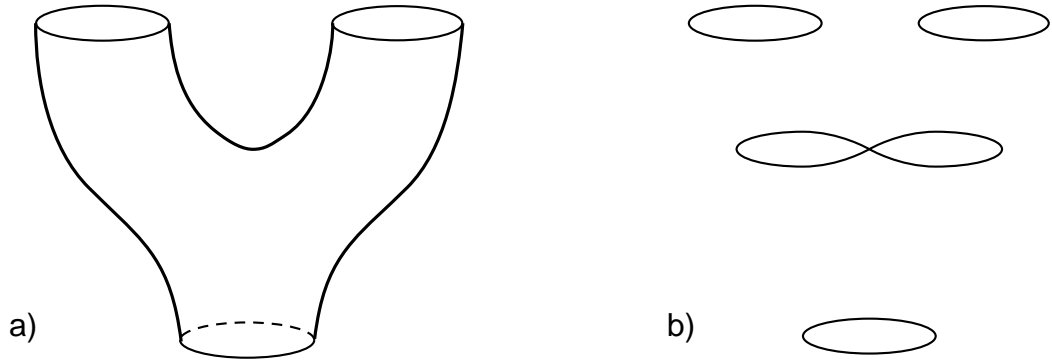}
\end{center}
\caption[]{a) World-sheet of closed string splitting.
b) Time slices of this process.}
\end{figure}
Figure~13 shows two strings scattering by exchange of one or two
strings.
\begin{figure}
\begin{center}
\leavevmode
\epsfbox{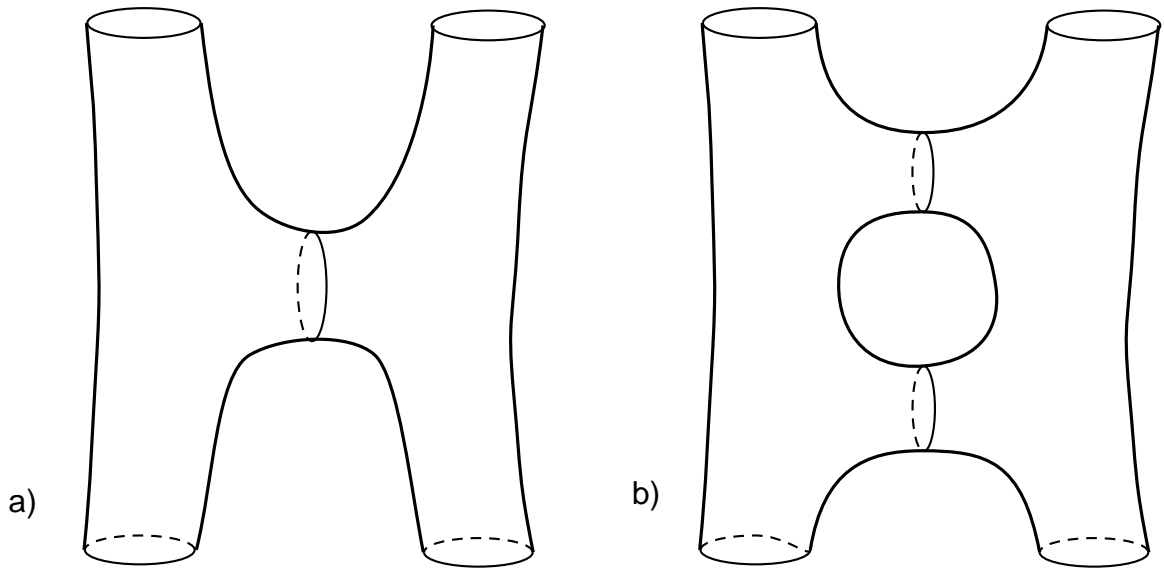}
\end{center}
\caption[]{String analogs of fig.~10bc: closed strings scattering by
exchange of one string (a) or two strings (b).}
\end{figure}
Since the string spectrum includes the graviton, these
amplitude include the gravitational processes discussed earlier.
But now there is no short-distance divergence because
the interaction is spread out.  I will be describing in somewhat more
detail how this works.

Now it is perhaps time to mention an additional term that can
appear in the Polyakov action.  This is $\Phi_0 \chi$, with $\Phi_0$
a parameter and
\begin{equation}
\chi = \frac{1}{4\pi} \int \sqrt{g}R,
\end{equation}
$R$ again being the world-sheet curvature.
This is not merely diff $\times$ Weyl
invariant, it is topologically invariant.  For a closed surface with
$h$ handles $\chi$ is the Euler number $\chi = 2 - 2h$.
Adding a handle, as in going from the one-string to the two-string
exchange in fig.~13, the path-integral weight $e^{- \Phi_0 \chi}$
changes by $e^{2\Phi_0}$.  So while this term does not affect
anything local like the world-sheet equations of motion, it does
affect the relative weights of surfaces of different topologies.
Adding a handle is like adding two trilinear closed string
interactions, so the closed string coupling $g_{\rm c}$ depends on
$\Phi_0$ as
$e^{\Phi_0}$.  For world-sheets with boundaries, as in open
string theory, diff $\times$ Weyl invariance requires also a surface
term
\begin{equation}
\chi = \frac{1}{4\pi} \int  d^2\sigma\, \sqrt{g} R +
\frac{1}{2\pi} \int_{\rm boundary} ds\, k
\end{equation}
where $k$ is the geodesic curvature of
the boundary; in terms of the tangent and normal vectors, $k = t^a n_b
\nabla_a t^b$.  Again this is purely topological, $\Phi_0$ times the
Euler number. For a compact surface with $h$ handles and $b$
boundaries the Euler number is $2 - 2h - b$.
For the open string amplitudes analogous to fig.~13, with a pair of open
strings exchanging one or two open strings, adding a strip increases $b$
by one and so the path integral weight changes by $e^{\Phi_0}$.  Thus the
coupling $g_{\rm o}$ of three open strings goes as $e^{\Phi_0/2}$.
This will play an important role at a later point, so let me
emphasize it: in theories with both open and closed strings, the
couplings are related
\begin{equation}
g_{\rm c} \sim g_{\rm o}^2 .
\end{equation}

For both particle physicists and condensed matter physicists, the
natural thing to try do at this point is to calculate a Green's
function, a propagator.  Since we are talking about one-dimensional
objects, this would be the amplitude to propagate from a given initial
configuration to a given final configuration, the sum over all
surfaces bounded by these two loops.  However,
while this seems like a very natural thing to do it is actually
extremely hard to carry out consistent with the diff $\times$ Weyl
invariance.  One already sees this in that the physical
state conditions require strings to be on-shell, $k^2 = - m^2$.
Any local source in spacetime would couple to all momenta, not
just those on the mass shell.  This should not be a surprise,
because this theory has spacetime gravity, so we have to have
observables which are spacetime coordinate-invariant.  It is
extremely clumsy to try to describe the position and shape of a loop
in a coordinate-invariant way.

What can be easily defined in an invariant way is the S-matrix, for
scattering from some set of incoming strings to some set of
outgoing strings.  Effectively the sources have been taken to
infinity.  Each of the external strings is a semi-infinite cylinder
in a world-sheet coordinate $w$.  We are
familiar with the mapping $z = e^{-i w}$ which takes this to the interior
of the disk, with $-i\infty$ mapped to the origin.
As shown in fig.~14, this mapping leaves a {\it compact}
surface.  It may seem that I am being careless in identifying
a long cylinder in spacetime with a long cylinder in the world-sheet
coordinate, but we will see that
long-distance propagation in spacetime
in fact comes from the latter.
\begin{figure}
\begin{center}
\leavevmode
\epsfbox{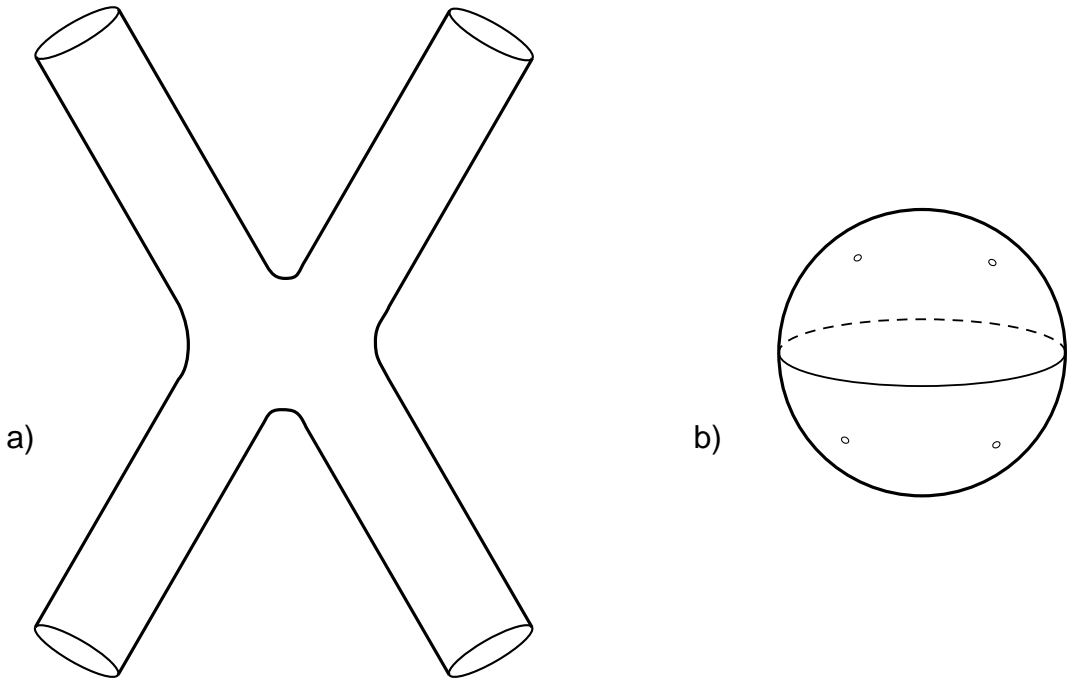}
\end{center}
\caption[]{a) $2 \to 2$ scattering process.  b) Conformally equivalent
picture, the cylinders reduced to small holes.}
\end{figure}
The source which creates the
incoming or outgoing string state is then a local operator,
known as a vertex operator ${\cal V}$.  We
are already familiar with the mapping between states and operators.
Thus, the tachyon $|0,k\rangle$ with $k^2 = 2$ is created by
${\cal V} = e^{i k \cdot X}$, the graviton, antisymmetric tensor, and
dilaton states $ \alpha_{-1}^\mu \tilde \alpha_{-1}^\nu |0,k\rangle$
with $k^2 = 0$ are created by ${\cal V} = \partial X^\mu \bar\partial
X^\nu e^{ik \cdot X}$, and so on.

In order to make the vertex operator diff-invariant we need to
integrate over the world-sheet,
\begin{equation}
V = 2 e^{\Phi_0}\int d^2 \sigma \,\sqrt{g} {\cal V}\ ,
\end{equation}
where we have included the coupling constant.
In flat gauge this becomes
\begin{equation}
 e^{\Phi_0} \int d^2z\, {\cal V}\ .
\end{equation}
Under conformal transformations the measure $d^2z$ transforms as a
$(-1,-1)$ tensor, so conformal invariance requires that ${\cal V}$
be a $(1,1)$ tensor.\footnote
{By the way, any vertex operator which is conformally invariant on
the flat world-sheet can be made diff $\times $ Weyl invariant by
appropriate coupling to the metric.}
This is precisely the OCQ physical
state condition~(\ref{ocq}), which gives us a simple way to
understand the value $a = 1$: it makes ${\cal V}$ a
(1,1) tensor.

We are led to the following expression for the S-matrix,
\begin{equation}
{\sf S} = \sum_{\rm compact \atop topologies}e^{(n - \chi)\Phi_0}
\int \frac{[dX\,dg]}{V_{\rm diff \times Weyl}} e^{-S_{\rm P} }\,
\prod_{i=1}^n \int d^2 \sigma_i \,{\cal V}_i\ . \label{sma}
\end{equation}
The product runs over the vertex operators incoming and outgoing
states, these being distinguished only by the sign of $k^0$.
(For a general CFT $S_{\rm P}$ is replaced by the appropriate
action).  We now need to understand how gauge fixing works
globally.  Locally the number of metric degrees of freedom (three)
matches
the number of gauge degrees of freedom, but globally there is a
small mismatch.  In fact, the space of equivalence classes, metrics
modulo diff $\times$ Weyl, is identical to a space described
earlier in these lectures, the space of Riemann surfaces.
Specifying the metric up to Weyl transformations singles out a
family of complex coordinates, namely those in which the metric is
proportional to $dz d\bar z$.  This is the definition of a Riemann
surface.  In the opposite direction, given a Riemann surface we can
construct a metric by taking $dz d\bar z$ in each patch and smoothing
between patches (this can always be done).  So this is an
isomorphism.

For example, take for the torus the fixed coordinate region
$0 \leq \sigma^1 \leq 2\pi$, $0 \leq \sigma^2 \leq 2\pi$ with
periodic boundary conditions, so the metric is a doubly periodic
function $g_{ab}(\sigma_1,\sigma_2)$.  Then by coordinate
transformations which preserve the periodicity and Weyl
transformations we can bring the metric to the form
\begin{equation}
ds^2 = |d\sigma^1 + \tau d\sigma^2|^2\ .
\end{equation}
for some $\tau$.  These are the same metrics described earlier in
terms of a fixed metric and $\tau$-dependent coordinate region.
So after gauge-fixing we are left with an integral over the moduli
space of Riemann surfaces, of complex dimension $0$ for  $h=0$, 1
for $h=1$ and $3h-3$ for $h \geq 2$.  For $h=0,1$ there is the
further complication of conformal Killing vectors, gauge symmetries
which remain after fixing the metric.  These can be fixed by fixing
the positions of some vertex operators, 3 at $h=0$ or $1$ at $h=1$.
In all, if we have $n$ vertex operators on a genus $h$ surface,
the total number of complex moduli (for the metric and the positions)
is $3h + n - 3$.

To carry out the Fadeev-Popov procedure we trade the original
integral $[dg]\,\prod_i d^2\sigma_i$ for
$[d\delta\sigma\,d\delta\omega]\,d\vec{t}\,\prod'_i d^2\sigma_i$
where $\vec{t}$ are the moduli for the surface and the
prime on the product denotes the omission of any fixed vertex
operators.  The Fadeev-Popov procedure can be carried out as before,
giving a mixture of functional and finite-dimensional determinants,
which again can be expressed in terms of a path integral over ghosts.
I will quote here only the result.  The S-matrix for $n$ external
strings is given by a sum and path integral
\begin{equation}
{\sf S} = \sum_{\rm compact \atop topologies}e^{(n - \chi)\Phi_0}
\int [dX\,db\,dc]\, e^{-S_{\rm P} - S_{\rm g} } \ \ldots\ .
\end{equation}
The insertions are of three types
\cite{FMS},~\cite{Msrs},~\cite{Gid_PR},~\cite{JBBS}:\\[3pt]
For each modulus (now divided into real parts $t^r$), the $b$-ghost
insertion
\begin{equation}
\frac{1}{4\pi} \int d^2 \sigma \,\sqrt{g} b^{ab} \frac{\partial
g_{ab}}{\partial t^r}\ .  \label{bins}
\end{equation}
For each vertex operator which is fixed, the insertion
\begin{equation}
c \tilde c {\cal V}_i\ .
\end{equation}
For each vertex operator which is integrated, the insertion
\begin{equation}
\int d^2\sigma\,\sqrt{g} {\cal V}_i\ .
\end{equation}

This can also be expressed in terms of the data which define the
Riemann surface, the transition functions.  It is convenient to fix
every vertex operator, put all the moduli in the transition
functions.  The vertex operators are then $c \tilde c {\cal V}_i$
as above, while the $b$-ghost insertions are
\begin{equation}
\frac{1}{2\pi i} \sum_{(mn)} \int_{C_{mn}}
\biggl\{ dz_m \frac{\partial f_{nm}}{\partial t^r} \biggr|_{z_n} b_{z_m
z_m} - {d\bar z_m} \frac{\partial \bar f_{nm}}{\partial t^r}
\biggr|_{z_n} b_{\bar z_m \bar z_m} \biggr\}\ , \label{bins2}
\end{equation}
where the sum runs over all pairs of overlapping patches and the
integral runs along any curve separating the patches.

Notice that for OCQ-type vertex operators the ghosts are in their
vacuum state $|\! \downarrow\downarrow \rangle$, which translates
into $c\tilde c$.  So it is the {\it fixed} vertex operators which
are given by the state-operator mapping.  The integrated vertex
operators can be understood as arising from the $b$-ghost insertions
for the position.  The insertion for translating a little patch
containing the vertex operator is $\tilde b_{-1} b_{-1}$, giving
\begin{equation}
\tilde b_{-1} b_{-1} \cdot c \tilde c {\cal V}_i
= {\cal V}_i\ ,
\end{equation}
which is the integrated form.  The rules in terms of Riemann
surfaces, by the way, apply to all BRST-invariant vertex operators,
not just the OCQ-type.

\subs{Trees and Loops}

To conclude this subject I work out one example and discuss some
general principles.  For four closed string tachyons on the sphere,
fully fixing the gauge invariance leaves one position integrated and
three fixed,
\begin{eqnarray}
&&{\sf S}(k_1,k_2,k_3,k_4)
\ =\  e^{2\Phi_0} C_{S_2}
\int d^2 z_4  \label{4p} \\
&&\qquad\qquad < \tilde{c} c e^{ik_1 \cdot X} (z_1,\bar z_1)\,
\tilde{c} c e^{ik_2 \cdot X} (z_2,\bar z_2)\, \tilde{c} c e^{ik_3
\cdot X} (z_3, \bar z_3) e^{ik_4 \cdot X} (z_4,\bar z_4)>_{S_2},
\nonumber
\end{eqnarray}
where the integral runs over the complex plane,
and $C_{S_2}$ is a numerical normalization factor for the path
integral on the sphere.  The expectation value can be obtained from
our earlier results, giving\footnote
{The factor of $i$, needed for unitarity, can be attributed to the
analytic continuation needed to define the $X^0$ integral.}
\begin{eqnarray}
{\sf S}(k_1,k_2,k_3,k_4)
&=& ie^{2\Phi_0} C_{S_2} (2\pi)^{26} \delta^{26}({\textstyle \sum_i
k_i}) \int d^2 z_4\,
|z_{12}|^2 |z_{13}|^2 |z_{23}|^2 \prod_{i < j}
|z_{ij}|^{2 k_i \cdot k_j} \nonumber\\
&\to& ie^{2\Phi_0} C_{S_2} (2\pi)^{26}\delta^{26}({\textstyle \sum_i
k_i}) \int d^2 z_4\, |z_4|^{2 k_1 \cdot k_4} |1-z_4|^{2 k_2 \cdot
k_4}. \label{vsint}
\end{eqnarray}
In the second line we have used the fact that the result is
independent of the fixed positions (as can be shown by a M\"obius
transformation) to move $z_1 \to 0$, $z_2 \to 1$, $z_3 \to \infty$.
The integral can be related to gamma functions with the result
\begin{eqnarray}
{\sf S}(k_1,k_2,k_3,k_4)
&=& ie^{2\Phi_0} C_{S_2} (2\pi)^{26} \delta^{26}({\textstyle \sum_i
k_i}) \\
&&\qquad\qquad \frac
{\Gamma(- s/2 - 1) \Gamma(- t/2 - 1) \Gamma(- u/2 -1)}
{\Gamma(- s/2 - t/2 - 2) \Gamma(- t/2 - u/2 -2)
\Gamma(- u/2 - s/2 -2)},
\nonumber
\end{eqnarray}
where $s = - (k_1 + k_2)^2$, $t = - (k_1 + k_3)^2$, and $u = - (k_1 +
k_4)^2$, and $u + t + s = -8$.

This is the {\it Virasoro-Shapiro amplitude}.  It has a single pole
when $s$, $t$, or $u$ takes a value $-2, 0, 2, 4, \ldots\ $.  These
are the masses of string states and correspond to processes where
two of the external strings join into one; the coefficient of the
pole is related to the square of the three-point amplitude, which
determines $C_{S_2} = 8\pi^2$.  The massless pole, for example, has
the correct form expected from graviton plus dilaton exchange (the
antisymmetric tensor doesn't contribute due to world-sheet parity
symmetry).

The notable feature of this amplitude is its soft high-energy
behavior.  Using Stirling's approximation for the exponentials, one
finds that at large $s$ and fixed $\theta$ (and so fixed $t/s$,
$u/s$) it behaves as
\begin{equation}
e^{- s f(\theta)} \label{hard}
\end{equation}
This is in contrast to
similar amplitudes in pointlike field theory, which fall as powers
of $s$, and arises from spread-out nature of the string (more on
this later).

Now we turn to the one loop amplitude.  The one-loop four-point
amplitude shown in fig.~15 contains the two-graviton exchange
described earlier, which had the severe divergence in field theory.
\begin{figure}
\begin{center}
\leavevmode
\epsfbox{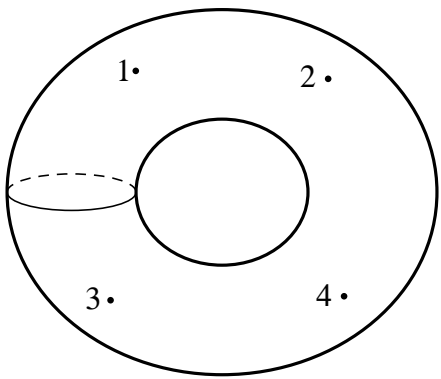}
\end{center}
\caption[]{One loop four-string amplitude.}
\end{figure}
Let us look at the integration over the momentum running around
the loop.  From the earlier discussion~(\ref{partit}), the path
integral translates into a Hamiltonian expression
involving
\begin{equation}
e^{2\pi i (\tau L_0 -\bar\tau \tilde L_0)}\ . \label{supp}
\end{equation}
For large loop momenta the dominant term here is
\begin{equation}
e^{-2\pi k^2 {\rm Im}(\tau)}\ .
\end{equation}
After Wick rotation of $k^0$, this is a convergent gaussian at
fixed $\tau_2$.  One might also worry about a divergence from the
sum over the string states running around the loop, but in spite
of the large number of states the sum is handily convergent
at fixed $\tau$ owing to the exponential suppression
factor~(\ref{supp}).  The region Im$(\tau) \to 0$ is the
potential danger, both for the integral over
momentum and for the sum over states.  But here we run into the
happy circumstance that this is not in the range of integration,
the moduli space of the torus, shown in fig.~6.  There is a
lower bound on~Im$(\tau)$, so the integral over momenta is
gaussian, as would be expected from the high energy
behavior~(\ref{hard}).  In contrast, in field theory one
could write the loop integral in a
Schwinger parameterization,
\begin{equation}
\int_0^\infty ds\, e^{-(k^2 + m^2)s}\ , \label{schwing}
\end{equation}
with $s$ being analogous to Im$(\tau)$, but here the integral
does indeed run down to $s=0$, leading in the end to a divergent
momentum integral.  (Cutting off the $s$ integral doesn't work,
as I will explain shortly).

So in this example we see that the would-be ultraviolet region of
moduli space is missing.  This is a general principle in string
theory: all the limits of moduli space can be interpreted as
{\it infrared} limits.  This can already be seen in the
Virasoro-Shapiro amplitude.  The moduli space has three limits,
$z_4 \to z_1$, $z_4 \to z_2$, or $z_4 \to z_3$.  The first is
shown fig.~16a; it is conformally equivalent to the long
cylinder in the fig.~16b, and to the plumbing-fixture
construction in fig.~16c with $q \sim z_4 - z_1 \to 0$.
\begin{figure}
\begin{center}
\leavevmode
\epsfbox{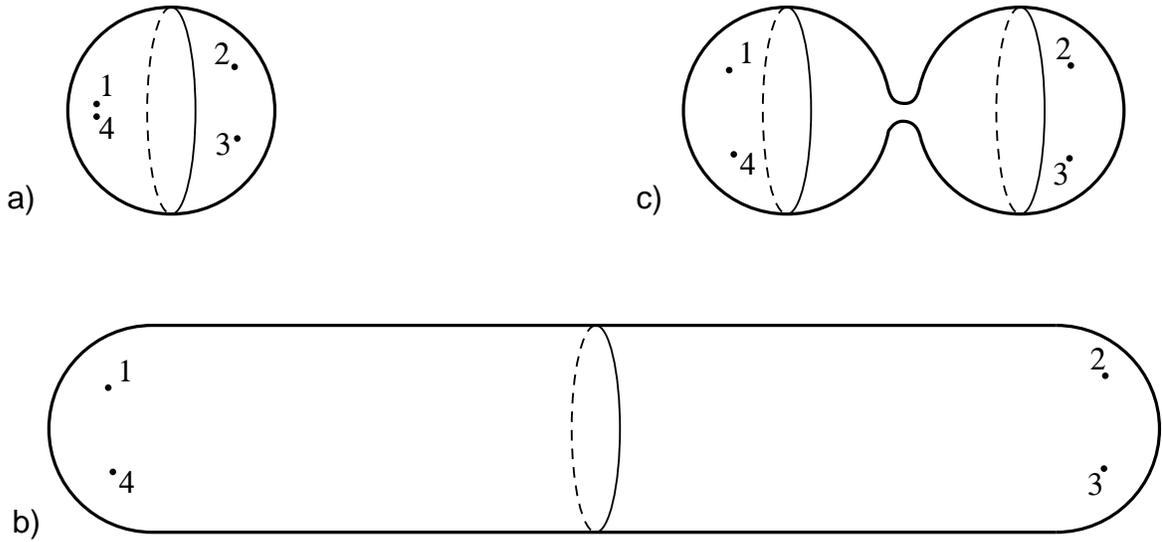}
\end{center}
\caption[]{Three pictures of the same limit of moduli space.
a) $z_4 \to z_1$ on the sphere.
b) Annular region around $z_{1,4}$ conformally transformed to a long
cylinder.  c) Long cylinder conformally transformed to a pinched
cylinder.}
\end{figure}
In the last
form we can use the sewing formula~(\ref{sew}) to express the
asymptotics in terms of a sum over intermediate string states.
The contribution of state $i$ is proportional to
\begin{equation}
\int_0 d^2q \, |q|^{k^2 + m_i^2 - 2}.
\end{equation}
The integral runs over some neighborhood of the origin.  The
behavior as $q \to 0$ is dominated by the lightest states.  The
integral converges when $k^2 + m_i^2$ is positive for all states
and can be defined elsewhere by analytic continuation.\footnote
{The divergence for $k^2 + m_i^2 < 0$ is uninteresting, an artifact
of the Schwinger-like integral representation, but the pole at $k^2 +
m_i^2 = 0$ is meaningful.}
The result has a pole proportional to
\begin{equation}
\frac{1}{k^2 + m_i^2}. \label{pole}
\end{equation}
This is the origin of the series of poles in the Virasoro-Shapiro
amplitude.  These poles correspond to long-distance propagation in
spacetime, so as in the discussion of fig.~14 this comes from a
degenerating (long) cylinder.

We could extract the asymptotics in this limit $z_4 \to z_1$
directly from the OPE; the sewing formula in this case reduces to
the OPE.  But the sewing formula is more general.  The surface
formed by sewing general surfaces ${\cal M}_1$ and ${\cal M}_2$
is conformally equivalent to a small copy of ${\cal M}_1$
inserted into ${\cal M}_2$ (and vice versa).  The sewing formula
related this to a sum of local operators inserted in
${\cal M}_2$, generalizing the OPE.

All limits of moduli space are of the same type as this one, with
one or more handles degenerating.\footnote
{Although this is well-known in the mathematics literature, I find
the most useful discussion for the purposes of physics to be that
in section~6 of ref.~\cite{Zsft}, where moduli space is explicitly
decomposed into a Feynman diagram-like sum.  We will return to ths
in section~4.1.}
They can thus be analyzed
by means of sewing in the same way.\footnote
{By the way, the $b$-ghost insertion for the moduli $q$ and $\bar
q$ is $b_0 \tilde b_0$.  This projects onto states which are
annihilated by $b_0$ and $\tilde b_0$, as claimed in our
discussion of the BRST cohomology.}
The asymptotics are
dominated by the lightest states, and the divergences all arise from
intermediate states being on the mass-shell.  This corresponds to
long-distance propagation in spacetime, and so is an infrared
effect.  There are no short-distance divergences.  As an example,
consider the limit of the one-loop four-point amplitude in which
all the vertex operators come together, fig.~17a.  This looks as though
it could produce a short-distance divergence.  But, in analogy to the
previous fig.~16, this is conformal to fig.~17b in which the
vertex operators are at the end of a long cylinder, and to
fig.~17c where a sphere and torus are sewn together.
\begin{figure}
\begin{center}
\leavevmode
\epsfbox{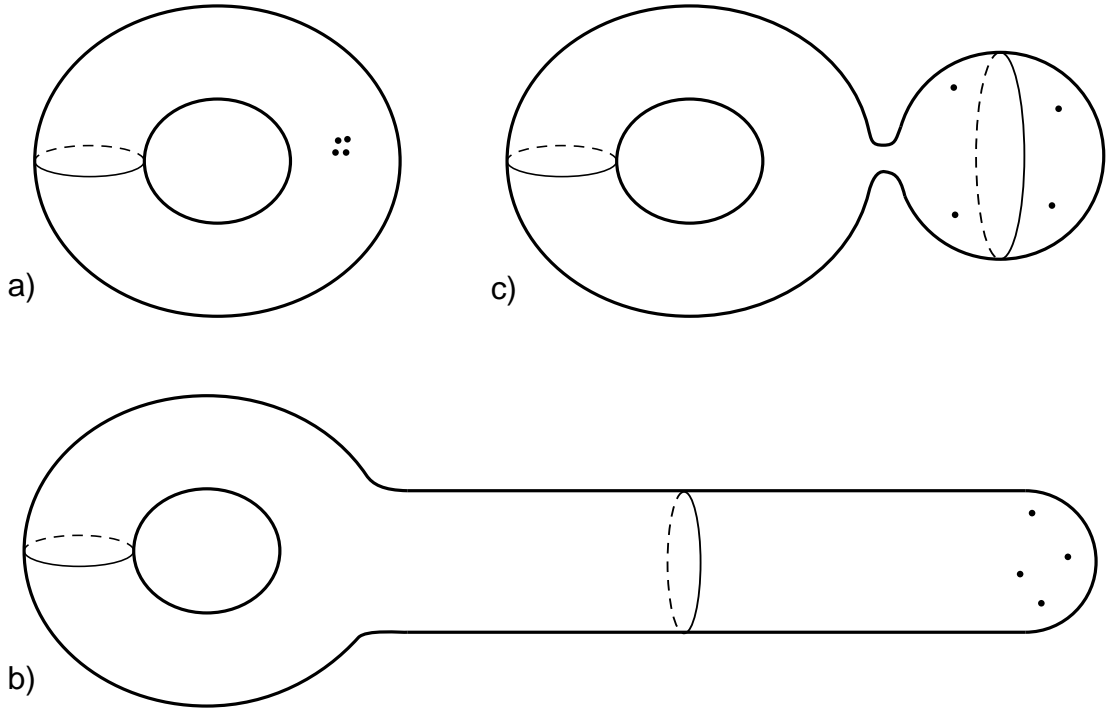}
\end{center}
\caption[]{Three pictures of the another limit of moduli space.
a) Four vertex operators coming together on the torus.
b) Annular region around operators conformally transformed to a long
cylinder.  c) Long cylinder conformally transformed to a pinched
cylinder.}
\end{figure}
There may indeed be a divergence, but it is an infrared effect,
dominated by the lightest states.  The momentum flowing through
the pinch must be zero by momentum conservation, so the
pole~(\ref{pole}) is 1/0 for any massless state.  The interpretation
is that the one-loop one-point function produces a source for the
massless field, so it is necessary to expand around a solution to
the loop-corrected equations.  This cancels the divergence, and is
known as the {\it Fischler-Susskind
mechanism}~\cite{FS},~\cite{Psew}.  More
generally the principle is that all divergences are from long
distances and so, as in field theory, go away if one asks the right
questions.

Finally a few comments about the general structure of amplitudes.
The interactions must respect the BRST invariance.  In particular,
the amplitude for an exact state $Q_{\rm B} \cdot {\cal V}_\chi$
must be zero---this is equivalent the the gravitational Ward
identity, together with the corresponding Ward identities for all
the higher levels.  To see the decoupling of the null state,
write $Q_{\rm B}$ as a contour integral around ${\cal V}_\chi$.
Expand the contour $C$ and contract it down around the other
insertions in the path integral.  The other vertex operators are
BRST invariant and give zero.  However, there is a nonzero residue
at the $b$ insertions, which by the OPE~(\ref{jbb}) is
$T^{\rm total}$.
In either form~(\ref{bins}) or~(\ref{bins2}), this
gives a total derivative of the path integral with respect to the
moduli.  Upon integration
over moduli space this becomes a surface
term\cite{FMS},~\cite{Msrs},~\cite{Gid_PR},~\cite{Mans}.  The
surface term vanishes under the same condition that the
(analytically continued) integral is finite---that is, unless we are
sitting on a pole (in which case the Fischler-Susskind mechanism
restores conformal invariance).

Notice that if we mutilate the theory by cutting out parts of
moduli space, the total derivative no longer integrates to zero and
the theory is inconsistent.  The same thing happens in field
theory if we try to cut off the Schwinger parameter
integral~(\ref{schwing}).  String theory manages to cut off the
modular integral while leaving a smooth moduli space without
spurious internal boundaries.

One can also arrive at the rules for the string S-matrix from
considerations of BRST invariance, rather than via gauge-fixing.
First, as found in the exercise below eq.~(\ref{nbnc}), ghost
number conservation requires the path integral to contain
insertions with net $N_b - N_c = 3h - 3$; our rules for the string
S-matrix are consistent with this.  These insertions
are BRST invariant only up to a total derivative with respect to the
moduli, so the path integral must be integrated over moduli space as
found from gauge fixing.  In trying to generalize string theory it is
often easiest to use this strategy directly to determine the form of
the amplitudes, rather than gauge-fixing from a locally-invariant
form. For example, superstring amplitudes can in this way be written
as an integral over supermoduli space.

This concludes our survey of the current perturbative answer to
the question, ``what is string theory?''  In particular we have
seen what are the ingredients that make the perturbation theory
consistent.

\section{Vacua and Dualities}
\setcounter{footnote}{0}

\subs{CFT's and Vacua}

In the last section we described the most general string theory as
defined by the world-sheet gauge algebra, by the world-sheet
topologies allowed, and by the particular world-sheet CFT or
super-CFT.\footnote{This was nicely laid out in the review
article~\cite{Sclass}.}
In this section we will try to get a better
understanding of the nature of the space of theories.  The main
theme is that what we have called different
string theories are in many (and possibly all) cases the {\it same}
theory expanded around a different vacuum state.  Moreover, in
some cases what appear to be different theories are in fact
the same theory in the {\it same} vacuum.  Some of
these connections are well-developed, while others are highly
conjectural.

Let us start by trying to understand better the idea of replacing
the spatial $X^\mu$'s with a more general CFT.
I will focus on the bosonic string, but the ideas all generalize
to the superstring.  For example,  replace the Polyakov action with
the non-linear sigma model~\cite{Fnlsm}-\cite{CFMP}
\begin{equation}
S_{\Sigma} = \frac{1}{8 \pi }\int d^2\sigma \, \biggl\{
\Bigl(\sqrt{g} g^{ab} G_{\mu\nu} (X) + i \epsilon^{ab} B_{\mu\nu} (X)
\Bigr) \partial_a X^\mu \partial_b X^\nu
+ 2 \sqrt{g} R \Phi(X) \biggr\},
\end{equation}
which is the most general coordinate invariant action we can make
with two derivatives.  This is the same theory~(\ref{nlsm}) that
we have already discussed on a flat world-sheet, and when the
$X$-dependence is slow the theory can be studied perturbatively
(this is now perturbatively in the world-sheet couplings).
This gives rise to a consistent string theory, to leading order
in the world-sheet perturbation theory, if\footnote
{To relate this to the flat world-sheet discussion, note that
on the flat world-sheet the $\Phi$ term in the action is trivial.
However, it affects the energy-momentum tensor, which is given by a
derivative with respect to the metric.  When the
equations~(\ref{beta}) are satisfied with a nontrivial $\Phi$,
$T_{z\bar z}$ given by eq.~(\ref{tzzb}) does not vanish, but
adding an `improvement term' produces a conserved energy-momentum
tensor with $T_{z\bar z} = 0$.  The last of line of
eq.(\ref{beta}) is the perturbatively-corrected matter plus ghost
central charge.}
\begin{eqnarray} 2 {\bf{ R}}_{\mu\nu}
 + 4 \nabla_\mu \nabla_\nu \Phi
-\frac{1}{2}H_{\mu \lambda \omega}H_\nu{}^{\lambda \omega}
& = & 0 \nonumber \\
- \nabla^\omega H_{\omega \mu \nu}
+ 2 \nabla^\omega \Phi H_{\omega \mu \nu} &=& 0
 \nonumber \\
D - 26 - 6\nabla^2 \Phi +
12 \nabla_\omega \Phi \nabla^\omega \Phi - \frac{1}{2} H_{\mu
\nu \lambda}H^{\mu\nu\lambda} &=& 0. \label{beta}
\end{eqnarray}

This conformal field theory has an obvious interpretation.  The
functions $G_{\mu\nu}(X)$, $B_{\mu\nu}(X)$ and $\Phi(X)$
correspond to nontrivial backgrounds of the string graviton,
antisymmetric tensor and dilaton.  One might have expected
something stringier-looking, since the graviton, etc., are
supposed to be strings.  But we found in our discussion of
scattering amplitudes that strings could be created by local
vertex operators; the sigma model action can then be thought of
from exponentiating the vertex operators, corresponding to a
coherent state of strings.\footnote{The dilaton actually involves
fluctuations both of $\Phi$ and the diagonal part of $G_{\mu\nu}$.
The curvature term in its vertex operator is of course not evident
on the flat world-sheet, but arises on a curved world-sheet when
the operator is renormalized in a coordinate-invariant way.}
Notice in particular that different values of the string coupling
$e^{\Phi_0}$ are now seen as corresponding to expanding around
different backgrounds in a single theory, namely different constant
values of the dilaton field $\Phi$.  The field equations~(\ref{beta})
arise from a spacetime action
\begin{equation}
{\bf {S}} = \frac{1}{2} \int d^D X\, \sqrt{-G} e^{-2 \Phi}
\biggl\{ - \frac{(D-26)}{3}
+ {\bf {R}} -\frac{1}{12} H_{\mu\nu\lambda} H^{\mu\nu\lambda}
+4 \nabla_\mu \Phi \nabla^\mu \Phi  \biggr\} . \label{stact}
\end{equation}
Higher orders in the world-sheet perturbation expansion produce
terms with more derivatives in the field equations and action.

Each static solution corresponds to a possible ground state.
For example, a background with four flat
dimensions and the remainder curved on a sufficiently small
distance scale $l_c$ would look like what we see around us.\footnote
{Direct experiment---the non-observation of the enormous number of
states with nontrivial dependence on the compact
dimensions---requires $l_c$ to be less than $10^{-16}$cm.  Other
considerations, however, require it to be right at the string scale
$10^{-32}$cm.  In particular, if there is a large hierarchy
between the string and compactification scales, the theory at
intermediate scales is a $D > 4$ field theory; this is badly
nonrenormalizable and breaks down well before the string
scale~\cite{Kscale}.} Even before string theory this was a promising
idea for unification.  It unites fields of different spins into a
single higher-dimensional field (Kaluza-Klein theory), and provides a
natural origin for the fermion generations as states with the same
gauge quantum numbers but different wavefunctions in the compact
dimensions.  By the way, we see from the loop correction to the
central charge in the last line of eq.~(\ref{beta})
 that the number of compact dimensions need not be
22.  Actually the expansion which gives rise to eqs.~(\ref{beta})
breaks down when the corrections are of order~1, but exact CFT's
with $D \neq c$ are known. An example is the WZW model,
correpsonding to strings propagating on a group
manifold~\cite{JSR},~\cite{GepW}.

For more general CFT's, the interpretation in terms of
compactification is not so clearcut.  But given the many equivalences
between two-dimensional field theories (the most familiar being
bosonization), it is quite possible that all can be understood in terms
of background fields, perhaps including massive string fields.  As one
example, the minimal models were first constructed in an abstract way,
but now several Lagrangian formulations are known.  One in particular,
the Landau-Ginsburg theory, can be regarded as the string moving in a
tachyon background~\cite{Zlg}.  So the general assumption
is that different CFT's should be regarded as different vacua.

Let us make a few general observations.  Just as we have noted that
the bosonic string always has a tachyon, it also has a graviton,
antisymmetric tensor, and dilaton.  The vertex operator
$e_{\mu\nu}\partial X^\mu \bar\partial X^\nu e^{ik\cdot X}$, with
$e_{\mu\nu}$ and $k_\mu$ lying in the noncompact flat directions, is
always a $(1,1)$ vertex operator for $k^2 = 0$.
The $(1,1)$ and $(1,0)$ superstrings also always include the graviton
and tachyon.

If the compact theory has a current algebra, the vertex
operator $\partial X^\mu \tilde j^a e^{i k \cdot X}$ or
$j^a \bar\partial X^\mu e^{i k \cdot X}$ is $(1,1)$ for $k^2 = 0$ and
corresponds to a gauge boson.  So the spacetime gauge symmetries are
in one-to-one correspondence with the global symmetries (current
algebras) of the compact CFT.  A similar result, a bit less
straightforward to derive, is that there is spacetime supersymmetry if
and only if the $N=1$ superconformal constraint algebra of the
superstring is embedded in a larger $N=2$ symmetry of the CFT, with
a quantization condition on the $U(1)$ charge of the $N=2$
algebra~\cite{BDFM}.

The number of CFT's or super-CFT's of appropriate central charge is
enormous, even restricting to those with exactly four noncompact flat
directions.  We are now interpreting these as different backgrounds
within a single theory, but the
effective four-dimensional physics is different in each.  The
problem of finding the right theory is thus transmuted in string
theory into the dynamical problem of finding the right background.
For backgrounds with spacetime supersymmetry, it is a theorem
that they remain stable to all orders of perturbation theory, but it
is also known, from the understanding of supersymmetry breaking in
field
theory, that most are destabilized by non-perturbative effects.
Spacetime supersymmetry of the effective field theory below the string
scale gives a great deal of information about the dynamics, a subject
currently in active development.  But it seems likely that dynamics
at the string scale will also play an important role.  It may also be
that there are in the end many stable ground states of string theory,
so that the choice between them will be determined in part by the
initial conditions.

\subs{Compactification on a Circle}

A flat spacetime with one dimension periodic,
\begin{equation}
X^{1} \cong X^{1} + 2\pi R
\end{equation}
is the simplest compactification of the bosonic string, but is
quite instructive.
(Equivalently, let the periodicity be $2\pi$ but $G_{11} = R^2$).  The
periodicity has two effects.  The first is that the string
wavefunctions must respect it, so \begin{equation}
k^{1} = \frac{n}{R} \label{k25}
\end{equation}
is quantized.  This is the same as for a field theory on this space.
The second effect is unique to string theory: a string can be wound
around the periodic dimension, so that
\begin{equation}
X^{1} (2\pi) = X^{1} (0) + 2\pi m R.
\end{equation}
Referring back to our mode expansion~(\ref{lex2}) for the free
scalar, we see that this means that the eigenvalues $k^{1}_{L,R}$
of $\alpha_0$ and $\tilde \alpha_0$ are no longer equal,
$k^{1}_L - k^{1}_R = m R$.  The total momentum~(\ref{k25}) is
$\frac{1}{2}(k^{1}_L + k^{1}_R)$, so
\begin{equation}
k^{1}_L = \frac{mR}{2} + \frac{n}{R}, \qquad
k^{1}_R = -\frac{mR}{2} + \frac{n}{R}. \label{klkr}
\end{equation}
Also from the expansion~(\ref{lex2}), the vertex operator for
such a state will be proportional to
\begin{equation}
e^{i k_L X_L(z) + i k_R X_R(\bar z)}. \label{wsvert}
\end{equation}
Although the $X_R X_R$ and $X_L X_L$
OPE's contain branch cuts, the OPE of any two vertex
operators~(\ref{wsvert}) in the spectrum~(\ref{klkr}) is
single-valued.

The $L_0 \pm \tilde L_0$ physical state conditions thus become
\begin{eqnarray}
&& M^2\ =\ k_0^2 - \sum_{\mu = 2}^{25} k_\mu^2 \ =\ \frac{m^2
R^2}{4} + \frac{n^2}{R^2} + {\sf L} + {\sf \tilde L} - 2 \nonumber\\
&& mn + {\sf L} - {\sf \tilde L} \ =\ 0\ .  \label{s1spec}
\end{eqnarray}
Looking at the massless spectrum, the  states
\begin{equation}
\alpha_{-1}^\mu \alpha_{-1}^\nu |0,m=n=0\rangle, \label{p67}
\end{equation}
with no compact momentum or winding, remain massless.  For $\mu$
and $\nu$ both in the range $0, 2\ldots, 25$, these are just the
graviton, dilaton, and antisymmetric tensor of the 25-dimen\-sional
theory.  When either $\mu = 1$ or $\nu = 1$ the state is a
vector in the noncompact dimensions, a gauge boson.  The
corresponding vertex operators are
\begin{equation}
(\partial X^\mu \bar\partial X^{1} \pm
\partial X^{1} \bar\partial X^\mu ) e^{i k \cdot X}. \label{gbvert}
\end{equation}
The plus sign comes from the 26-dimensional metric, the
Kaluza-Klein mechanism.  The minus sign comes from the
26-dimensional antisymmetric tensor, a generalization of the
Kaluza-Klein mechanism; call this an $H$ gauge boson.  The
operator product of the gauge boson vertex operator~(\ref{gbvert})
with a general vertex operator of momentum $l$ is proportional to
$l^\mu (l_L^{1} \pm l^{1}_R)$.  The Kaluza-Klein and $H$ gauge
bosons thus couple to the compact momentum and winding number
respectively.  Finally, the state~(\ref{p67}) with $\mu=\nu=1$
is a 25-dimensional scalar, and is the metric component
corresponding to the radius $R$ of the compact dimension.

So far this is the same as would be found just from the low energy
field theory~(\ref{stact}), but this simple theory has some
interesting stringy physics.  Consider the four sets of states with
$|m|= |n| = 1$ and ${\sf L} + {\sf \tilde L} = 1$,
\begin{equation}
\tilde\alpha_{-1}^\mu | 0, m=n= \pm 1 \rangle, \qquad
\alpha_{-1}^\mu | 0, m=-n= \pm 1 \rangle.
\end{equation}
Their masses are
\begin{equation}
M = |R^2 - 2|/2R.
\end{equation}
Precisely at the radius $R = \sqrt{2}$ these states are massless,
and so are gauge bosons.  At this radius, the spectrum~(\ref{klkr})
includes the four currents
\begin{equation}
:\! e^{\pm i X^{1}_L(z) \sqrt{2} }\! :, \quad
:\! e^{\pm i X^{1}_R(\bar z) \sqrt{2} }\! :\ .
\end{equation}
Together with the two $U(1)$ currents
\begin{equation}
\partial X^{1}(z), \qquad \bar\partial X^{1}(\bar z), \label{zcur}
\end{equation}
these form an analytic and an antianalytic $SU(2)$ current
algebra.\footnote
{This $SU(2)$ is likely familiar to those in condensed matter physics
who have looked at the Luttinger liquid or other one-dimensional
quantum systems.  At general radii, these currents have dimension
$h + \tilde h = 1 + M^2$ and spin $h - \tilde h = \pm 1$, and are
no longer conserved.  See ref.~\cite{KFP} for a recent example of a
(dirty) system where $R$ flows to $\sqrt 2$ at long distance,
producing an
$SU(2)$ symmetry in the long distance theory which is not present in
the underlying theory.}  This $SU(2) \times SU(2)$ symmetry has
nothing to do with the Lorentz invariance of the flat-spacetime
theory.  Its emergence at the critical radius is an example of the
large symmetry that is hidden in string theory, almost completely
broken.  By the way, the mass of the gauge boson, as $R$ moves away
from the critical value, comes from the ordinary Anderson-Higgs
mechanism.  The vertex operator for a small change in $R$,
\begin{equation}
\partial X^{1} \bar\partial X^{1}, \label{radmod}
\end{equation}
is built out of the $SU(2)$
currents~(\ref{zcur}), and so transforms as the $z$-component of a
vector under each $SU(2)$, breaking $SU(2) \times SU(2)$ down to
the Kaluza-Klein and $H$ $U(1) \times U(1)$.

There is another stringy phenomenon here.  As $R
\to \infty$, the states with $m \neq 0$ go to infinite mass, while
the states with $m = 0$, $n \neq 0$ form a continuum.  This is
simply because it costs an energy of order $R$ to wind around the
large dimension, while the momentum in that direction becomes
continuous in the limit.  This is all as in field theory.  But look
at $R \to 0$.  States with $n \neq 0$ become very massive because of
the large compact momentum, just as in field theory.  But the states
with $n= 0$, $m \neq 0$ are now forming a continuum, something
which has no analog in field theory.  In fact, the spectrum is
invariant under {\it R-duality}, also known as {\it T-duality},
\begin{equation}
R \leftrightarrow \frac{2}{R}, \qquad m \leftrightarrow n,
\end{equation}
which interchanges large and small radius, and interchanges compact
momentum with winding number~\cite{KY}; for recent
reviews see~\cite{GPR},~\cite{Tsdual}.

This is not only a symmetry of the spectrum, but also of the
interactions.  $R$-duality takes $k^{1}_L \to k^{1}_L$, $k^{1}_R
\to -k^{1}_R$.  If we extend this to
\begin{equation}
\alpha_m^{1} \to \alpha_m^{1},\qquad
\tilde\alpha_m^{1} \to - \tilde\alpha_m^{1} \ ,
\end{equation}
so that
\begin{equation}
X_L^{1} \to X_L^{1}, \qquad  X_R^{1} \to -X_R^{1}\ ,
\end{equation}
it is a symmetry of the OPE, and so by the sewing principle holds
for all Riemann surfaces.  Finally, a transformation of the dilaton
is needed to make the loop expansions the same.  After integrating
the spacetime action~(\ref{stact}) over $X^{1}$, the effective
25-dimensional action is weighted by $2\pi R e^{- 2\Phi(X)}$.  In
order that this be invariant we need $2\pi R' e^{- 2\Phi'(X)} =
2\pi R e^{- 2\Phi(X)}$~\cite{Busch},
\begin{equation}
\Phi'(X) = \Phi (X) - \ln(R/\sqrt{2}).
\end{equation}
The $R$ and $2/R$ theories are then physically
identical.\footnote{To restore units, the statement is that
$R' = \alpha'/R$, where $\alpha'$ is again of order
$10^{-32}$cm.}

This can also be seen in a very different, and deeper,
way~\cite{DHS}. Notice that the self-dual radius is also the point of
enlarged gauge symmetry.  We have observed that the vertex operator
corresponding to a change in the radius transforms non-trivially
under $SU(2) \times SU(2)$.  In fact, a rotation by $\pi$ around the
$x$-axis of {\it one} of the $SU(2)$'s takes this operator into minus
itself. So increasing $R$ is {\it gauge-equivalent} to decreasing it.
This implies that $R$-duality is a symmetry not only in perturbation
theory (which is all that we can conclude from the argument above)
but is in fact an {\it exact} symmetry.  We can say this, even
though we know nothing about non-perturbative string theory,
because we do know that any violation of gauge symmetry would make
the low energy theory inconsistent.  We can also say that the
$R$ and $2/R$ vacua are not just identical states, they are {\it the
same} state.  This is not just a semantic distinction.  It means
that there can exist defects in spacetime, such that as one
encircles them $R$ changes continuously from its original value
to the dual value.

Duality is a striking indication that strings do not sense
spacetime in the same way as particles, and that our
notions of spacetime geometry and even topology break down at short
distance.\footnote{We have already seen another example of this in the
level~1 $SU(2)$ WZW model, which can be described in terms of three
coordinates or in terms of one.}  Thus, we want to
think of $X^{1}(z,\bar z) = X^{1}_L(z) + X^{1}_R(\bar z)$ as the
location of some world-sheet point in spacetime, while the dual
coordinate $X_{\rm d}^{1}(z,\bar z) = X^{1}_L(z) - X^{1}_R(\bar z)$
is a much more complicated and nonlocal object.  Yet if the theory
is compactified at some radius near $\sqrt{2}$ these are equally
physical, with the $R \to \infty$ physics being simple in
terms of $X^{1}$, and the $R \to 0$ physics simple in terms of
$X_{\rm d}^{1}$.  Duality suggests that there is a minimum spacetime
length scale---we can restrict to $R > \sqrt{2}$.\footnote
{Though see ref.~\cite{GPS} for a situation where it
appears that some dimensions are actually becoming much smaller than
this.}  So if spacetime breaks
down at the string scale, what is to replace it?  A given CFT may
have many different Lagrangian representations, each giving a
different picture of spacetime. It is the OPE coefficients
$c^k\!_{ij}$ which as we have discussed are common to all
representations of the theory, and which determine the string
amplitudes.  But these seem to me rather abstract to be the
fundamental description.

\subs{More on $R$-Duality}

Duality has been a source of great fascination, and has many
extensions, discussed in the
reviews~\cite{GPR},~\cite{Tsdual}.  I will mention only a few.
First, it can be extended to any translationally invariant
direction~\cite{Busch}.  Consider a world-sheet action involving
fields
$V_a$,
$\theta$, and
$X^0, X^2, \ldots X^{25}$,
\begin{equation}
S = \frac{1}{4\pi} \int d^2z\, \Bigl\{
G^{-1}(X) V \tilde V +
\theta (\partial \tilde V - \bar\partial V)
+\ldots \Bigr\}\ . \label{step1}
\end{equation}
Here, the ellipsis stands for terms involving the other world-sheet
fields $X$ but not $\theta$ or $V$, and $G(X)$ is an arbitrary
function.  Integrating out $V$ by completing the square leaves
\begin{equation}
S' = \frac{1}{4\pi} \int d^2z \,\Bigl\{
G(X) \partial \theta \bar\partial \theta
+\ldots \Bigr\}\ .
\end{equation}
On the other hand, integrating out $\theta$
forces $\partial \tilde V - \bar\partial V = 0$, so $V_a$ is a
gradient
\begin{equation}
V = \partial \theta_{\rm d}, \qquad \tilde V = \bar\partial
\theta_{\rm d} \label{grad}
\end{equation}
for some function $\theta_{\rm d}$.  The action then becomes
\begin{equation}
S'' = \frac{1}{4\pi} \int d^2z \, \Bigl\{
G^{-1}(X) \partial \theta_{\rm d} \bar\partial \theta_{\rm d}
+\ldots \Bigr\}\ . \label{stepn}
\end{equation}
The two actions with reciprocal kinetic terms are thus equivalent.
Noting that the equation of motion for $V$ is
\begin{equation}
V = -\partial \Bigl( G^{-1}(X) \theta \Bigr), \qquad \tilde V =
\bar\partial \Bigl( G^{-1}(X) \theta \Bigr)\ ,
\end{equation}
this reduces for constant $G(X)$ to the earlier duality
transformation, with $\theta$ a constant times $X^{1}$.
In particular, a careful treatment of the measure
produces the transformation of the dilaton, $\Phi'(X) = \Phi (X) -
\ln \sqrt{G(X)}$.
The above is readily extended to nonzero $G_{1\mu}$ and
nonzero $B_{1\mu}$.\footnote{A loose end: the constraint from
$\theta$ only forces $V_a$ to be a gradient locally, so around
a closed curve $\theta_{\rm d}$ need not be single-valued.  Thus we
have
related the $\theta$ theory at $R=\infty$ to the $\theta_{\rm d}$
theory at $R=0$.  With careful attention to surface terms one can
extend this to finite $R$.}  In this form, duality can be applied to
many interesting string backgrounds.

With more than one periodic dimension there is much more structure.
I briefly summarize the case of two compact directions $X^1$ and
$X^2$~\cite{du2}.  Let each be periodic with period
$2\pi$, with
$G_{11}$,
$G_{12}$, $G_{22}$, and $B_{12}$ constants.  Parameterize these
four fields in terms of two complex parameters $\tau = \tau_1 +
i\tau_2$ and $\rho = \rho_1 + i \rho_2$ as
\begin{eqnarray}
ds^2 &=& \frac{\rho_2}{\tau_2} |dX^1 + \tau dX^2|^2
\nonumber\\
B_{12} &=& 2\rho_1. \label{2dpar}
\end{eqnarray}
There is a large discrete group of equivalences.  The
reparameterization
\begin{equation}
X^1 = {X^1}' d + {X^2}' b, \qquad X^2 = {X^1}' c + {X^2}' a
\end{equation}
for $a,b,c,d$ integers such that $ad-bc = 1$ preserves the
periodicity.  The transformed background is
\begin{equation}
\tau' = \frac{a \tau + b}{c \tau + d},\qquad \rho' =  \rho\ .
\end{equation}
Notice that this is exactly the same as the modular
transformation~(\ref{sl2z}) of the torus, but now acting on
spacetime rather than the world-sheet.  This equivalence is just
a change of basis vectors for the spatial periodicity.  It is
not at all stringy---it also holds for a field theory in this
spacetime.

There are other, stringier, equivalences.  Consider the term in
the action involving $B_{12}$,
\begin{equation}
\frac{B_{12}}{4\pi} \int d^2z\, \Bigl\{ \partial X^1
\bar\partial X^2 - \bar\partial X^1 \partial X^2 \Bigr\} \ .
\label{bact}
\end{equation}
The integrand is a total derivative, $\partial (X^1 \bar\partial
X^2) - \bar\partial (X^1 \partial X^2 )$.  This would seem to
imply that the theory is independent of $B_{12}$, but we have to
be careful because $X^{1,2}$ need not be periodic on the
world-sheet.  Consider a toroidal world-sheet wound once on the
toroidal spacetime, $X^1 = {\rm Re}(z)$, $X^2 = {\rm
Im}(z)/{Im}(\tau_w)$ ($\tau_w$ being the world-sheet modulus).
The action~(\ref{bact}) becomes $i \pi B_{12}$, so the path
integral weight becomes $e^{i \pi B_{12}}$.
This is invariant under {\it discrete} shift
\begin{equation}
B_{12} \to B_{12} + 2 \ \Rightarrow\ \rho \to \rho + 1\ .
\end{equation}
In addition there is simultaneous duality on the $X^1$ and $X^2$.
This takes $E_{\mu\nu} = G_{\mu\nu} + i B_{\mu\nu}$ to
its inverse.  In terms of the parameterization~(\ref{2dpar})
this is simply $\rho \to -1/\rho$.  Again this and the shift
$\rho \to \rho + 1$ generate the full $SL(2,Z)$
\begin{equation}
\rho' = \frac{a \rho + b}{c \rho + d},\qquad \tau' =  \tau\ .
\end{equation}
There are a few other transformations.  Duality on the separate
axes takes $(\tau,\rho) \to (\rho,\tau)$.  Spacetime parity,
$X^1 \to - X^1$ takes $(\tau,\rho) \to (-\bar\tau, -\bar\rho)$,
and world-sheet parity takes $(\tau,\rho) \to (\tau, -\bar\rho)$.
(The last two are not symmetries of the heterotic string).
In all, the full set of dualities is $SL(2,Z) \times SL(2,Z)$,
up to some $Z_2$ factors.  The space of backgrounds,
which for the single dimension was the half-line $R \geq
\sqrt{2}$, is here given by two copies of the modular region of
the torus, with some additional $Z_2$ identifications.

Besides duality, the other equivalence of CFT's which has
attracted a great deal of attention is mirror symmetry~\cite{stY}.
This is an equivalence of $N=2$ super-CFT's arising from
compactification on smooth manifolds, with the
distinguishing feature that it flips the sign of the
$U(1)$ current of the $N=2$ algebra.\footnote {Mirror symmetry is
not the same as
$R$-duality, except in special cases~\cite{GivW}.  In general, the
Lagrangians of mirror-symmetric theories cannot be directly
transformed into one another in the way we have done for
$R$-duality.}  The equality of the Yukawa couplings (OPE
coefficients) on a manifold and its mirror give relations between
previously unconnected mathematical structures.  The most
interesting physical phenomenon is the existence of examples where
a continuous change in the background fields of one manifold
(and so a continuous change in the CFT)
maps to its mirror passing through a singular configuration and
changing topology.

The open string cannot wind around a periodic dimension, so one
does not expect it to be dual~\cite{DLP}.  The normal open string has
the Neumann boundary condition $n^a \partial_a X^\mu = 0$.  The
duality transformation is $\partial_a X^1 = \epsilon_a\!^b \partial_b
X^1_{\rm d}$, so the boundary condition is
\begin{equation}
n^a \epsilon_a\!^b \partial_b X_{\rm d} = t^b \partial_b X_{\rm d}\ .
\end{equation}
The {\it tangential} derivative vanishes, so $X^1_{\rm d}$ is constant
along the boundary.  Moreover if we consider points $p_1$ and
$p_2$ on two {\it different} boundaries, we have
\begin{equation}
X^1_{\rm d}(p_1) - X^1_{\rm d}(p_2) = \int_C
(dz \,\partial X^1_{\rm d} +
d\bar z\, \bar\partial X^1_{\rm d}) = \int_C (dz \,\partial X^1 -
d\bar z\, \bar\partial X^1) = 4\pi k^1 = \frac{4\pi n}{R} = 2\pi
R_{\rm d} n.
\end{equation}
Here $C$ is any curve connecting $p_1$ and $p_2$.  We can imagine
cutting the path integral open along $C$ in terms of the open
string Hilbert space.  The mode expansion~(\ref{osmode}) then
relates this to the momentum $k^1$ in the Neumann picture.  The result
is that the points differ by precisely a multiple of the
periodicity of the dual space---they are at the same point.
It also follows from the mode expansion~(\ref{osmode})
that flipping the sign of the right-moving part of $X$ converts
Neumann to Dirichlet boundary conditions.
Taking $R \to 0$, and so $R' \to\infty$, we have a space in which all
string endpoints are constrained to move on a hyperplane of fixed
$X^1_{\rm d}$.  Open strings are thus found only at this hyperplane,
while
the closed strings (which as we have noted are present in any open
string theory) are free to move everywhere.
This
hyperplane is actually a dynamical object, the
D(irichlet)-brane~\cite{DLP},~\cite{Ldir}. The open string state
$\alpha^1_{-1}\ket{0}$ is a massless excitation along the D-brane,
whose couplings are just those of a transverse ripple of the
hyperplane.

In the classification of string theories by CFT,
world-sheet topology, and constraint algebra, we have
explored the idea that the first of these corresponds to
the vacuum of the theory.  Different world-sheet topologies,
however, would seem to be truly distinct theories.  Remarkably,
duality suggests that this is not the case: translating the
D-brane off to infinity, one obtains in the limit a theory of
closed strings only.  A similar result holds for unoriented
theories---the dual theory has an extended object (the
`orientifold'), away from which there are only oriented
world-sheets~\cite{DLP}.

\subs{$N=0$ in $N=1$ in $\ldots$?}

The third part of the Schwarz classification is the world-sheet
constraint algebra.  A recent argument of Berkovits and
Vafa~\cite{BV} suggests that this too is determined by the vacuum.
Let $T^{\rm m}$ be the $c = 26$ CFT for any bosonic string theory.
Add in a
$\lambda = \frac{3}{2}$ $bc$ system $(b_1,c_1)$, which has
central charge $-11$.  Then the following energy-momentum tensor
and supercurrent form a $c = 15$ $N=1$ super-CFT:
\begin{eqnarray}
T &=& T^{\rm m} - :\! b_1 \partial c_1  \! :
- \frac{1}{2} \partial :\! b_1 c_1 \! : + \frac{1}{2} \partial^2
(c_1 \partial c_1) \nonumber\\
T_F &=& b_1 + c_1 T^{\rm m} +:\! c_1 \partial c_1 b_1 \! :
+ \frac{5}{2} \partial^2 c_1 .  \label{bv}
\end{eqnarray}
This super-CFT can be used as the matter CFT for the $N=1$
superstring.  It is not in the general class described in
section~\ref{gensec} because the non-unitary part is not of the
standard $X^0 \psi^0$ form (the $b_1 c_1$ theory must be
nonunitary because the central charge is negative).  But it might
arise from backgrounds of fields with nontrivial time
components.

Now, what seems remarkable is that the BRST cohomology of this
$N=1$ theory is identical to that of the $N=0$ theory based on
$T^{\rm m}$, the constraints from $T_F$ removing the $b_1 c_1$
degrees of freedom.  Moreover, the amplitudes of the two theories
are the same, the $b_1 c_1$ path integral canceling the $\beta_1
\gamma_1$ path integral in a nontrivial way~\cite{BV}.\footnote
{According to Distler (private communication) there may be a problem on
higher-genus surfaces.}
So the bosonic
string theories would indeed seem to be vacua (in the broad sense
of super-CFT's) of the $N=1$ string.

This can be carried further---a general $N=1$ string can in this way
be embedded in $N=2$, which can be embedded in $N=3$, and so on
indefinitely~\cite{BOP}.
It can also be embedded in the $W_3$ string
and generalizations~\cite{BFW}.
Another chain of embeddings is possible
in a series of linear higher spin algebras $w_N$~\cite{KST}.
This is true even though the $N >
2$ superconformal algebras and $N > 2$ $w_N$ algebras do not have flat
spacetime realizations of the ordinary sort.  This begins to seem
like too much of a good thing.

Indeed,
the following example may help to put this in perspective.  Consider
a field theory with some complex scalars $\phi_i(x)$ and a Lagrangian
density $L(\phi_i)$ of no special symmetry.  We will make this look as
though it has a local $U(1)$ symmetry $\phi_i \to e^{i q_i \theta(x)}
\phi_i$ for arbitrary choice of $q_i$.  First, add a scalar field
$\chi(x)$ and define
\begin{equation}
L'(\phi_i,\chi) = L(\phi_i e^{-iq_i\chi}) - \frac{1}{2} \partial_\mu
\chi \partial^\mu \chi\ .
\end{equation}
This is invariant under the global symmetry
\begin{equation}
\chi'(x) = \chi(x) + \theta, \qquad \phi'_i(x) = e^{i q_i \theta}
\phi_i(x).
\end{equation}
Now add a gauge field $A_\mu$ without a kinetic term,
\begin{equation}
L''(\phi_i,\chi) = L(\phi_i e^{-iq_i\chi}) - \frac{1}{2}
(\partial_\mu \chi - A_\mu) (\partial^\mu \chi - A^\mu)\ .
\end{equation}
This is invariant under the {\it local symmetry}
\begin{equation}
\chi'(x) = \chi(x) + \theta(x), \qquad \phi'_i(x) = e^{-i q_i
\theta(x)} \phi_i(x), \qquad A'_\mu(x) = A_\mu(x) + \partial_\mu
\theta(x)
\end{equation}
But in fact $L''$ describes the same theory as the original $L$.
By the gauge choice $\theta(x) = - \chi(x)$ we can set $\chi' = 0$.
Then $A'_\mu$ decouples and has a trivial gaussian path integral,
leaving $L$.

One has to think that this symmetry does not mean anything, since it
is completely independent of the original theory.  What we have done
is to gauge a nonlinearly realized symmetry.  Non-linear in this
context refers to the term $\theta(x)$ in the transformation of
$\chi$, which is zeroth order in the fields; nonlinear terms of order
field-squared and higher would not have the same effect.
We see that a gauged nonlinear symmetry is like no symmetry at all.
This is the second time (out of three) that we encounter this theme:
that a gauge symmetry is after all just a redundancy, though
sometimes a very useful one, and we can always be more redundant.

Supersymmetry also has nonlinear realizations (in fact they were
discovered quite early in the subject).  Kunitomo~\cite{Knlr} argues
the above procedure for world-sheet supersymmetry, applied to the
bosonic string, gives the Berkovits-Vafa construction (there are
some quantum corrections in the currents~(\ref{bv}) that have to be
found by hand).  Indeed, the linear $b_1$ term in $T_F$ means that
the superconformal symmetry is nonlinearly realized on $c_1$.
It is then not so surprising that one can make the bosonic string look
like it has all these extra symmetries.

It is not clear what the moral is.  This makes the embedding seem
rather trivial, but it is a reminder that the amount of gauge
symmetry in a theory can be somewhat arbitrary, and so the
classification by gauge algebra
is not so absolute.  It would be telling if
one could reach the theory~(\ref{bv}) by turning on background
fields from some more familiar vacuum of the superstring.  By the
way, it is somewhat odd to expect the $N=0$ string as a ground state
of the
$N=1$ string, since the latter has fewer degrees of freedom (smaller
matter central charge), and even odder to get $N=1$ from $N=2$ since
the latter has just a scalar.  Earlier attempts to go the other way,
getting $N=1$ as a ground state of $N=0$, did not seem to lead
anywhere.

\subs{$S$-Duality}

The last equivalence we have to discuss is $S$-duality, a conjectured
equivalence between weakly coupled and strongly coupled string
theory~\cite{FILQ}. Since the string coupling is $e^{\Phi}$, this is
to say that the vacua $\Phi$ and $\Phi'$ are actually the same
state, where $\Phi'$ runs from $-\infty$ to $\infty$ as $\Phi$ does
the reverse.

This idea is much more far-reaching than anything we have
discussed thus far.  All the previous equivalences that we discussed
held order-by-order in perturbation theory.  $S$-duality
certainly does not hold order by order---it relates the perturbative
expansion, around zero coupling, to an expansion around infinite
coupling.
Thus it is a statement about the exact amplitude, and involves
nontrivial relations among all orders of perturbation theory.  My
purpose in the second half of these lectures was to try to go beyond
string perturbation theory, and $S$-duality is our first example.  It
is a subject of great current interest, several major papers having
appeared just since I began writing these lectures.  Unfortunately it
is a hard subject to present, because it is rather intricate and
because it is not one I have worked on in detail.  So I will just try
to summarize some of the main ideas.\footnote{A recent review by
Sen~\cite{Senrev} and a seminar given by Jeff Harvey at UCSB were
very helpful.}

To begin, consider the free Maxwell equations,
\begin{eqnarray}
\partial_\mu F^{\mu\nu} = 0 \nonumber\\
\partial_\mu F_{\rm d}^{\mu\nu} = 0,
\end{eqnarray}
where $F_{\rm d}^{\mu\nu} = \frac{1}{2}\epsilon^{\mu\nu\alpha\beta}
F_{\alpha\beta}$.  These are invariant under
\begin{equation}
F^{\mu\nu} \leftrightarrow F_{\rm d}^{\mu\nu}, \label{abdu}
\end{equation}
which interchanges the
electric and magnetic fields.  Note that the first equation is an
equation of motion, derived from the action, while the second is a
Bianchi identity, which follows from $F_{\mu\nu} = \partial_\mu A_\nu
- \partial_\nu A_\mu$ independent of the action.
One can instead write
$F_{\rm d}^{\mu\nu}$ as the curl of a dual vector potential
$A_{{\rm d} \mu}$, and
the equation of motion and Bianchi identity again change roles.

The classical Maxwell equations remain invariant if we add both
electric and magnetic sources.  This can be extended to the quantum
theory provided the Dirac quantization condition is
satisfied~\cite{Dirac}. That is, if a particle of electric and
magnetic charge $(Q_e, Q_m)$ exists, and another of charges $(Q'_e,
Q'_m)$, then
\begin{equation}
Q_e Q'_m - Q_m Q'_e \in 2\pi {\bf Z}\ .
\end{equation}
The simplest solution to this is that states be restricted to the
lattice
\begin{equation}
Q_e = e n_1, \qquad Q_m = \frac{2\pi}{e} n_2 \label{emlat}
\end{equation}
for integer $n_1$ and $n_2$.  A theory with an electrically charged
field $(e,0)$ and a magnetically charged field $(0,2\pi/e)$ is thus
invariant under the electric-magnetic duality~(\ref{abdu}) if we also
take $e \to 2\pi/e$, interchanging weak and strong coupling.
Unfortunately this is rather formal, because either the electric
or magnetic coupling is always strong, and we do not know how to make
sense of the resulting quantum field theory.

The charge lattice~(\ref{emlat}) is not the most general solution to
the Dirac quantization condition.  If a
$\theta$-parameter is added for the gauge field, the electric charges
shift by an amount proportional to the magnetic
charge~\cite{Wtheta},
\begin{equation}
Q_e = e n_1 + e n_2 \frac{\theta}{2\pi} , \qquad Q_m = \frac{2\pi}{e}
n_2.
\label{witlat}
\end{equation}
This is invariant under $\theta \to \theta+2\pi$, with $n_1 \to n_1 -
n_2$.  Electric-magnetic duality generalizes to nonzero $\theta$.
It is useful to form the combination
\begin{equation}
\tau = \frac{\theta}{2\pi} + i \frac{2\pi}{e^2}\ .
\end{equation}
Under electric-magnetic duality and $\theta \to \theta+2\pi$ we have
respectively
\begin{eqnarray}
\tau \to -\frac{1}{\tau}, && n_1 \leftrightarrow n_2 \nonumber\\
\tau \to \tau + 1, && n_1 \to n_1 - n_2, \quad n_2 \to n_2.
\label{sl2zem}
\end{eqnarray}
These do not commute, and they generate the familiar $SL(2,Z)$, found
earlier as the modular symmetry~(\ref{sl2z}) of the torus and in
section~3.3 as a duality symmetry with two compact
dimensions.

The {\it allowed} spectrum~(\ref{witlat}) is invariant under the
$SL(2,Z)$ duality, but it is a stronger statement for the {\it
actual} spectrum to be dual.  In the example above, for example,
only the states $(n_1, n_2) = (1,0)$ and $(0,1)$ appeared and the
spectrum was invariant under electric-magnetic duality (in this
Abelian example we are putting in fields by hand, but in the
non-Abelian case below the dynamics will determine the spectrum).
Now consider what happens to the magnetic state $(0,1)$ as we
increase $\theta$ and so its electric charge.  One scenario is that
at $\theta = \pi$, where its electric charge is $\frac{1}{2}$, it
becomes degenerate with the state $(n_1, n_2) = (-1,1)$ with charge
$-\frac{1}{2}$, and that at somewhat larger $\theta$ it is unstable
to decay into the latter state plus the $(n_1, n_2) = (1,0)$ electric
state.  The $SL(2,Z)$, if a symmetry, relates the theory at any
$\tau$ to that at some $\tau$ in the standard fundamental region. In
the case just described, there are two stable states everywhere, but
at the boundaries of the standard fundamental region there is a phase
transition in the spectrum and the quantum numbers of the {\it
stable} states change.

To be precise, the discussion assumed
that the state $(n_1, n_2) = (1,0)$ was massless at $\theta = \pi$.
If it is extremely light there is a narrow region where both
magnetic states are stable, and the phase transition separates into
two.  On the other hand, if it is heavy enough both states are
stable for all $\theta$.
In fact, in the supersymmetric case described below, the latter
situation holds and states cannot become unstable, at least for
$n_1$ and $n_2$ relatively prime.  The states $(n_1, n_2) = (0,1)$
and $(-1,1)$ must both be stable everywhere, as well as a large
number of others related by further $SL(2,Z)$
transformations---namely, all states with $n_1$ and $n_2$ relatively
prime.

Now let us go on to the case of interest, the pure non-Abelian
gauge theory.  The gauge field is itself electrically, not
magnetically, charged, which seems to introduce an essential
asymmetry.  This shows up in the field equations and Bianchi
identities,
\begin{eqnarray}
D_\mu F^{\mu\nu} = 0 \nonumber\\
D_\mu F_{\rm d}^{\mu\nu} = 0,
\end{eqnarray}
where $D_\mu$ is a covariant derivative, containing $A_\mu$ and
{\it not} $A_{{\rm d} \mu}$.  These equations can no longer be
written in terms of a dual vector potential $A_{{\rm d} \mu}$, and any
attempt to rewrite the action in terms of a dual vector potential
seems to lead rapidly to a mess.

But maybe we just haven't looked hard enough.
Non-Abelian theories do have magnetic monopole configurations,
some of which become stable when the symmetry is spontaneously broken
to an Abelian group.\footnote{As in the Ising model, duality would be
expected to exchange such topological objects with the fundamental
ones.}
There is some circumstantial
evidence for duality, especially in $N=4$ supersymmetric
gauge theories (and in some $N=2$ theories with matter).
Namely,\\[3pt]
1. The lattice of allowed electric and magnetic charges is invariant
under duality (or in some cases is related to that of a dual gauge
group)~\cite{MO},~\cite{GNO}.\\[3pt]
Also, when the non-Abelian symmetry is spontaneously broken to an
Abelian symmetry,\\[3pt]
2. The low energy (Abelian) theory is invariant under duality.\\[3pt]
3. The spectrum of massive gauge bosons and massive monopoles is
invariant under duality.\\[3pt]
4. The long-ranged forces between electric and magnetic charges
(from gauge fields and also massless scalars) are invariant under
duality.

It may seem remarkable that one can make such statements, since we
know so little about strongly coupled theories.  Actually, point~1
is just based on topology and should not depend on the interactions.
Points~2 to~4 use the supersymmetry in an essential way.  The low
energy theory consists of Abelian gauge bosons and neutral scalars
(and their superpartners) without any renormalizable interactions,
so is free at low energy even though the underlying theory is strongly
coupled.  Moreover its form is fixed by the $N=4$ supersymmetry,
so we don't need to solve the underlying theory to find it.  What is
left is the familiar Abelian duality.

Points~3 and~4 use the supersymmetry in a richer
way~\cite{WO}.\footnote{As an aside, spacetime supersymmetry and its
breaking are key issues in nonperturbative string theory, but ones
that I will have to largely neglect because of restrictions of time
and emphasis.  This is the one point where I will discuss
supersymmetry in any detail.}  Let us consider the following algebra,
\begin{eqnarray}
&&\{ Q_1, Q_1^\dagger \} = H + Q, \qquad \{ Q_2, Q_2^\dagger \}
= H - Q \nonumber\\
&&\{ Q_i, Q_j \} = \{ Q_i, H \} = \{ Q_i, Q \} =
\{ Q^\dagger_i, Q^\dagger_j \} = \{ Q^\dagger_i, H \}
= \{ Q^\dagger_i, Q \} = 0.
\end{eqnarray}
Here $Q_i$ are two supercharges, whose anticommutator includes the
Hamiltonian $H$ (as always in supersymmetry) and also a conserved
charge $Q$ (as is often the case).  You can think of this example as
living in zero space dimensions so
that there is no momentum or spin to worry about.  Going to
eigenspaces of $H$ and $Q$ with eigenvalues $h$ and $q$, the
supercharges form two fermionic oscillators, with the standard
representation in terms of $2^2$ states,
\begin{equation}
Q_1 |\! \downarrow\downarrow \rangle =
Q_2 |\! \downarrow\downarrow \rangle = 0, \qquad
Q^\dagger_1 |\! \downarrow\downarrow \rangle =
\sqrt{h+q}\, |\! \uparrow\downarrow \rangle, \qquad
Q^\dagger_2 |\! \downarrow\downarrow \rangle =
\sqrt{h-q}\, |\! \downarrow\uparrow \rangle, \label{srep}
\end{equation}
and so on, with the normalization factors $h \pm q$ coming from the
algebra.  Note the inequality
\begin{equation}
h + q = \langle \psi | \{ Q_1, Q_1^\dagger \} | \psi \rangle
= \| Q_1^\dagger  | \psi \rangle \|^2 +
\| Q_1 | \psi \rangle \|^2 \geq 0,
\end{equation}
and the same for $h - q$, so
\begin{equation}
h \geq |q|.
\end{equation}
Now, when the inequality is saturated, one of the terms in
(\ref{srep}) vanishes and a {\it small representation} with {\it 2
states} becomes possible.
\begin{figure}
\begin{center}
\leavevmode
\epsfbox{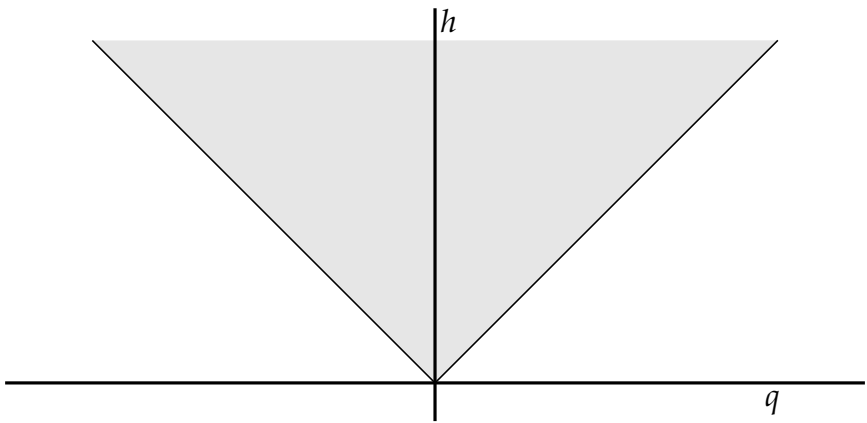}
\end{center}
\caption[]{Allowed states in the $h$-$q$ plane (shaded).  Small
representations occur on the boundary.}
\end{figure}
This idea plays a key role
throughout supersymmetry.  Now note that if we have a small
representation and we make continuous changes in the parameters
in the Hamiltonian, the energy of the states cannot change---it must
stay at $h = |q|$ because to move to $h > |q|$ we would need four
states.\footnote{This would be possible if there happened to be
another small representation, of opposite fermion number, at the same
value of $q$.}

The $N=4$ supersymmetry algebra has regular representations of $2^8$
states and small representations of $2^4$ states.  The masses,
electric and magnetic charges satisfy the Bogomol\'nyi-Witten-Olive
bound~\cite{WO}
\begin{equation}
M^2 \geq v^2 (Q_e^2 + Q_m^2), \label{bwo}
\end{equation}
where $v$ is the symmetry-breaking expectation value (this is for a
single $U(1)$ group).  Small representations saturate the bound.
In the perturbative,
small $g$, theory, both the charged gauge bosons and the
stable monopole solutions are in small representations, and moreover
one finds the same spin spectrum in each case.  Now vary the
coupling $g$.  Small representations have to stay small, and saturate
the bound, so by the time we get to $g' = \sqrt{4\pi}/g$ the spin and
mass spectrum is the same as we started with but with electric and
magnetic charges interchanged.

I have never known what to make of this---is it a simple consequence
of supersymmetry restricted to the Bogomol\'nyi sector, or is it
evidence for a duality of the full theory?  I have always been
skeptical, again because the duality seems to lead to a mess at
the Lagrangian level.
However, notice that while the BWO bound determines the {\it allowed}
spectrum of small representations, but it is a further non-trivial
fact that the {\it actual} spectrum found is dual~\cite{Osb}.  The
final bit of evidence for duality, the long-ranged force, is also a
consequence of the BWO bound.

Everything I have just said was known before 1980 for the
electric-magnetic duality.  Sen recently observed that the $SL(2,Z)$
would have additional implications (\cite{Senrev} and references
therein).  In particular, it requires stable states of monopole
charge
$n_2$ greater than one, which must appear as bound states of the
$n_2 = 1$ states.  Some of these have now been
found~\cite{Senbound}.  Recently, long papers by Seiberg
and Witten~\cite{SW1},~\cite{SW2} and Vafa and Witten~\cite{VafW} on
$N=2$ and
$N=4$ supersymmetric gauge theories have appeared, which seem to
present further evidence for duality, though I have not absorbed
these. All of this evidence refers in a sense only to the
Bogomol\'nyi sector of the spectrum, but so much is accumulating
that it is harder to believe that it is not a symmetry of the full
spectrum.

Thus far the discussion has involved only field theory, $N=4$
supersymmetric Yang-Mills.  For the heterotic string compactified on
a six-torus (that is, $\mu = 4, \ldots 9$ periodic) the low energy
theory contains $N=4$ Yang-Mills, and the conjecture is that this
string theory also is
self-dual~\cite{FILQ},~\cite{Senrev}.\footnote{For the heterotic
string theory on backgrounds of lower symmetry,
$S$-duality might also change the background.}  The circumstantial
evidence is of the same type as the above.  Let me just note a few
important differences.  The coupling constant $g^2$ is now a field,
being proportional to
$e^{\Phi}$. But so also is the topological angle $\theta$: it is
proportional to the axion field $a$ obtained from the antisymmetric
tensor.\\[3pt]
{\bf Exercise:} Show that for the two-index potential $B_{\mu\nu}$, if
one interchanges the Bianchi identity for the field strength
(vanishing curl) with the equation of motion (vanishing gradient), in
{\it four} dimensions one obtains a massless scalar field.  Show that
in {\it ten} dimensions one obtains a 7-index antisymmetric tensor
field strength which is the curl of a 6-index potential, and which is
invariant under a 5-index gauge transformation.\\[3pt]
The low energy field theory for the dilaton, axion, and gauge fields,
and their supersymmetric partners, is $SL(2,Z)$ invariant.

The other difference is that the BWO sector, which in the field
theory case consisted of a few small representations, is now very much
larger, with infinite numbers of electrically charged string states
and magnetically charged soliton states, of various types.  The duality
of the allowed states follows from supersymmetry as above, but
duality of the actual spectrum implies much more,
including `stringy'
monopoles which involve fields not in the low energy effective field
theory.  At this point it is much less clear whether the actual
spectrum is $S$-dual.  Incidentally,
Sen has also argued that $S$-duality, like $R$-duality, must be a
gauge symmetry in string theory.

Finally, another conjecture is that string theory is dual to a theory
of fundamental five-dimensional objects~(ref.~\cite{Ds5d} and
further references in ref.~\cite{Senrev}).  The evidence is of the
same type as the above---duality of the low energy field theory, and
interchange of the spectra of `fundamental' objects and solitonic
ones.  These arguments are made even though it is not known if it is
possible to quantize the fundamental five-brane---the world-sheet
theory is a six-dimensional field theory and so non-renormalizable.
They are made on the basis of presumed low-energy field theory of the
five-brane (it couples naturally to a six-index antisymmetric tensor
gauge field; see the above exercise), on its classical soliton
solutions, on scaling arguments based on the world-sheet and
spacetime theories, and on the possible winding states of a
five-dimensional object.  An interesting connection between this and
$S$-duality is that
\begin{equation}
S_{\rm string} =
(\mbox{5-brane $\to$ string}) (R_{\mbox{\scriptsize 5-brane}})
(\mbox{string $\to$ 5-brane})\ ,
\end{equation}
so string/5-brane duality would imply $S$-duality, given the more
straightforward $R$-duality.

\section{String Field Theory or Not String Field Theory}
\setcounter{footnote}{0}

Thus far, we have defined string theory only through its
perturbation expansion, the analog of Volume One of Bjorken and Drell.
Now we would like to find Volume Two.  We know that in field theory
there are many important phenomena that cannot be seen in perturbation
theory, and string theory will have all of these and probably more.
Also, we have found that string theory contains an enormous amount of
spacetime gauge invariance, in the form of spacetime Ward
identities satisfied by the scattering amplitudes.  But again, we
know from Yang-Mills theory and general relativity that this is a
very clumsy way to think about spacetime gauge invariance.  These
theories of course have a geometrical interpretation which is
essential to understanding the physics, and which is
disguised in the perturbation theory but is evident when they are
written as field theories.  So it seems that we should try something
similar in string theory, to introduce some sort of string field
$\Psi$, to find an action with the appropriate gauge symmetries, and
then to recover the perturbation theory from a gauge-fixed path
integral.  This section consists of a few assorted remarks about this
idea, focusing first on some very attractive features, and then
on some indications that it may not be quite the right
thing to do.

\subs{String Field Theory}

A closed string field will create or destroy a string along some
closed curve in spacetime, so it is a functional $\Psi[X]$ of such
paths.  An open string field will be a functional of open
curves.  In the earlier discussion, the closed and open string
wavefunctions $\psi[X]$ were similarly functionals of the
configuration of the string in spacetime, so this is like second
quantization, `promoting' the one-particle wavefunction to an operator
(or, in the path integral formalism, to a variable of integration).
It turns out that things work very much more nicely if one first goes
to the BRST-invariant form of the theory, and then second
quantizes~\cite{Sbrst},~\cite{Wsft}.\footnote
{String field theory is a large subject.  I will make a few
appropriate references, but the reader should consult the
reviews~\cite{Tsft},~\cite{Ssft},~\cite{Zsft} for extensive
references.} In this form the wavefunction also includes the state
of the ghosts. Since the $b$ and $c$ ghosts are conjugate, we can
for example regard
$b$ as the momentum and $c$ are the coordinate, so the wavefunction
would be a functional $\psi[X,c]$, and the corresponding field a
functional $\Psi[X,c]$.  Actually, it is still convenient to use the
bra-ket notation, writing the string field as $| \Psi \rangle$, which
is just the abstract notation for the functional $\Psi[X,c]
= \langle X,c | \Psi \rangle$.
We would have to use a different notation for the {\it states}
of the theory, something like {\boldmath $| \Phi )$}.
That is, {\boldmath $| \Phi )$}
is a state with any number of
strings, and $\Psi[X,c]$ or $| \Psi \rangle$ act as an operators on
these states.

We can think about this another way by expanding the functional
$| \Psi \rangle$ in terms of a complete set of such functionals.  For
the open string, for example,
\begin{equation}
| \Psi \rangle = \int \frac{d^{26}k}{(2\pi)^{26}}\,
\Bigl\{ T(k) |0,k\rangle + i A_\mu(k) \alpha_{-1}^\mu |0,k\rangle
+ B(k) b_{-1} c_0 |0,k\rangle + \ldots \, \Bigr\}. \label{sfe}
\end{equation}
The functions $T(k)$, $A_\mu(k)$, $B(k),\ldots$ or their
Fourier transforms $T(X)$, $A_\mu(X)$, $B(X),\ldots$ are the arbitrary
coefficients in the expansion.  There is an obvious interpretation of
$T(X)$ as the spacetime tachyon field and $A_\mu(X)$ as the
spacetime gauge field, while $B(X)$ will turn out to be an auxiliary
field.\footnote
{In the expansion~(\ref{sfe}), I have for brevity kept only terms
with the same ghost number as the ground state; more on this later.}

Now we need an invariance principle.  Since wavefunctions $Q_{\rm B}
| \chi \rangle$ are equivalent to zero, the natural guess is
\begin{equation}
| \Psi \rangle \ \to\ | \Psi \rangle + Q_{\rm B}
| \Upsilon \rangle \label{sginv}
\end{equation}
for any $| \Upsilon \rangle$.  Taking
\begin{equation}
| \Upsilon \rangle = \int \frac{d^{26}k}{(2\pi)^{26}}\,
\Bigl\{ \lambda (k) b_{-1} |0,k\rangle + \ldots \, \Bigr\},
\label{sfeg}
\end{equation}
(so that $Q_{\rm B} | \Upsilon \rangle$ has the same ghost number as
$| \Psi \rangle$) one finds from the expansion~(\ref{qex}) of $Q_{\rm
B}$ that the invariance~(\ref{sginv}) becomes
\begin{equation}
\delta A_\mu(X) = - 2 \partial_\mu \lambda(X)
\end{equation}
which is indeed the linearized gauge invariance of electromagnetism.

There is an obvious free field equation,
\begin{equation}
Q_{\rm B} | \Psi \rangle \ = 0, \label{sffe}
\end{equation}
which is invariant under~(\ref{sginv}) because $Q_{\rm B}^2 = 0$.
In components, this becomes
\begin{eqnarray}
&&\partial^2 T(X) = - \frac{1}{2} T(X), \qquad B(X) = - \partial_\mu
A^\mu(X) \nonumber\\
&& \partial^2 A_\mu(X) = - \partial_\mu B(X) =
\partial_\mu \partial_\nu A^\nu(X).
\end{eqnarray}
We obtain the appropriate Klein-Gordon equation for the tachyon and
the gauge-invariant free Maxwell equation for $A_\mu$.  Note that
the field equation~(\ref{sffe}) has the same form as the physical
state condition which earlier gave us the mass-shell conditions.
The one difference is that we required $b_0$ to annihilate physical
states, which at this level leads to $B(X) = 0$ and so
$\partial^2 A_\mu(X) = 0$.  So the $b_0$ condition is a stringy
generalization of Feynman gauge.

We can readily write down an invariant free action as well.  It is
simply
\begin{equation}
S_{\rm open}
 = \frac{1}{2} \langle \Psi | Q_{\rm B} | \Psi \rangle. \label{sfa2}
\end{equation}
The ghost number works out so that the action for
the fields~(\ref{sfe}) is non-zero.  Writing the inner product
in terms of the path integral on the disk, the vertex operators
${\cal V}_{\Psi}$ each have ghost number~1, as does $Q_{\rm B}$,
adding up to~3 as required by the same calculation~(\ref{nbnc}) as on
the sphere.  Not surprisingly, the action for $A_\mu$ is the free
Maxwell action, after $B$ is integrated out.  We have used the
bilinear inner product $\langle\ |\ \rangle$, but in fact the string
field must satisfy a reality condition
\begin{equation}
\langle \Psi | = \langle\!\langle \Psi | .
\end{equation}
This is necessary for unitarity, generalizing the familiar fact that
the metric and Yang-Mills fields must be real.

This all seems very beautiful to me.  The familiar spacetime gauge
invariance is embedded as the lowest component of a much larger
symmetry acting on all higher levels of the string, and the BRST
formalism allows this to be done in an extremely compact and elegant
way.  In fact, the action~(\ref{sfa2}) has the same structure as the
Abelian Chern-Simons action $\int d^3 X \, AdA$~\cite{Wsft}, where
$Q_{\rm B}$ is analogous to the exterior derivative $d$: note that
both are nilpotent. Incidentally, the components of other ghost
number, not written in the expansion~(\ref{sfe}), just play the role
of Fadeev-Popov ghosts for the spacetime gauge symmetry after the
gauge is fixed, and one obtains a BRST-invariant string field
theory.  To be precise, there are several complications relative to
the earlier discussion of BRST symmetry on the world-sheet, so a
generalization of BRST, the Batalin-Vilkovisky formalism,
emerges~\cite{Tsft}.

This all generalizes to the closed string, with a slight complication.
It is necessary to impose conditions $(b_0 - \tilde b_0) | \Psi \rangle
= (L_0 - \tilde L_0) | \Psi \rangle = 0$ on the closed string field
and on the gauge parameter $| \Upsilon \rangle$,
and the action is
\begin{equation}
S_{\rm closed}
 = \frac{1}{2} \langle \Psi | (c_0 - \tilde c_0) Q_{\rm B} | \Psi
\rangle.
\end{equation}
The ghost number is then correct.  The invariance~(\ref{sginv})
is now an infinite-dimensional generalization of linearized coordinate
invariance.

Now the question is whether we can find an appropriate nonlinear
generalization of the string gauge symmetry and action.  Let us
transform this into a related question.  After gauge fixing, the path
integral over the string field can be expanded perturbatively in terms
of Feynman diagrams built from string propagators and vertices.  Is it
possible to represent the amplitudes, which were described earlier in
terms of a sum over world-sheets, in this way?

I am going to make a distinction here between what I will call an
{\it effective string field theory} and a {\it bare string field theory,}
and describe the former first.  What I am call an effective string
theory is described in ref.~\cite{Zsft}.  It is always possible by
brute force to write the string amplitudes in terms of Feynman
diagrams.  Let us introduce an arbitrary three-closed-string vertex,
as depicted in fig.~19a.  Figure~19b shows a four-string amplitude
built from two such vertices.
\begin{figure}
\begin{center}
\leavevmode
\epsfbox{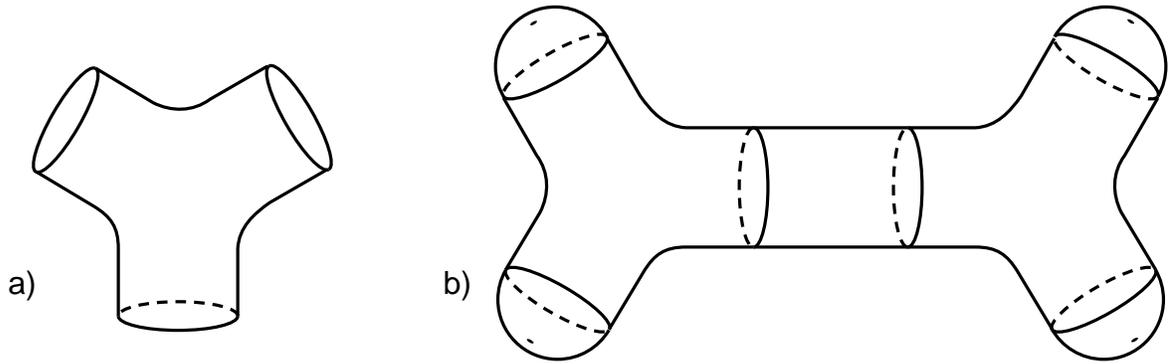}
\end{center}
\caption[]{a) Three-string vertex. b) Four string amplitude.}
\end{figure}
The propagators are cylinders, with lengths
integrated from zero to infinity (generalizing the Schwinger
representation of the propagator in field theory).
There are three such graphs, from the three channels.  Now recall
from the Virasoro-Shapiro amplitude~(\ref{4p}) that this amplitude is
supposed to be given by an integral of one vertex operator position
$z_4$ over the complex plane, the other three operators being fixed.
The graphs of fig.~19b cover three round regions centered on
the three fixed positions, fig.~20a, but this inevitably leaves a
region between uncovered.\footnote{
It also could be that the round regions overlap and double-cover
some part of the $z_4$-plane; one can avoid this by making the
`stubs' on the three-string vertex long enough.
Also, by taking a more complicated propagator it may be possible to
cover the four-point amplitude correctly, but this will fail for the
five-point amplitude.}
\begin{figure}
\begin{center}
\leavevmode
\epsfbox{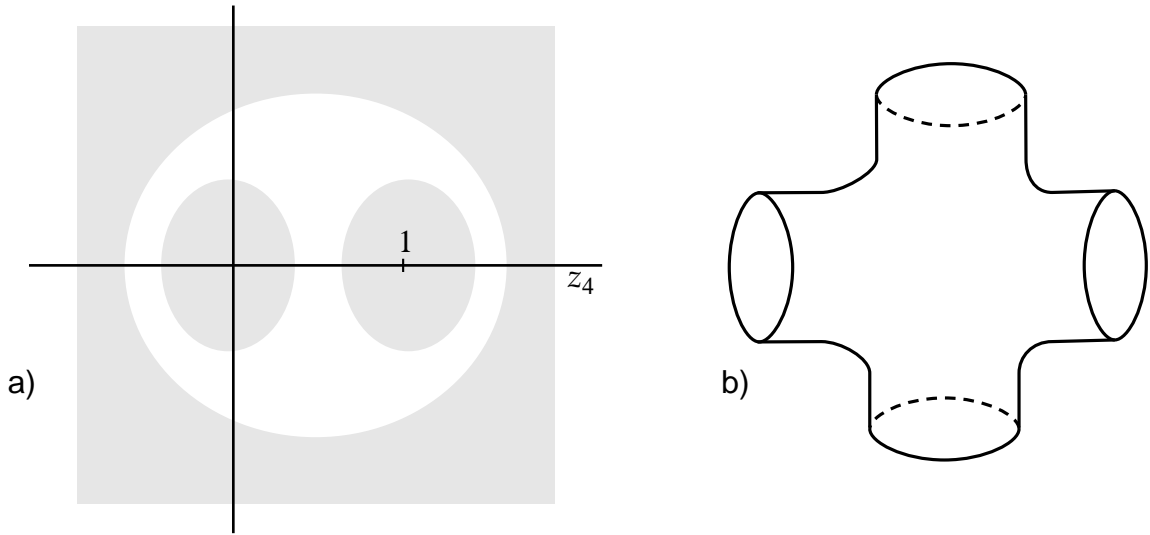}
\end{center}
\caption[]{a) Graph of fig.~19b covers three round regions in the
complex plane. b) Four-string vertex needed to cover unshaded region in
fig.~20a.}
\end{figure}
So it is necessary to introduce a four-string vertex as shown in
fig.~20b.  An integration over shape in included in the definition
of the vertex so as to cover exactly the missing region of
fig.~20a.  Going on to higher amplitudes, one must further introduce
$n$-string vertices for all $n$.
At tree level the vertices have the topology of spheres with $n$
holes.  However, when one goes on to loop amplitudes, the tree level
action again does not cover moduli space, so it is necessary to add
additional vertices containing internal loops.  This procedure can be
carried out iteratively, and in the end the full perturbation series
is written in terms of Feynman graphs.  Ref.~\cite{Zsft} gives an
explicit construction of one possible set of vertices.
All of this can also be
applied to theories of open plus closed strings.

Now one can work backwards, writing the $n$-string vertex as a term
in the action with a product of $n$ string fields.  The resulting
action does indeed have a non-linear generalization of the string
gauge symmetry.  This is not surprising for the following reason.
In the discussion of spacetime Ward identities in section~2.7,
I emphasized that these would hold if and only if the integration
ran over the correct moduli space.  The iterative construction of the
vertices does this, so the spacetime Ward identities hold and a
corresponding invariance should be present in the action.  The
condition that moduli space be properly covered is equivalent to
a set of identities for the vertices, and these same identities
imply the nonlinear invariance of the action~\cite{Zsft}.

\subs{Not String Field Theory}

I have called the above construction an {\it effective string field
theory} because it is very similar to a Wilsonian effective field
theory.  The long-distance physics is explicitly represented, while
the short-distance physics is already integrated out.
(A similar analogy was made in refs.~\cite{BdA}).  In
particular, we have seen that long-distance propagation in
spacetime, producing the poles in string amplitudes, comes from long
cylinders or strips at the boundaries of moduli space.  In the
effective string field theory, the vertices include an integration
over moduli, but only an {\it interior} region of moduli space, as
in the example of fig.~20.  The degenerating cylinders appear
explicitly as propagators. On the other hand, we will see at two
points later on that stringy physics comes from the interior of
moduli space, which  are hidden inside the vertices.

Effective string field theory is very useful for some purposes.
By describing explicitly the boundaries of moduli space it is a
useful tool for demonstrating the finiteness and unitarity of string
perturbation theory---see the discussion at the end of section~2.
But it
would not seem to be the right tool for studying nonperturbative
string theory, since the stringy physics is already integrated out.
Can we do better?  For the open string we
certainly can~\cite{Wsft}.  The simple Witten vertex shown in
fig.~21, in which three open strings are joined by gluing their
halves together in pairs, exactly covers moduli space.
\begin{figure}
\begin{center}
\leavevmode
\epsfbox{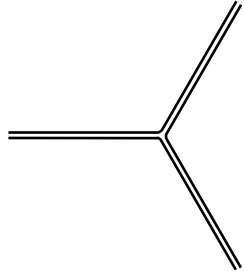}
\end{center}
\caption[]{Witten vertex for three open strings.}
\end{figure}
The interacting action thus has a
nonlinear generalization of the string gauge invariance, and
has the same structure as the non-Abelian Chern-Simons action.
I will call this a `bare string theory,' since the action has a closed
form without corrections from all orders in the loop expansion.

The perturbative formulations of open and closed string theories were
quite parallel.  For example, the spectrum of the closed
string was essentially the tensor product of two copies of the open
string spectrum.  But now we run into a real asymmetry, for there does
not seem to be any covariant closed string field theory simpler than the
effective theory that I described earlier.\footnote
{I exclude here the non-covariant light cone string field theory.
Possibly this is the right approach, but there are many things about it
that I do not understand---subtleties with contact terms, and with the
vacuum structure.  Also, there was some development of light-cone-like
covariant string field theories, but these seem to have had
difficulties.}
The existence of $n$-point
tree level interactions would not have been surprising, since general
relativity is non-polynomial while Yang-Mills theory is polynomial.
But the fact that the action receives corrections from all orders in
the quantum loop expansion means that a great deal of non-trivial
physics is already integrated out.

There is another, rather remarkable, asymmetry between open and closed
strings.  The Witten open string theory covers the moduli
space of Riemann surface with boundaries.  But as I have emphasized at
the end of section~2.5, this will include processes with intermediate
closed strings.  The simplest example of this is in fig.~22, the
one loop vacuum amplitude, with one propagator and no
vertices.
\begin{figure}
\begin{center}
\leavevmode
\epsfbox{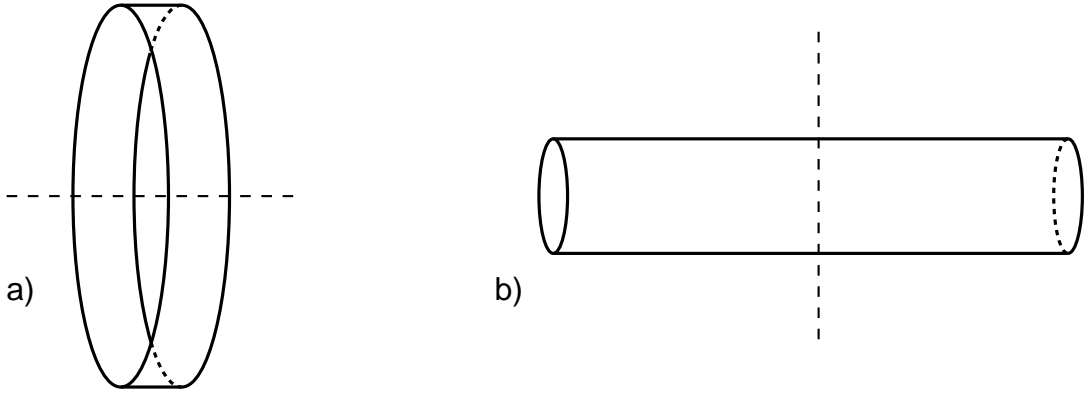}
\end{center}
\caption[]{a) Annulus at large $t$.  Cutting on dashed line gives an
open string pair. b) Annulus at small $t$.  Cutting on dashed line
gives a single closed string.}
\end{figure}
This has a modulus $t$, the ratio of the circumference to
the length, which runs from $0$ to $\infty$.
For large $t$,
fig.~22a, this looks like a vacuum fluctuation consisting of a pair
of open strings.  But for small $t$, fig.~22b, it looks like a
single closed string appearing and then disappearing.  Indeed, the
string graph contains both processes.  The same thing happens at for
higher order open string amplitudes.
While we can think of the Witten theory as a path integral over
open string fields, it contains the full open plus closed string
physics~\cite{GMW}.

It is not clear what the logical relation is between the closed and
open strings here.  It is not that the closed strings are bound states
of open strings---there are no interactions in fig.~22.  In some
sense the closed strings are singular configurations of the open
string field.  Siegel has suggested recently that the relationship is
similar to bosonization, though I do not grasp all his
arguments~\cite{Socb}.   Perhaps the lesson is that open string field
theory is all we need, with the closed string emerging as in
fig.~22.  This idea was pursued for a while, but it is
not clear where to go with it~\cite{Scsft}.  There is also the
difficulty that the most promising theory, the heterotic string,
does not have an open string version.

The graph in fig.~22 is the same as the
one-loop vacuum amplitude in field theory, with $t$ the Schwinger
parameter, and it corresponds to summing
$\frac{1}{2} \omega$ times the spacetime volume over all open string
modes.  It is interesting to contrast this with closed string theory.
Figure~22 is obtained by taking an open string propagator, a strip of
length/width$=t$, gluing the ends and integrating over $t$.  The
analogous construction in closed string theory would be to take a
cylinder, gluing the ends, and integrating over lengths and also
twist angles.  This would correspond to integrating over tori, with the
modulus $\tau$ running over the full region Re$(\tau) < \frac{1}{2}$,
Im$(\tau) > 0$.  This is not the integration region, fig.~6, for
closed string theory.  So the closed string vacuum amplitude is not
given by summing $\frac{1}{2} \omega$ over closed string frequencies,
a strong indication that closed
string theory is not a field theory~\cite{Ptor}.
The open string vacuum amplitude {\it is} given by summing $\frac{1}{2}
\omega$ over open string frequencies, though it includes also the
closed string process of fig.~22b.

There is one more important non-field-theoretic property of closed
strings, having to do with the large-order behavior of the
perturbation series.  Consider a quantum field theory where the
coupling constant $g$ appears only as an overall factor $g^{-2}$ in
the action, so that $g^2$ is the loop counting parameter.  Rather
generally, the large order behavior of the perturbation
series $\sum_{h = 0}^\infty g^{2h} T_h$ (reviewed in
ref.~\cite{GZ-J}) is
\begin{equation}
T_h\ \sim\ h! h^A C^{-h} \label{hopt}
\end{equation}
with $A$ and $C$ constants.  The dominant factor here is the factorial.
Generically this arises simply because the number of Feynman graphs
with $h$ loops is of order $h!$---if one cuts off the momentum
integrations in the IR and UV the propagator and vertex factors just
enter into the constants $A$ and
$C$.\footnote{But in some cases there are also contributions of order
$h!$ coming from the momentum integrations in a small subset of graphs.
I do not know if there is any deep reason why two such different
sources give effects of the same order.}
The perturbation series is thus divergent.  The ratio of successive
terms is
\begin{equation}
\frac{g^{2h} T_h}{g^{2h-2} T_{h-1}} \ \sim\ h C^{-1} g^2,
\end{equation}
so the smallest term is for $\tilde h \sim Cg^{-2}$.  From
Stirling's approximation,
\begin{equation}
A_{\tilde h} \sim e^{-C/g^2}. \label{nptyp}
\end{equation}
This is the smallest term, so represents the maximum accuracy of
perturbation theory as an asymptotic series.  Indeed, many
non-perturbative effects, effects which do not occur at any order of
perturbation theory, are of this magnitude, including confinement,
chiral symmetry breaking, supersymmetry breaking in supersymmetric
gauge theories, instantons, even BCS superconductivity.

Open string theory is much the same~\cite{GPer}.  The Witten field
theory represents moduli space as a sum of Feynman graphs, so the
factorial growth of the number of graphs implies a factorial volume
of moduli space.  The integrand can be bounded in the interior of
moduli space, so again barring infrared divergences the same
estimate~(\ref{hopt}) holds, and nonperturbative effects of order
$e^{-C/g_{\rm o}^2}$ are expected.

The same argument does not apply to closed string theory, because the
effective string field theory has complicated vertices containing
integrals over moduli space.  Shenker estimates the large order
behavior as follows~\cite{Slo}.  He uses the fact that Witten's open
string field theory generates the amplitudes of the full open plus
closed string theory.   He then argues that the purely closed-string
amplitudes are a non-negligible portion of the full amplitudes,
so that the perturbation theory for them grows as rapidly.
Nonperturbative effects are then of order
\begin{equation}
e^{-C/g_{\rm o}^2} = e^{-C/g_{\rm c}}.
\end{equation}
In terms of the closed string perturbation theory, this translates
into large order behavior proportional to $(2h)!$, as had been
discovered in the matrix models.

These $e^{-C/g_{\rm c}}$ effects have no analog in field theory, and
their nature is not known.  We will see what they are in the matrix
model, but it has been hard to guess how to generalize the result.
At small $g_{\rm c}$ these stringy effects are larger than the familiar and
important $e^{-C/g_{\rm c}^2}$ nonperturbative effects in
the low energy field theory.
Moreover, they are likely to involve phenomena which are unique to
string theory.  Of all the things I am covering in these lectures, this
is the one where I most wanted to be able to say something new.
I have made one observation~\cite{Pdir}, which is that if one
includes boundaries with Dirichlet conditions, $X^\mu =$constant,
these act as instantons but with a weight of the desired form
$e^{-C/g_{\rm c}}$. The conjecture
is that the stringy nonperturbative effects make their appearance as a
sum over various kinds of boundaries; I am trying to test this in the
matrix model.

In lieu of anything solid to say, let me make a conjecture: that
closed string field theory is simply wrong nonperturbatively, and that
closed strings themselves are collective excitations of some other
degrees of freedom, in terms of which the theory should be formulated.
Let me list some evidence for this.
\begin{itemize}
\item  The rapid growth of perturbation theory, suggesting that the
perturbative description is rather far from the exact formulation.
\item  The fact that closed string field theory does not seem to work,
except as an effective theory with the stringy physics integrated out.
Note that the rapid growth of perturbation theory implies that the
interior of moduli space, the part which does not look like Feynman
graphs and is integrated into the vertices, is very large.
\item  The appearance of closed strings in open string field theory.
\item  Matrix models, where closed strings are collective (bosonized)
excitations of free fer\-mions.
\item  The closed string is {\it almost} the product of
right- and left-moving theories, suggesting that it might be useful
to regard it as a bound state of right-moving and left-moving
strings.  In the flat space theory,
there are two things which glue the two sides together.  The first is
the equality of the zero modes eigenvalues, $\alpha_0 =
\tilde\alpha_0$.  The second is the global structure of moduli space.
Locally, the moduli space has a natural complex structure
corresponding to the right-left separation---for example, the modulus
$\tau$ of the torus appears as $\tau L_0 - \bar\tau \tilde L_0$.
But the region of integration is not any sort of product.
\item  Too much gauge symmetry.
Here I am being contrary, since earlier I told you that it was
wonderful that string theory embedded the spacetime gauge symmetries
in a much larger structure.  There I was following the usual particle
theory paradigm that local symmetry is holy, and that as one goes to
higher and more fundamental energies one expects to see more and more
of it, as in $SU(3) \times U(1) \in SU(3) \times SU(2) \times U(1)
\in SU(5)$ of the Standard Model and GUTS.  But this need not be the
case.  As I have tried to emphasize, gauge symmetry is just a
useful redundancy, and there are examples where one emerges at low
energy even though there is no sign of it in the underlying theory.
The familiar example from particle theory is the $CP(n)$ sigma
model~\cite{DDL}. In condensed matter physics, this has been proposed
to occur in theories of strongly coupled electrons, where the
electron separates into a `spinon' and `holon,'
\begin{equation}
\psi_e(x,t) = \psi_s(x,t) \phi(x,t).
\end{equation}
This decomposition is redundant,
the transformation
\begin{equation}
\psi'_s(x,t) = e^{i \lambda(x,t)} \psi_s(x,t), \qquad
\phi'(x,t) = e^{-i \lambda(x,t)} \phi(x,t)
\end{equation}
leaving the physical field, the electron, invariant.  It is plausible
that under some conditions this redundancy is elevated to a dynamical
symmetry (for a review see ref.~\cite{Frad}).  As far as I know there
is no reason in principle that this cannot happen.\footnote
{Ref.~\cite{WW} shows that in some circumstances a global symmetry of
the underlying theory cannot be promoted to a local symmetry
at long distance,
but it imposes no restriction on the promotion of a redundancy, which
acts trivially in the underlying theory.}
So perhaps the short distance theory, rather than exhibiting an
enormous gauge symmetry, should be formulated entirely in terms of
invariants.

Recall from the discussion in section~3.4 that one way to distinguish
a useless from a useful redundancy is to see whether the fields
transform inhomogeneously.  The string gauge symmetry is
\begin{equation}
| \Psi \rangle \ \to\ | \Psi \rangle + Q_{\rm B}
| \Upsilon \rangle + O(| \Psi \rangle )
\end{equation}
The linear term is as before, eq.~(\ref{sginv}).  We do not know the
higher terms in any simple form but we do not need them, because it
is precisely the term $Q_{\rm B} | \Upsilon \rangle$, of zeroth
order in
$| \Psi \rangle$, that is relevant here.  If this vanishes (at least
for some momentum), the gauge symmetry is real.  So we are looking for
solutions of
\begin{equation}
Q_{\rm B} | \Upsilon \rangle = 0.
\end{equation}
This looks like the physical state condition, but it is different
because the $| \Upsilon \rangle $ have ghost number one less than the
states, so it is a different cohomology.  In fact, the only solutions
for a flat background are $\Upsilon = \partial X^\mu \tilde c$ or
$c\bar\partial X^\mu$, which are just the translations, so the only
`real' gauge symmetries are those of the graviton and antisymmetric
tensor.  But this need not be conclusive---there are other backgrounds
where other parts of the string gauge symmetry are unbroken, as we
will see in section~5, so it may be that it is useful to keep the full
redundancy.

\item  The observation that holes in the world-sheet naturally give
effects of order $e^{-C/g_{\rm c}}$
\cite{Pdir}, suggesting a breakdown of
the world-sheet.
\end{itemize}
This is the set of ideas I play with, though it does not yet add up
to anything coherent.

The idea that we should look for more fundamental degrees of freedom
in string theory was put forward in ref.~\cite{AtW}, in a study
of the high-temperature behavior of string theory.
This discusses a number of other non-field-theoretic properties of
closed strings, and also suggests breakdown of the world-sheet by way
of holes.  Another idea~\cite{Tbit},~\cite{Sbit} is that
the string should be thought of as a
collection of bits; the string picture breaks down when the
density of bits becomes large.  It may be that these ideas are
connected, holes appearing in the world-sheet because neighboring bits
unbind.

\subs{High Energy and Temperature}

In particle physics the traditional way to find out what things are
made of is to bang them together.  I will briefly describe three
different high-energy regimes, each of which gives a different
picture.  We might hope that in one limit or another the theory will
simplify enough to allow us to go beyond perturbation theory.

The first limit is scattering at high center-of-mass energy, $E =
s^{1/2}$, and fixed angle.  This is where Rutherford found the
atomic nucleus, and where SLAC found the partonic constituents of
hadrons, quarks and gluons.  In relativistic field theory this process
probes distances of order
$E^{-1}$.    We have already seen in the
tree-level Virasoro-Shapiro amplitude that the amplitude is very
soft in this limit.  To get further insight\cite{GM} consider
the path integral over $X^\mu$ from which it was obtained,
\begin{equation} \int [dX]\, e^{-S_{\rm P} + i \sum_{i=1}^4 k_i \cdot
X(\sigma_i)}\ ,
\end{equation}
the second term in the exponent being from the tachyon vertex
operators.  The limit is essentially
obtained by scaling up the $k_i$ uniformly.  The action $S_{\rm P}$ is
quadratic in $X$, so the path integral is determined by a saddle point
$X^\mu_{\rm cl} \ \propto\ E$, and the amplitude is $e^{-O(E^2)}$.
This same saddle point dominates if the
tachyons are replaced with other states of the string; in particular,
the masses of the external states drop out in the limit.   Also the
$z_4$ integral in the resulting
amplitude~(\ref{vsint}) is dominated by a saddle point, at
\begin{equation}
z_4 = \frac{k_1 \cdot k_4}{(k_1 + k_2)\cdot k_4}\ .
\end{equation}
Note that this is in the interior of moduli space, another example
of the idea that stringy behavior comes from the interior.

In contrast
to field theory, the size of the interaction region {\it grows} with
energy~\cite{GM}. Combining the low energy field theoretic behavior
with this high energy result, the effective uncertainty is
\begin{equation}
{\mit \Delta} X \ \sim\ \frac{1}{E} + \alpha' E. \label{strun}
\end{equation}
This is further evidence for an effective minimum distance in string
theory, as found earlier from duality.  It appears that the string
can carry only of order one string unit of energy per unit length, so
to transfer a much larger energy $E$ many bits are needed---the
effective number of partons is proportional to the energy.

Gross and Mende were able to find the dominant saddle point at every
order of perturbation theory.  It is just an
$n$-fold cover of the tree-level saddle point, scaled by a factor
$n^{-1}$, with $n-1=h$ being the number of loops.  The exponential in
the scattering amplitude~(\ref{hard}) is suppressed by $n$,
\begin{equation}
c_n e^{- s f(\theta) / n} \label{hardn}
\end{equation}
so at large $s$ high orders dominate.
This can be interpreted in terms of dividing the scattering into $n$
softer scatterings of angle $\theta/n$.
There has been some discussion of the summation of this
series~\cite{MOog}, and of a large symmetry in the high energy
limit~\cite{Ghesym}, assuming that it is correct simply to sum the
leading behavior from each order.\footnote{There has also been a
complementary study, of the all-orders summation of small-angle
scattering~\cite{ACV}.} It is interesting to contrast this with the
earlier discussion of the large-order behavior.  The
amplitude~(\ref{hardn}) comes from a single saddle point in moduli
space, so is unrelated to the large volume noted earlier.  However,
it is enhanced by the kinematics and actually grows much faster than
the volume of moduli space, $c_n$ being proportional to $(9h)!$.

The second limit is soft scattering, high energy and small angle,
holding fixed
the momentum transfer $q \sim
(-t)^{1/2}$.  Whereas the hard scattering is an
occasional rare process where many bits of string move together,
the soft scattering is a picture of the typical string configuration.
{}From
Stirling's approximation, the Virasoro-Shapiro amplitude is of order
\begin{equation}
A_{\rm VS}\ \sim\ s^{2 + t} t^{-1} = \frac{s^2}{t} e^{-q^2 (\ln s)}.
\end{equation}
This is the gravitational amplitude $s^2/t$ modified by a form-factor
which corresponds to an object with a size of order $\sqrt{\ln
E}$~\cite{Slor}. There is another way to see this same result.  Let
us calculate the root-mean-square size of the string ground state.
The mode expansion gives
\begin{equation}
\langle 0 | (X^1(\sigma) - \overline {X^1})^2 | 0  \rangle
\ =\ \sum_{m = 1}^\infty \frac{1}{m^2}
\langle 0 | (\alpha_m \alpha_{-m} + \tilde\alpha_m \tilde\alpha_{-m})
| 0 \rangle\ = 2 \sum_{m = 1}^\infty \frac{1}{m}.
\end{equation}
This is divergent but has no direct physical significance.  A
measurement on a time-scale $\delta t$ is sensitive only up to modes
of frequency less than $\delta t^{-1}$,
essentially $m\, \delta t < 1$.
So the log divergence becomes $\ln (\delta t^{-1})$, and the size is
the square root of this~\cite{Slor}.  This agrees with the form
factor, where the soft scattering probes time scales $\delta t \sim
{1/E}$.

Just as the hard scattering~(\ref{strun}) is interpreted as a
stringy uncertainty principle, Susskind interprets this root-log-$E$
growth as a stringy Lorentz transformation, transverse sizes not
being
constant as in classical physics.  A similar calculation shows that
the longitudinal size of the string goes to a constant of order the
string scale, not the usual contraction as $1/E$.

The square root of a logarithm is a very slow function, and
ordinarily would be of little importance.  But there is one
situation where enormous boosts are encountered---a black hole.
 From Strominger and Verlinde you have heard two very different
points of view on whether short distance physics, such as string
theory, can be relevant to the information problem.  What makes this
so controversial is that in a black hole there are no large local
invariants, such as would ordinarily be needed for the low energy
effective field theory to break down, but there is a very large
nonlocal invariant, the relative boost between an infalling and an
asymptotic observer.  The external observer `sees' the infalling
observer slow down and sit forever (or until the hole decays) on the
horizon.  One tick of the infalling clock takes longer and longer as
seen from the outside, the ratio going as $e^{t/4M}$ with $M$ the
mass of the hole and $t$ the time measured by the external observer.
So the external observer is seeing the internal motion of infalling
strings slowed down, a fixed time scale $\delta t_{\rm ext}$
corresponding to a time interval $e^{-t/4M}\delta t_{\rm ext}$
for the infalling object.  As time goes on, the external observer
`sees' more and more of the modes of the string, and it appears to
grow, as
\begin{equation}
\sqrt {\ln(e^{t/4M}\delta t_{\rm ext}^{-1}) }\ \propto
t^{1/2}.
\end{equation}
Before the black hole evaporates, the string grows to macroscopic
sizes, and low energy field theory no longer applies.
Also, because of its longitudinal behavior
the string, rather than contracting closer and closer to the
horizon, remains a finite thickness above it.  Again, this would have a
profound effect on the Hawking radiation.

However, I have put `sees' in quotes, because to see one needs
light, and the black hole is black.  Of course there is Hawking
radiation, but the conventional wisdom is that this is produced
outside the horizon---the relevant modes are in their ground states
near the horizon---and so cannot reveal all of the marvelous things
that the string is doing, and the string physics is irrelevant.
At least, I have not been able to imagine a real dynamical
calculation which shows the stringy effects to be relevant.  But
this picture of the essentially diffusive growth of the string is
very simple and appealing~\cite{MPT}, and the stringy properties do
dovetail very nicely with what would be needed to solve the
information problem.  It may be that the low energy field theoretic
treatment is internally consistent but not correct, because the
large nonlocal invariant makes string theory important even before
the low energy field theory breaks down internally due to a large
local invariant. This would be a satisfying resolution of the
various points of view; the problem is to find the right calculation!

The previous limits involved two strings, each in a state of
low excitation but with large center of mass energy.  The final
limit I will discuss is that of a single string in a very high state
of excitation.  The density of single-string states per unit energy
$n(\varepsilon)$ is related to the free energy in the noninteracting
limit,
\begin{equation}
\beta F(\beta) = \int_0^\infty d\varepsilon\,n(\varepsilon)
e^{-\beta\varepsilon},  \label{freeen}
\end{equation}
where I have used the fact that $e^{-\beta\varepsilon}$ will be much
less than~1 for the relevant states.
The density of states grows exponentially, as $e^{\beta_{\rm c}
\varepsilon}$, so the integral converges at low temperature but
diverges at $\beta < \beta_{\rm c}$.
It is tempting to interpret this as a transition
(the Hagedorn transition) to a phase where the conjectured
fundamental degrees of freedom will be evident.  As yet, there
is little understanding of the high-temperature phase.

There is a large literature on this subject, but
here I just wish to note that there is a very simple picture of the
typical high-energy string state, which gives a surprisingly good
quantitative account of the density of
states~\cite{SSk}-\cite{LoTh}. First let me
tell you more precisely what the density of states is.  In the limit
$\varepsilon \to \infty$, at fixed volume,
it is
\begin{equation}
n(\varepsilon) = \frac{1}{\varepsilon} e^{\beta_{\rm c}\varepsilon}.
\label{den1}
\end{equation}
This is true for any finite volume spatial background, such as a
torus, or a group manifold.  If instead one takes the linear size to
infinity faster than $\varepsilon^{1/2}$, one finds
\begin{equation}
n(\varepsilon) = \frac{V}{\varepsilon^{(1 + D)/2}}
e^{\beta_{\rm c}\varepsilon}.
\label{den2}
\end{equation}
where $D$ is the number of spacetime dimensions and $V$ the volume.

The results~(\ref{den1}) and~(\ref{den2}) follow if one assumes that
the typical highly excited string is a random walk, with its length
proportional to its energy.  First, the exponential factor
comes from the fact that the number of steps is proportional to the
length, with a choice to be made at each step.  The Hagedorn
transition thus arises from a competition between the energy and entropy
of a long string.
The power-law prefactor is important for the details of
the transition.  In the density~(\ref{den2}), the factor of $V$ is
counts the number of places a random walk might start.  We must then
divide by the typical volume of the random walk, because we can make
a closed string only if the final point and initial point coincide;
this gives a factor of $\varepsilon^{(1-D)/2}$.  Finally, this
overcounts by a factor of the length of the string, since we can
start at any point along the closed string, so we need an additional
factor of $\varepsilon^{-1}$, giving the result~(\ref{den2}).
In finite volume, the random walk will eventually fill the space
so that its volume is just $V$, giving instead~(\ref{den1}).
Remarkably, the density of states is completely independent of the
size and shape of the space in this limit.
With such a simple picture of the free theory we might hope that the
interacting theory will be tractable.

\section{Matrix Models}
\setcounter{footnote}{0}

\subs{$D=2$ String Theory}

Returning to the theme of my introduction,
one of the things I learned from Wilson's lecture is the extent to
which he was influenced by the pion-nucleon static model.  This is
an essentially one-dimensional free pion field in interaction with a
static nucleon with some internal states.\footnote{To the condensed
matter audience this will sound very similar to the Kondo model.  In
fact it was only some time later that Wilson learned about the Kondo
model and realized that his ideas would apply.}  In spite of having
only one interacting degree of freedom, this model has the essence
of quantum field theory, a nontrivial renormalization
group flow.

String theory has an enormous number of degrees of freedom and a
rather rigid structure.  It is not easy to find models which
are simple enough to deal with and yet retain the essential
features of the theory.  In the previous section we mentioned
various high energy limits where things seem to simplify a bit, but
as yet it has not been possible to do anything beyond perturbation
theory.  The one limit which has turned out to be solvable is the
limit of small numbers of spacetime dimensions,
$1 \leq D \leq 2$ (for reviews see
refs.~\cite{GMtasi},~\cite{Krev}).

A simple solution to the background field equations~(\ref{beta}) for
arbitrary $D$ is~\cite{Mdil}
\begin{equation}
G_{\mu\nu} = \eta_{\mu\nu}, \qquad B_{\mu\nu} = 0, \qquad
\Phi = Q X^1/2 \label{lindbg}
\end{equation}
with
\begin{equation}
Q^2 = \frac{26 - D}{3}.
\end{equation}
The field equations~(\ref{beta}) were derived in world-sheet
perturbation theory and are generally true only when the gradients
of the background fields are small, but the solution~(\ref{lindbg}) is
a special case because the path integral remains gaussian.  In fact,
we have already constructed the CFT.
On a flat world-sheet the
world-sheet action reduces to the ordinary free $X^\mu$, but with
energy-momentum tensor as in eq.~(\ref{ldcft}),
\begin{equation}
T = -\frac{1}{2} :\! \partial X^\mu \partial X_\mu \!: -
\frac{1}{2} Q \partial^2 X^1
\end{equation}
giving $c = D + 3 Q^2 = 26$.

What is the physics of this background?  Recall that $e^\Phi$ plays
the role of the string coupling, so the coupling is
position-dependent, as $e^{Q X^1/2}$.  For $X^1 \to -\infty$ the
coupling goes to zero, but for $X^1 \to \infty$ the coupling diverges
and string perturbation theory breaks down.  Adding a tachyon
background produces a theory in which string
perturbation theory is valid.  The vertex operator $e^{\alpha X^1}$
has weight
\begin{equation}
h = \tilde h = \frac{-(2\alpha - Q)^2 + Q^2}{8},  \label{lvert}
\end{equation}
so for $Q^2 \geq 8$, which is $D \leq 2$, there are two positive
real solutions to the mass-shell condition $h = \tilde h
= 1$. It is the lesser,
$\alpha_1 \leq Q/2$, which is appropriate,\footnote{This is a
subtle point~\cite{Sliou}, and I do not want to
go off on this tangent, but I would like in this footnote to state
my understanding~\cite{Pliou}.  The tachyon has a second order field equation,
and the background is a linear combination of {\it both}
solutions.  At large
$X^1$, the nonlinearities due to the tachyon self-interaction become
large and the linearized solution no longer holds.  The appropriate
linear combination of the two solutions is determined by a condition
of nonsingularity in the nonlinear region.  In the linear region,
it is the more slowly decaying solution, the
lesser value of $\alpha_1$, which dominates. Incidentally, for $D=2$
the two roots are equal and the dominant term is $X^1 e^{X^1
\sqrt{2}}.$  The linear part is important in understanding details
of the amplitudes.}
\begin{equation}
T(X^1) \sim e^{\alpha_1 X^1}\ .  \label{tachback}
\end{equation}
Adding this
to the action gives
\begin{equation}
S = \frac{1}{8 \pi } \int d^2\sigma \, \Bigl\{
\sqrt{g} g^{ab} \eta_{\mu\nu} \partial_a X^\mu
\partial_b X^\nu  + Q \sqrt{g} R X^1 + \mu e^{\alpha X^1} \Bigr\}.
\label{diltach}
\end{equation}
The $X^1$ action is the Liouville theory.  The tachyon background
(`Liouville wall')
suppresses the path integral at large $X^1$ where the coupling is
strong, the exponential factor in the acion dominating the linear factor
from the growth of the coupling.

This string theory is solvable at $D =1$ or 2, or for one Liouville
dimension plus a minimal model (recall from eq.~(\ref{cdof})
that the central charge counts the number of degrees of freedom,
so a minimal model is like less than one whole dimension).
I will concentrate on
$D=2$, which has the most physics.  As we have discussed, the
physical state conditions remove two sets of oscillators, leaving
in this case none.  Only the center of mass motion, the tachyon,
remains.\footnote {Actually, this counting breaks down at discrete
momenta, where there are extra physical states, an important point
to which I will return.}  For real momentum the vertex
operator~(\ref{lindvo}) is $e^{i k \cdot X + QX^1/2}$.  The factor
$e^{QX^1/2}$ we now recognize as the string coupling, which
multiplies the tachyon wavefunction. The weight, given that $Q^2 =
8$ in $D=2$, is $1 +
\frac{1}{2} k^2$, so the mass-shell condition is $k^2 = 0$.  The
tachyon is misnamed in $D=2$: it is massless!\\[3pt]
{\bf Exercise:} Show in the same way that $M^2 = -\frac{1}{12}(D -
2)$ in general
$D$. This can be interpreted as the negative Casimir energy of the
$D-2$ transverse modes.

The theory thus reduces to a massless scalar field moving in one
time and one space dimension.  The spatial dimension is not
translation-invariant. Rather, the coupling goes to zero at $X^1 \to
-\infty$, and grows with $X^1$ until we reach the `Liouville wall,'
the tachyon background which cuts off further
propagation~\cite{Pcrit}.
\begin{figure}
\begin{center}
\leavevmode
\epsfbox{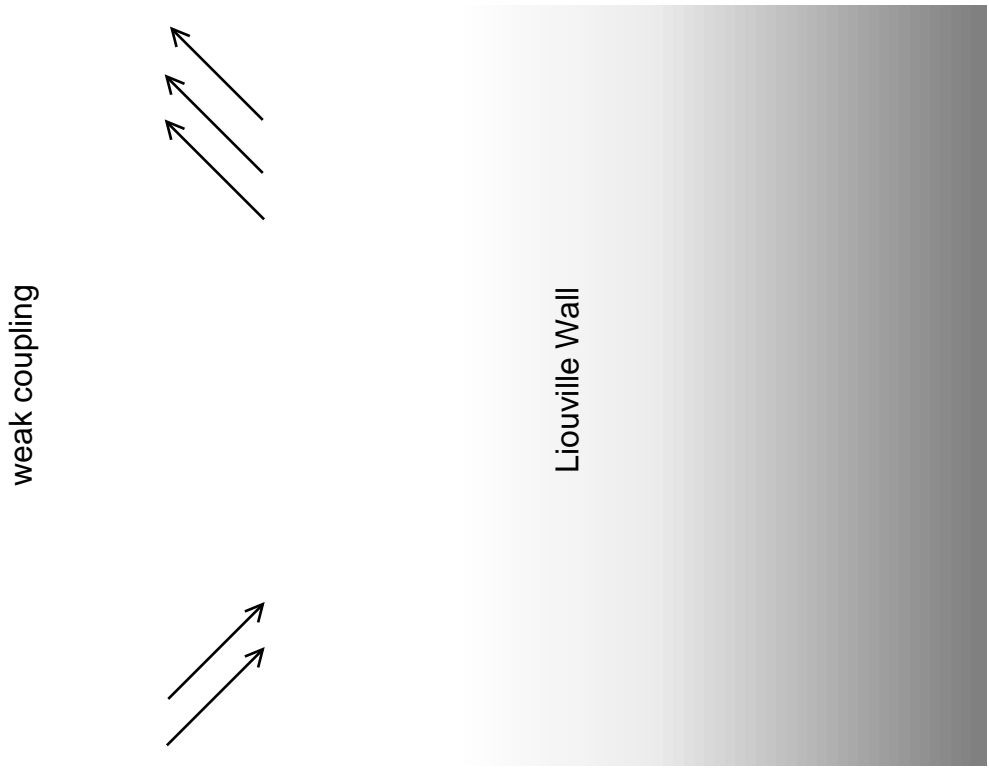}
\end{center}
\caption[]{Two dimensional spacetime with dilaton and tachyon
backgrounds.  The S-matrix describes transitions from a state
with $m$ incoming tachyons to a state with $n$ outgoing
tachyons.}
\end{figure}
The basic physics is
to throw tachyons in from $-\infty$; near the wall they interact
with each other and reflect back to $-\infty$.  This is about as
rich as throwing pions at a static nucleon.  Just as the latter
proved to contain the essence of field theory, the $D=2$ string
has taught us at least one important lesson about the full theory,
the existence of the large nonperturbative effects.  I will review
the solution, what has been learned, and what more we might hope to
learn.

\subs{The $D=1$ Matrix Model}

The method by which the $D=2$ string theory was solved is completely
orthogonal to what we have discussed thus far.  It begins with the
quantum mechanics of an $N \times N$ Hermitean matrix
{\bf M}(t)~\cite{BIPZ}.  The action is
\begin{equation}
S = \beta N \int dt \, \biggl\{ \frac{1}{2} {\rm Tr}\,(\dot {\bf
M}^2)  + {\rm Tr}\, V({\bf M})\biggr\}\ .
\end{equation}
Here $\beta$ is a parameter and $V$ a general potential.
You have already heard about this sort of thing from one or more of
Ambjorn, Ginsparg, and Moore, so I will go through the solution
without dwelling too much on the details.
A graph with given numbers of propagators $P$, vertices $V$, and
loops $L$ will depends on $\beta$ and $N$ as
$\beta^{n_V - n_P} N^{n_V - n_P + n_L} = \beta^{-n_L}
(\beta N)^{\chi},$ where $\chi$ is Euler number of the surface we
get by filling in the index loops in the graphs.  Thus, for any
amplitude
\begin{equation}
A(\beta, N) = \sum_{n_L, \chi} A_{n_L, \chi} \beta^{-n_L}
(\beta N)^{\chi},
\label{bN}
\end{equation}
where $A_{n_L, \chi}$ is the contribution from graphs with given
numbers of loops and given topology.
The Hamiltonian is
\begin{equation}
H = -\frac{1}{2\beta N} \sum_{i,j} \frac{\partial}{\partial M_{ij}}
\frac{\partial}{\partial M_{ji}} + \beta N {\rm Tr}\, V({\bf M}).
\end{equation}
The action is invariant under the $U(N)$ which takes ${\bf M} \to
{\bf UMU}^{-1}$.  We will only be
interested in singlet states---when one cuts open a graph, the states
will always be traces of {\bf M}'s, and so invariant.
A singlet state is a function only of the eigenvalues $\lambda_i$.
Acting on such a state one finds (use first and second order
perturbation theory to take the derivatives of eigenvalues)
\begin{eqnarray}
H { \Psi}(\lambda) &=& \sum_k \biggl\{ -\frac{1}{2\beta
N}  \frac{\partial^2}{\partial \lambda_k^2}
+ \frac{1}{\beta N} \sum_{l \neq k} \frac{1}{\lambda_l - \lambda_k}
\frac{\partial}{\partial \lambda_k}
+ \beta N V(\lambda_k) \biggr\} { \Psi}(\lambda) \nonumber\\
&=& \Delta^{-1}(\lambda)H' \Delta(\lambda) { \Psi}(\lambda)
\label{hprime}
\end{eqnarray}
where
\begin{equation}
H' = \beta N \sum_k \biggl\{ -\frac{1}{2\beta^2 N^2}
\frac{\partial^2}{\partial \lambda_k^2}
+ V(\lambda_k) \biggr\}, \label{ffh}
\end{equation}
and $\Delta(\lambda) = \prod_{k < l}(\lambda_k - \lambda_l)$.
We will absorb the factor of $\Delta$ into the wavefunction and work
with ${ \Upsilon}(\lambda) = \Delta(\lambda) { \Psi}(\lambda)$.
The inner product works out in a simple way: after the angular
integrations,
\begin{equation}
\int d^{N^2}{\bf M}\, { \Psi}^*(\lambda) { \Psi}(\lambda)
\ \propto \int d^N \lambda \, { \Upsilon}^*(\lambda)
{ \Upsilon}(\lambda),
\end{equation}
so ${ \Upsilon}$ is the probability amplitude for the eigenvalues.

The Hamiltonian~(\ref{ffh}) describes $N$ decoupled
coordinates.  The factor of $\Delta$ in $\Upsilon$ makes the
wavefunction antisymmetric, so this is a system of $N$ free
fermions~\cite{BIPZ}.  It can also be written in terms of a
second-quantized spinless fermion field $\zeta(\lambda)$,
\begin{equation}
H' =  \beta N \int d\lambda\, \biggl\{ \frac{1}{2\beta^2 N^2}
\partial_\lambda
\zeta^\dagger \partial_\lambda \zeta
+ V(\lambda) \zeta^\dagger \zeta \biggr\}\ .
\label{2q}
\end{equation}

 From the $N$-dependence~(\ref{bN}), the $N \to \infty$ limit is
dominated by graphs of spherical topology.  This limit is easily
taken in the free-fermion form.  Although we set the original
$\hbar$ to 1, we see that if we define $H' = \beta N H''$ then
$\beta N$ appears in $H''$ precisely the way $1/\hbar$ appears
in the Hamiltonian for nonrelativistic quantum mechanics.
The $N \to \infty$
limit at fixed $\beta$ is then
classical.  Each fermion occupies a volume $2\pi$`$\hbar$' =
$2\pi/\beta N$ in phase space.  Levels with single-particle energy
$E'' = \frac{1}{2} p^2 + V < \varepsilon_F$ are
filled.  The total number of fermions is $N$, so
\begin{equation}
\frac{1}{\beta} =
\frac{N}{\beta N} = \int \frac{dp\,d\lambda}{2\pi}\,
\theta\biggl( \varepsilon_F -
\frac{p^2 }{2} - V(\lambda) \biggr)\ . \label{nfcon}
\end{equation}

Now let us, to be specific, take
\begin{equation}
V(\lambda) = \frac{1}{4} \lambda^2 (2 - \lambda)^2. \label{2well}
\end{equation}
For large $\beta$, the phase space integral~(\ref{nfcon}) is small,
and the fermions sit near the quadratic minimum at
$\lambda = 0$.  One finds $\varepsilon_F \sim \beta^{-1/2}$.
There are no fermions in the second minimum at $\lambda = 2$:
the Feynman rules represent perturbation theory around ${\bf M} =0$,
and the second minimum is invisible in perturbation theory.
Figure~24 shows the potential, with occupied states shaded.
\begin{figure}
\begin{center}
\leavevmode
\epsfbox{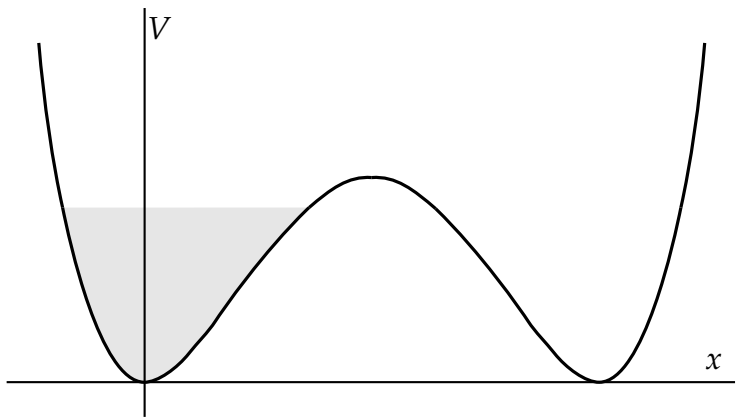}
\end{center}
\caption[]{Potential energy for fermions, with filled states
shaded.}
\end{figure}
As $\beta$ decreases, the occupied phase space volume grows.
At some critical value $\beta_{\rm c}$ the Fermi level reaches the
local maximum, $\varepsilon_{\rm c} = \frac{1}{4}$ at $\lambda =
1$, and there is a phase transition (see~\cite{BIZ} for a review).
Quantities such as \begin{equation}
\langle 0 | H' | 0 \rangle
= N^2 \int \frac{dp\,d\lambda}{2\pi}\, \left\{ \frac{p^2 }{2} -
V(\lambda) \right\} \theta\biggl( \varepsilon_F -
\frac{p^2 }{2} - V(\lambda) \biggr)  \label{g0h}
\end{equation}
are analytic for $\beta > \beta_{\rm c}$ and have a singularity
at $\beta_{\rm c}$.\\[3pt]
{\bf Exercise:} Show that in terms of $\delta = \varepsilon_{\rm
c} - \varepsilon_F$,
\begin{equation}
\beta - \beta_{\rm c} \sim  \delta \ln \delta, \qquad
\langle 0 | H | 0 \rangle \sim \delta^2 \ln \delta,
\end{equation}
up to terms analytic in $\delta$~\cite{KMs2}.
This takes a bit of care,
because the analytic terms in these two quantities are larger
than the singular ones,
though they are irrelevant because they do not come from large
surfaces.
Show from this that the
contribution $A_{n_L, 2}$ defined in eq.~(\ref{bN}) goes as
$n_L^{-3} \beta_{\rm c}^{n_L}$ for $\langle 0 |
H | 0 \rangle$ at large $n_L$.

Now recall, from the counting~(\ref{bN}), that $\beta^{-1}$ is the
loop-counting parameter for perturbation theory.  So for a
potential with a local maximum, the perturbation series has a finite
radius of convergence, diverging beyond $\beta^{-1} = \beta_{\rm
c}^{-1}$. The non-analyticity in $\beta - \beta_{\rm c}$ then
arises from large orders of perturbation theory, large graphs.
Here is where the connection with string theory comes in.  Assign a
length $a$ to each propagator.  As $n_L \to \infty$, one might hope
to take $a \to 0$ in such a way that the sum over graphs approaches
a sum over smooth surfaces, so the critical behavior in $\beta -
\beta_{\rm c}$ is given by a string theory~\cite{KMs2}.  There are
other possibilities---the typical graph might instead look like a
branched polymer---which can arise in the limit of large discrete
surfaces.  But in this case we are lucky, and the limit is a string
theory.  We can verify this by comparing various quantities that
can be calculated both in the matrix model and in the continuum
theory, and then use the matrix model to do many calculations that
we cannot yet do in the continuum.

We have taken the large $N$ limit, so we have only surfaces of
spherical topology.  Thus, the ground state energy~(\ref{g0h}) is
explicitly $O(N^2)$.  We can do better~\cite{c1refs}.
Let us take $N \to\infty$
and $\beta \to \beta_{\rm c}$ together, holding fixed
\begin{equation}
\overline\mu = \beta N (\varepsilon_{\rm c} - \varepsilon_F) .
\end{equation}
This is the {\it double scaling limit.}\footnote
{This limit was carried out first for the $D=0$ matrix
model~\cite{c0refs}.
Although this theory has fewer degrees of freedom, the solution and
the double scaling limit are perhaps a bit more difficult.}
Since $(\beta N)^{-1}$
plays the role of $\hbar$, the splitting between fermionic levels is
of order $1/\beta N$, and the ratio of this splitting to the
distance
$\varepsilon_{\rm c} - \varepsilon_F$ from the critical Fermi level
remains constant.  The splitting thus remains non-negligible in the
double scaling limit.  Since the splitting is a $1/N$ effect,
vanishing in the spherical limit, the double scaling limit is
keeping nontrivial contributions from all topologies,
with Euler number
$\chi$ being weighted by $\overline\mu^\chi$.  That is, each
additional handle brings in $\overline \mu^{-2}$.

The double-scaling limit is easily taken directly in the
Hamiltonian.  Define
\begin{equation}
\lambda - 1 = (\beta N)^{-1/2} x, \qquad
\zeta = (\beta N)^{1/4} \psi.
\end{equation}
Thus, for fixed $x$, the eigenvalue
$\lambda$ approaches the local maximum as $N
\to \infty$.  Only the quadratic behavior near the maximum survives
in this limit.  The Hamiltonian becomes
\begin{equation}
H' - \beta N \varepsilon_F =  \int dx\, \biggl\{ \frac{1}{2}
\partial_x \psi^\dagger \partial_x \psi
-\frac{x^2}{2} \psi^\dagger\psi + \overline\mu \psi^\dagger\psi
\biggr\}\ .
\label{2qds}
\end{equation}
The result is simply a Fermi sea in an inverted oscillator
potential.

\subs{Matrix Model $\leftrightarrow$ String}

What do we expect for the continuum string theory corresponding to
this critical behavior?  The number of links separating two vertices
gives a measure of distance, so we expect a world-sheet metric
$g_{ab}(\sigma)$.  And, each vertex occurs at some time $t$, so this
should become a world-sheet field $t(\sigma)$ in the continuum limit.
The most relevant action for these fields is
\begin{equation}
S_{nc} = \int d^2\sigma \, \sqrt{g} \biggl\{
-c_1 g_{ab} \partial_a t \partial_b t + c_2 + c_3 R \biggr\}\ .
\end{equation}
There is no reason to expect Weyl-invariance.  Consequently the
world-sheet cosmological constant term $c_2$ has been included.
We have earlier discussed briefly the world-sheet theories without
Weyl invariance.  We can go to the conformal gauge
\begin{equation}
g_{ab}(\sigma) = e^{\varphi(\sigma)} \hat g_{ab}(\sigma),
\end{equation}
in which the metric is reduced to the degree of freedom
$\varphi(\sigma)$. Taking into account the Fadeev-Popov
determinant and the Weyl anomaly from the $t$-integration, the
action becomes~\cite{Poly1}
\begin{equation}
\int d^2\sigma \, \sqrt{\hat g} \biggl\{
-c_1 \hat g_{ab} \partial_a t \partial_b t + c_2 e^{\varphi} + c_3
\hat R + \frac{25}{96\pi}
( \hat g^{ab} \partial_a \varphi \partial b \varphi + 2 \hat R
\varphi ) \biggr\}\ . \label{tvar}
\end{equation}

We started with one spacetime dimension, but as we have noted
the Weyl factor in the metric behaves very much like an extra
embedding dimension.  So we can think of this as a
{\it non-critical
string} in one dimension with metric $g_{ab}$, or a {\it critical
string} in two dimensions with metric $\hat g_{ab}$.  The former is
appropriate if one is interested in two-dimensional quantum
gravity on the world-sheet; the latter is appropriate if one is
interested in two-dimensional string theory in the embedding space.
Ambjorn's lectures focus on the former, and mine
on the latter.  Although the theories are the same, one tends to
ask different questions in the two cases.  In the former case,
the focus is on various measures of the world-sheet geometry, while
in the latter it is on scattering processes in the embedding
space.

Comparing the result~(\ref{tvar}) with our
expectation~(\ref{diltach}), we see precisely the same terms
with $\varphi \propto X^1$, but the coefficients cannot be made to
match up---the `25' needs to be `24' and the exponent in the
cosmological term is wrong.  This has a simple explanation.  The
derivation of~(\ref{tvar}) implicitly defined the $\varphi$ path
integral using the metric $g_{ab}$, while the latter used the metric
$\hat g_{ab}$, and the difference is accounted for by the
corresponding renormalization of the measure~\cite{ddk}.
This is presumed now to be well-understood~\cite{DHddk} and is not
something I want to dwell on, so I use the action in the
form~(\ref{diltach}) with the relative normalization of $X^0$ (in
the string theory) and $t$ (in the matrix model) yet to be
determined.

To summarize, we expect that the free fermion theory~(\ref{2qds}) is
equivalent to $D=2$ string theory in the dilaton and tachyon
background~(\ref{tachback}).  In the string theory we had tachyons
bouncing off a Liouville wall.  In the matrix model we have
fermions bouncing off the inverted harmonic potential.  So things
look very close, the string theory being a bosonization of the matrix
model.

Let us carry out the bosonization in
detail~\cite{DJcol}-\cite{Wclass}.  In the classical limit we can
describe the collective motions of the fermions in terms of a time
dependent Fermi surface, separating the filled and empty phase space
regions.  Fermions on the surface on the surface move freely in the
inverted harmonic potential, \begin{equation}
D_t p = x, \qquad D_t x = p, \label{ihoem}
\end{equation}
where $D_t$ denotes the co-moving derivative, following one
phase-space point on the surface.  The solution is
\begin{eqnarray}
p = - a \sinh (t - b), \qquad x = -a \cosh (t - b)
\end{eqnarray}
with integration constants $a$ and $b$.  In contrast, the position
$p$ of the Fermi surface at fixed $x$ satisfies $\partial_t p|_x =
D_t p - \partial_x p|_t D_t x = x - \partial_x p|_t p$.  For
perturbations that are not too large (and we will return to these
later), the Fermi sea can be described by the positions of its
upper lower surfaces at each time, $p_\pm(x,t)$, and each satisfies
\begin{equation}
\partial_t p_{\pm}(x,t) = x - p_\pm(x,t) \partial_x p_\pm(x,t).
\label{eom}
\end{equation}
For example, from the Hamiltonian~(\ref{2qds}) it follows that the
static Fermi level is given by $\frac{1}{2}(p^2 - x^2) + \overline
\mu = 0$ or
\begin{equation}
p_{\pm}(x)_{\rm static} = \pm \sqrt{x^2 - 2\overline\mu}\ ,
\label{statlev}
\end{equation}
which satisfies the equation of motion~(\ref{eom}).
\begin{figure}
\begin{center}
\leavevmode
\epsfbox{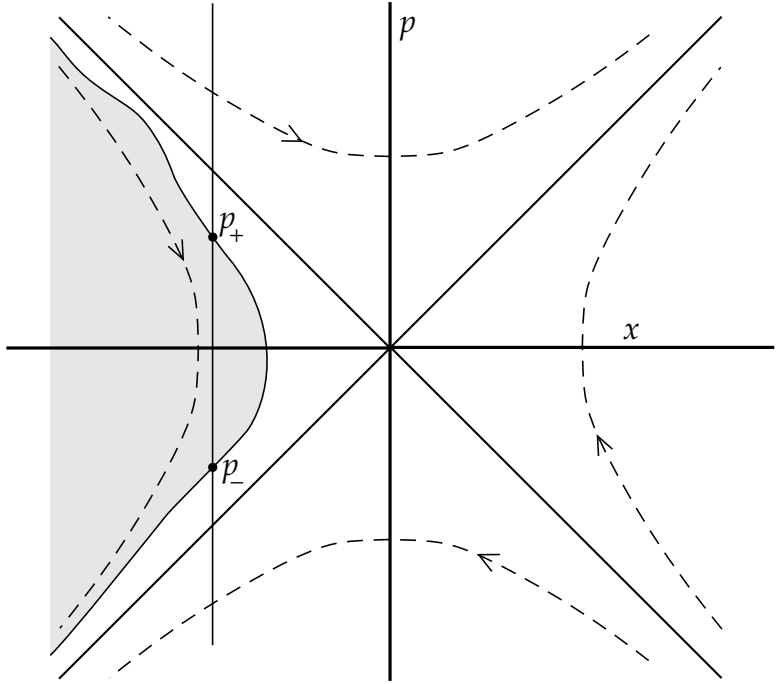}
\end{center}
\caption[]{Fermi sea: filled states in the $x$-$p$ plane are
shaded.  The state of the sea is given by the upper and lower
surfaces $p_{\pm}(x,t)$, shown at a representative value of $x$.
Points on the Fermi surface move along
hyperbolic orbits.}
\end{figure}
The Hamiltonian in terms of $p_\pm$ is obtained by integrating
the single-particle energy over the Fermi sea,
\begin{eqnarray}
H' &=& \frac{1}{2\pi} \int_{-\infty}^\infty dx
\int_{p_-}^{p_+} dp\, \frac{1}{2} (p^2 - x^2) \nonumber\\
&=& \frac{1}{2\pi}\int_{-\infty}^\infty dx\,
\biggl\{ \frac{1}{6}(p_+^3 - p_-^3)
-\frac{x^2}{2}(p_+ - p_-) \biggr\}\ . \label{ham}
\end{eqnarray}
Also, from the equation of motion and the Hamiltonian one
can deduce the commutator
\begin{equation}
[ p_\pm(x), p_\pm(y) ]  = - 2\pi i \partial_x \delta(x-y),
\qquad [ p_+(x), p_-(y) ] = 0.
\end{equation}

To write this in terms of a massless scalar, define the coordinate
$q = - \ln(-x)$, which runs from $-\infty$ to $\infty$ as
$x$ runs from $-\infty$ to 0, and define a scalar field
$\overline{\cal S}(q,t)$.
\begin{eqnarray}
p_{\pm}(x,t) &=& \mp x \pm \frac{1}{x} \epsilon_\pm(q,t)  \label{ppm}
\nonumber\\
\pi^{-1/2} \epsilon_\pm(q,t) &=& \pm \overline\pi_{\cal S}(q,t) -
\partial_q \overline {\cal S}(q,t).
\end{eqnarray}
The bar is to distinguish
this scalar from a slightly different one to appear later.  The
Hamiltonian takes the form
\begin{equation}
H' = \frac{1}{2} \int_{-\infty}^\infty
 dq \Bigl\{ \overline\pi_{\cal S}^2 + (\partial_q \overline {\cal
S})^2 + e^{2q} O(\overline {\cal S}^3) \Bigr\}.
\end{equation}
The quadratic part is the canonically normalized free scalar
Hamiltonian; the equation of motion is $(\partial_t^2 -
\partial_q^2) \overline {\cal S} = e^{2q} O(\overline {\cal
S}^2)$.

The collective motion of the matrix model fermions has the same
qualitative behavior as the $D=2$ string theory discussed earlier.
Asymptotically it is a free massless scalar, with interactions
that grow exponentially.  In the string theory, strings are
reflected from the Liouville wall before the coupling diverges,
while in the matrix model they are reflected from the end of the
eigenvalue distribution.  Thus, our guess that the $D=2$
critical string describes the critical behavior of the $D=1$ matrix
model appears to be correct.  The coupling goes as $e^{\sqrt{2}
X^1}$ in string theory and $e^{2q}$ in the matrix model, which
determines the relative normalization of these.  Also, the velocity
is 1 in both the $q$-$t$ and $X^1$-$X^0$ planes, so we have
\begin{equation}
X^1 \sim q \sqrt{2},\qquad X^0 \sim t \sqrt{2}. \label{mmst}
\end{equation}
We had earlier concluded that $\overline \mu^{-2}$ is the
loop-counting parameter, which would make $\overline \mu^{-1}$
the three-string coupling.  We can now check this.  From the
static distribution~(\ref{statlev}), the turning point in the
eigenvalue distribution is at $x_t \ \propto\ \overline\mu^{1/2}$.
The dominant
interactions occur at the Liouville wall $x_t$, where the coupling
$g_{\rm s}$ goes as $e^{2q_t}= x_t^{-2} \ \propto\
\overline\mu^{-1}$ as expected.
More detailed comparisons are possible as
well, and in particular the relation between the string
coordinate $X^1$ and the matrix model eigenvalue $q$ is more
subtle than the scaling argument~(\ref{mmst}) suggests.
We will return to this,
after discussing some general issues.

Two asides before we go on.  First, it is interesting to
contrast the bosonization we have just done with relativistic
bosonization.  The latter relates the collective motion of a
free massless relativistic fermion to a {\it free} massless boson.
The difference is that the relativistic fermion has a linear
dispersion relation, $\omega \ \propto\ k$, while the matrix
model fermions have $\omega \ \propto\ k^2$.  Looking at the
Hamiltonian, we see that the cubic scalar interaction comes from
the term quadratic in momentum.  We can think about this as
follows.  Relativistic fermions have no dispersion, the
velocity $\partial_k \omega$ being a constant.  A collective
pulse propagates without changing its shape.
The matrix model fermions have a $k$-dependent velocity,
so that a pulse deforms as it propagates.  This translates
into interactions among the bosons.
\begin{figure}
\begin{center}
\leavevmode
\epsfbox{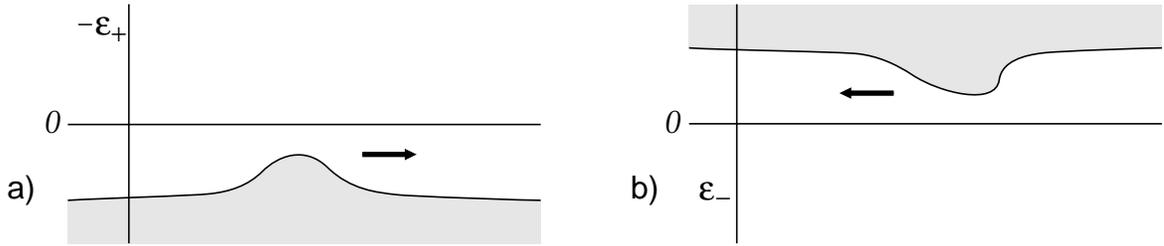}
\end{center}
\caption[]{a) Incoming pulse on top of the static Fermi sea.
For clarity the variable $-\epsilon_+(q,t)$ is shown.
b) Later outgoing pulse, with shape distorted by dispersion.
Fermions closer to $|p| = |x|$ travel more deeply into the
potential and take longer to return.}
\end{figure}

Second, the endpoint of the distribution,
where $p_+ - p_- \to 0$, causes difficulties.  At this point the
dynamics is determined not by the Hamiltonian but by the
constraint that the fermion density $p_+ - p_-$ be positive.
Further, perturbation theory in the collective Hamiltonian
suffers various divergences at the endpoint, which need to be
appropriately regulated.  It is sometimes useful to work with
the Fourier transformed fermion $\tilde \psi(p,t)$, in terms
of which the Hamiltonian looks almost the same~\cite{Kcomm},
\begin{equation}
H' - \beta N \varepsilon_F =  \int dp\, \biggl\{ -\frac{1}{2}
\partial_p \tilde\psi^\dagger \tilde\partial_p \psi
+ \frac{p^2}{2} \tilde\psi^\dagger\tilde\psi + \overline\mu
\tilde\psi^\dagger\tilde\psi
\biggr\}\ .
\label{xtop}
\end{equation}
The collective motion can now be described in terms of the
surface $x(p, t)$.  In this variable the sea has only an upper
edge, and there is no endpoint to worry about.

\subs{General Issues}

So what does one learn from all this?
The object of study is the amplitude for $m$
incoming strings of energies $\omega_1, \ldots, \omega_m$ to evolve
into $n$ outgoing strings of energies $\omega'_1, \ldots,
\omega'_n$.
The matrix model allows
the exact calculation of these amplitudes~\cite{MPR}.  One starts with
the incoming strings, translates into an incoming fermion state
using bosonization (asymptotically we have noted that the bosons
become free, so one can use the familiar relativistic bosonization),
evolve the free fermions in the inverted harmonic potential, and
then translate back using bosonization.  The final closed form
from combining all these steps is a little
complicated,\footnote{The
expressions do simplify in the leading order, string tree level.
We will work this out later.}
but for given $m$ and $n$ one can follow through the
steps and find the exact answer, without resorting to expansion
in the closed string coupling $\overline{\mu}^{-1}$.
The same calculations remain
exceedingly hard within the continuum string theory.  One must in
principle carry out the path integral over $X^\mu$, integrate over
moduli space, and sum over genus.  Even the first of these is
nontrivial, because the path integral is not gaussian, due to the
tachyon background.  The path integral on the sphere has been carried
out using difficult analytic arguments, and for some amplitudes by
taking advantage of the large symmetry of the problem, to be
developed later.

Undoubtedly the main physical lesson learned thus far is the
discovery of the large non-perturbative effects, which as we have
seen turned out to be a general feature of string theory~\cite{Slo}.
In the matrix model, there is an obvious nonperturbative process, the
tunneling of a single fermion through the potential barrier.  Now
we have to consider an issue that we have ignored so far, namely
what is going on on the other side of the barrier. There are many
possibilities---that the other side is empty (so the state we are
considering is unstable); that it is filled to the same Fermi
level as the first well; that the potential is modified,
say by an infinite barrier at some $x \geq 0$.  None of these
would change the perturbation theory, so they represent a
nonperturbative ambiguity.  Later we will examine the problem of
the nonperturbative definition of the theory, but for now we will
just use this ambiguity to estimate the inaccuracy of perturbation
theory.  The fermion tunneling amplitude is $e^{-B}$ with $B$ given
in the WKB approximation by
\begin{equation}
B = \int_{-\sqrt{2\overline\mu}}^{\sqrt{2\overline\mu}} dx\,
\sqrt{2\overline\mu - x^2} = \pi\overline\mu.
\end{equation}
So the tunneling amplitude is indeed $e^{-O(\overline\mu)} =
e^{-O(g_{\rm s}^{-1})}$.  From the discussion in section~4, one
would expect this to be associated with a $(2h)!$ growth of the
perturbation theory.  This growth is indeed found when the
amplitudes are expanded in powers of $\overline\mu^{-2}$~\cite{c1refs}.

In this case one can see what the
nonperturbative effect associated with the large order
behavior is, namely single-fermion tunneling.  The
matrix model is an example of this picture in which the
strings are collective excitations of something simpler, the free
fermions.  The bosonic collective Hamiltonian can presumably be
used to generate all orders of perturbation theory, but it is
not at all clear that one can make sense of it non-perturbatively.
The description of the single-fermion tunneling in terms of the
collective Hamiltonian is extremely clumsy and ad hoc.  More
generally, the collective description seems to break down near the
end of the eigenvalue distribution, or for processes involving
interference between two edges of the Fermi surface.  For
example, the raising operators of the $\overline{\cal S}$ field
create a fermion-hole pair.  An operator with momentum greater
than $p_+ - p_-$ tries to create a hole below the lower edge $p_-$
of the Fermi sea, which is impossible.  So the algebra of the
bosonic raising and lowering operators---the enumeration of
states---breaks down.  A very similar thing happens in the case of
$D=2$ Yang-Mills theory.  The
string representation can be obtained as a bosonization of a
non-relativistic free fermion representation just as here, and
again the enumeration of string states breaks down
non-perturbatively, when the number of boxes in a Young tableau
column exceeds $N$.

One would of course like to use this to get some insight into the
nature of the nonperturbative effects and of the hypothetical
fundamental degrees of freedom in higher dimensions.  One can try
looking at limits in the higher
dimensional theory as I have discussed, but another avenue is to
try to make as explicit as possible the connection between the
free fermion picture and the string picture here in the $D=2$
theory.  At present the connection is quite roundabout, going
through the matrix model and double scaling limit.\footnote
{Something I have tried without success is to go directly from the
string path integral to a spacetime Hamiltonian by a good choice
of gauge.  Unfortunately, the obvious gauge choice, $\sigma^0 =
X^0$ so as to identify world-sheet and embedding times, does not
seem to work even for the simpler problem of a relativistic
particle.  The gauge choice assumes that $X^0$ is a monotonic
function of $\sigma_0$, so that if the bath backtracks one must
build it by combining monotonic paths.  Unfortunately, the
typical relativistic particle path backtracks an infinite number of
times in the continuum limit.  Light cone quantization avoids this
backtracking problem in flat spacetime, but has been rather hard to
implement with the linear dilaton
background~\cite{Smithlc}.  Kawai and collaborators have
made some progress along slightly different
lines~\cite{Ktemp}.}  In fact, the connection is rather
subtle.  The string theory tachyon is related in an obvious way to the
collective motion of the fermions, but in fact there is more in the
string theory.  The graviton-dilaton sector of the $D=2$ string
theory is the same as the dilaton gravity theories discussed by
Strominger and Verlinde.  So although there is no propagating
graviton or dilaton, there are still physical effects from gravity,
including a black hole solution.  These are not evident in the matrix
model.  It is important to clarify this, both to develop the matrix
model
$\leftrightarrow$ string connection, and because
the gravitational effects are of interest in their own right.

We have been discussing
the $D=1$ (also called $c=1$) matrix model. The $D=0$ model, based
on a single matrix ${\bf M}$ without time, leads in a similar way
to the $D=1$ string (pure world-sheet gravity = one Liouville
dimension).  Multi-matrix models
${\bf M}_i$ lead to $0 < c < 1$, the minimal models plus a
Liouville dimension.  Also, fine-tuned multicritical points
in the $D=0$ matrix model lead to
non-unitary minimal models plus the Liouville dimension.  The
extension to $c > 1$ is problematic.  The matrix model can no
longer be reduced to eigenvalues, and no solution is known.  Also,
as we have noted the tachyon really is tachyonic beyond $D > 2$, so
the vacuum is unstable and presumably does not correspond to an
attractive fixed point of the discrete theory.

Even at $D \leq 2$
it would be useful to have a greater variety of models, so
as to get some perspective.
The matrix model can be generalized to open plus closed
strings~\cite{Yopen} (a different approach is given in
ref.~\cite{KKomm}).  This reduces to a theory of {\it interacting}
fermions.  It is similar to the
$U(N)$ Calogero-Sutherland model recently considered by Haldane and
others, but has not been solved.  Even short of a full solution, it
would be useful to understand in the classical limit the connection
between the collective excitations and the open and closed strings.
Another extension is spacetime supersymmetry.  Supersymmetry breaking
is after all one of the most important non-perturbative phenomenon
in string theory; spacetime supersymmetry would also help with the
tachyon instability of the $c>1$ string.  There is nothing useful
here yet.  A straightforward supersymmetrization of the $D=1$
matrix model exists and has a
double-scaling limit~\cite{MPar}. But Shyamoli Chaudhuri and I have
recently shown that the naive world-sheet theory is not conformally
invariant, and the theory flows to some exotic fixed point that
does not seem to be connected with the string physics that is of
interest~\cite{CPmp}.

To conclude these lectures, I will first develop some details
of the tree-level scattering amplitudes.  I will then discuss the
gravitational physics of the $D=2$ string, including the black hole
solution, as well as the
generalization to higher string levels.  Finally, I look for the
corresponding physics in the matrix model.  I show how
gravitational physics emerges in the weak field limit, then look at
physics in strong fields---an (as yet unsuccessful) search for the
black hole solution, some exotic strong-field physics that arises
in the matrix model, and the nonperturbative ambiguity.

\subs{Tree-Level Scattering}

Consider a small incoming pulse on top of the static
solution $p_{\pm,\rm static}$, eq.~(\ref{statlev}).  As we have
discussed, dispersion changes the shape of the pulse: higher points on
the pulse travel deeper into the potential and emerge later.  Let us
work this out at the classical level.  The equation of
motion~(\ref{ihoem}) implies that
\begin{equation}
v = (-x-p) e^{-t}, \qquad w = (-x+p) e^t
\end{equation}
are
constant for a particle moving in the inverted harmonic potential.
Then so also are the integrals
\begin{equation}
v_{mn} = e^{(n-m)t} \int_{F - F_0} \frac{dp\, dx}{2\pi}\, (-x-p)^m
(-x+p)^n . \label{vmn}
\end{equation}
The integral runs over the Fermi sea $F$, with the integral over the
static Fermi sea $F_0$ subtracted, so this will converge for a pulse
of finite width.
Evaluating this in the limits $t \to -\infty$
and $t \to \infty$ relates the incoming and outgoing pulses.  As $t
\to -\infty$ points approach the line $p = -x$, so
\begin{equation}
v \to e^{q-t} \epsilon_{+}(q - t), \qquad w \to 2 e^{- q + t}\ .
\end{equation}
I have used the expansion~(\ref{ppm}), with the free field equation
of motion $\epsilon_+(q,t) \sim \epsilon_{+}(q - t)$ for the incoming
wave.  Similarly,
as $t \to \infty$,
\begin{equation}
v \to 2e^{-q-t}, \qquad w \to e^{q+t} \epsilon_{-}(q + t)
\end{equation}
with $\epsilon_-$ the outgoing wave.  Thus,
\begin{eqnarray}
v_{mn} &=& \frac{2^n}{2\pi(m+1)} \int_{-\infty}^\infty dt\,
e^{(n-m)(t-q)}
\Bigl\{ (\epsilon_+(t-q))^{m+1} - \overline\mu^{m+1} \Bigr\}
\nonumber\\
&=& \frac{2^m}{2\pi(n+1)} \int_{-\infty}^\infty dt\,
e^{(n-m)(t+q)}
\Bigl\{ (\epsilon_-(t+q))^{n+1} - \overline\mu^{n+1} \Bigr\}.
\label{vasym}
\end{eqnarray}

Applying this for $m = 0$, $n = i\omega$ gives the
Fourier transform of the incoming wave as a nonlinear function of the
outgoing wave~\cite{MPless},~\cite{Pcon}.  Expanding around the static
background, $\epsilon_{\pm}(q_\mp) = \overline\mu +
\delta\epsilon_{\pm}(q_\mp)$, gives
\begin{equation}
\int_{-\infty}^\infty dq_-\, e^{i \omega q_-}\delta\epsilon_{+}(q_-)
= (2\overline\mu)^{i \omega} \int_{-\infty}^\infty dq_+\, e^{-i
\omega q_+}
\sum_{k=1}^\infty \frac{\overline\mu^{1-k}}{k!}
\frac{\Gamma(1+i\omega)}{\Gamma(2 - k + i\omega)}
\delta\epsilon_{+}(q_+)^k.
\end{equation}
In terms of the Fourier modes,
\begin{equation}
\epsilon_{\pm}(q_{\mp}) = \overline\mu + \frac{1}{\sqrt{2\pi}}
\int_{-\infty}^{\infty} d\omega\,
\overline\alpha_{\pm}(\omega) e^{\pm i
\omega q_\mp}
\end{equation}
this becomes
\begin{eqnarray}
\overline\alpha_{+}(\omega) &=& \sum_{k = 1}^\infty
\frac{(\sqrt{2\pi}\overline\mu)^{1-k}}{k!}
\frac{\Gamma(1+i\omega)}{\Gamma(2 - k + i\omega)} \label{inout}\\
&&\qquad \int d\omega_1 \ldots d\omega_k\,
\overline\alpha_{-}(\omega_1)
\ldots \overline\alpha_{-}(\omega_k)
\delta(\omega_1 + \ldots + \omega_k - \omega)\ .
\nonumber
\end{eqnarray}

It is a useful fact that classical scattering is the same as tree level
quantum scattering.  The quantum operators satisfy the same equation of
motion as the classical field, so the solution~(\ref{inout}) also
holds, except that we would have to be careful about operator ordering.
But commutators are of order $\hbar$ and so do not matter to leading
order.
The modes have been normalized to satisfy
\begin{equation}
[ \overline\alpha_{\pm}(\omega),
\overline\alpha_{\pm}(\omega')]
= -2\pi\omega \delta(\omega - \omega'). \label{modecc}
\end{equation}
Defining the in-states
\begin{equation}
| \omega_1, \ldots, \omega_n; \overline{\rm in} \rangle
= \overline\alpha_{+}(\omega_1) \ldots
\overline\alpha_{+}(\omega_n) |0\rangle, \qquad \omega_i > 0,
\end{equation}
and similarly for the out-states, the classical S-matrix~(\ref{inout})
plus the canonical commutator~(\ref{modecc}) give the tree level
S-matrix.  For example, the $1 \to n$ amplitude is~\cite{MPless}
\begin{equation}
\langle \omega_1, \ldots, \omega_n;
\overline{\rm out}| \omega; \overline{\rm in} \rangle
= \left( \frac{\sqrt{2\pi}}{\overline\mu} \right)^{n-1}
\frac{\Gamma(1+i\omega)}{\Gamma(2 - n + i\omega)}
2\pi \delta(\omega_1 + \ldots + \omega_n - \omega).  \label{sbar}
\end{equation}

This same calculation can also be done by string methods, but only with
great effort and ingenuity.  The main complication is the exponential
term in the string action, from the tachyon background.  Far from the
Liouville wall this is small, so for so-called `bulk' processes which
happen far away from the wall we can use the free action or at least
expand in powers of $\overline\mu$.  (One can think about using
wavepackets to separate out the bulk processes, as I will describe
later).  We then have a free-field string calculation of the type
that gave the Virasoro-Shapiro amplitude.  In $D > 2$ this reduces
to $\Gamma$-functions only for four particles, but in $D = 2$ the
kinematic restrictions make it possible to evaluate for any number.
Also, an analytic continuation makes it possible to deduce the full
amplitude from the bulk amplitude.  The result is almost the same as
the matrix model result~(\ref{sbar}).  That is, comparison of the
string and matrix model S-matrices shows that the latter is related
to the former by~\cite{dFK}-\cite{GKs1c1}.
\begin{eqnarray}
\alpha_+(\omega) &=&
\frac{\Gamma(i \omega )}{\Gamma(-i \omega )}
\overline\alpha_+(\omega) \nonumber\\
\alpha_-(\omega) &=&
\frac{\Gamma(-i \omega )}{\Gamma(i \omega )}
\overline\alpha_-(\omega).  \label{tachren}
\end{eqnarray}
The ratio of gamma functions, known as the `leg pole' factor, is a
pure phase for real frequencies, so this is a unitary transformation
on the states.  Such a redefinition is not surprising.  The matrix
model gives a discrete approximation to the local vertex operator
$e^{i q \cdot X + X^1 \sqrt 2}$.  One expects to find a
renormalization for any such cutoff construction.  Following this
line of thought one can, purely within the matrix model, deduce the
poles and zeroes of the leg pole factor.\footnote
{To be precise, the correspondence between the string and matrix
model S-matrices has only been verified explicitly at tree level.
However, there is strong reason to believe that the leg pole factor
is the same to all orders of perturbation theory. The
renormalization comes from small distances on the world sheet, and
the string coupling is a relevant interaction.  That is, short
distance is the free asymptotic region of the critical string
picture.}

 I will not pursue this,
but simply take the leg pole factor from the comparison of
amplitudes.  Although a pure phase, it will play an essential role
soon.

\subs{Spacetime Gravity in the $D=2$ String}

Now let us take a closer look at the graviton, dilaton,
and antisymmetric tensor states in the $D=2$ string,
\begin{equation}
|e,k\rangle = e_{\mu\nu} \alpha^\mu_{-1} \tilde\alpha^\nu_{-1}
|0,k\rangle. \label{lev1l}
\end{equation}
Recall from section~1.6 the Virasoro generators
\begin{eqnarray}
L_0 &=& \frac{1}{2} k^2
+ \frac{a^2}{2} + \sum_{n=1}^\infty \alpha_{-n}
\cdot \alpha_{n}
\nonumber\\
L_m &=& \frac{1}{2} \sum_{n=-\infty}^\infty \alpha_n
\cdot \alpha_{m-n} + i (m+1) a \cdot \alpha_m ,
\qquad m \neq 0,
\end{eqnarray}
where $a^\mu = (0, \sqrt{2})$.  Also recall $\alpha_0 = k^\mu - i
a^\mu$. Acting on the state~(\ref{lev1l}), the physical state
conditions become (we use the convenient OCQ form)
\begin{eqnarray}
L_0, \tilde L_0:&& k^2 = - a^2 \nonumber\\
L_1, \tilde L_1:&&  (k + i a)^\mu e_{\mu\nu} =
(k + i a)^\nu e_{\mu\nu} = 0.
\end{eqnarray}
Also, acting with $L_{-1}$ on the state
$f_\mu \alpha^\mu_{-1} |0,k\rangle$ and with $\tilde L_{-1}$ on
$\tilde f_\nu \tilde\alpha^\nu_{-1} |0,k\rangle$ we find the following
spurious states
\begin{equation}
\mbox{spurious:}\qquad e_{\mu\nu} = (k - i a)_\mu \tilde f_\nu
+ f_\mu (k - i a)_\nu\
\end{equation}
for any $f$, $\tilde f$.

The $L_1$ and $\tilde L_{1}$ conditions require $e_{\mu\nu}$ to be
orthogonal to $k + ia$ on both indices, or
$e_{\mu\nu} = n_\mu n_\nu$ where $n_\mu$ is in the one-dimensional
space orthogonal to $k + i a$.
The $L_0$ conditions gives $(k + ia) \cdot (k - ia) = 0$,
so $n\ \propto\ k - ia$.  But this means that $e_{\mu\nu}$ is
spurious, so there are no `observable' states.

This is correct at generic momenta but breaks down at special
points~\cite{Plp}. If $k = -ia$, the $L_{1}$ and $\tilde L_{1}$
conditions are empty.  States with polarizations proportional to
$a^\mu$ are null in this case, so we can remove the spacelike
polarizations and be left with the observable state
\begin{equation}
\alpha^0_{-1} \tilde\alpha^0_{-1} |0, - ia \rangle\ .
\end{equation}
If $k = ia$ the spurious state vanishes.  The physical state
condition requires $e_{\mu\nu}$ to be orthogonal to $a^\mu$, so the
observable state is
\begin{equation}
\alpha^0_{-1} \tilde\alpha^0_{-1} |0, + ia \rangle\ .
\end{equation}
The corresponding vertex operators are respectively
\begin{equation}
:\! \partial X^0 \bar\partial X^0 e^{2\sqrt 2 X^1} \!:\ , \qquad
:\! \partial X^0 \bar\partial X^0  \!:\ .
\end{equation}

What does this mean physically?  The absence of observable states at
general momenta means that there are no particle-like states---one
cannot make a wavepacket.  The vertex operator
$:\! \partial X^0 \bar\partial X^0  \!:$ has an obvious
interpretation---it corresponds to an infinitesimal rescaling of the
flat metric $G_{00}$, and can be absorbed into a rescaling of
$X^0$.\footnote
{We are considering $X^0$ to be noncompact, so this rescaling has no
effect.  If $X^0$ were periodic, it would correspond to a change
in the periodicity.}
The other operator is more problematic because it corresponds to
$\delta G_{00}\ \propto  e^{2\sqrt 2 X^1}$,
diverging at infinity.
In fact, you have seen this before, in the talks by Strominger and
Verlinde.  The low energy graviton-dilaton action for the $D=2$
string,
\begin{equation}
{\bf {S}} = \frac{1}{2} \int d^D X\, \sqrt{-G} e^{-2 \Phi}
\biggl\{ - 8 + {\bf {R}}
+4 \nabla_\mu \Phi \nabla^\mu \Phi  \biggr\}  \label{gdact}
\end{equation}
is the same as they considered.  As they discussed, this has
a black hole solution~\cite{Wbh},~\cite{MSWbh}, which in one set of
coordinates is
\begin{eqnarray}
G_{11} &=& 1,\qquad G_{00} = -\tanh^2
\biggl(X^1\sqrt{2} + \frac{1}{2} \ln M \biggr)
\nonumber\\
\Phi &=& -\frac{1}{2}\ln (M/2) + \ln \cosh \biggl( X^1\sqrt{2} +
\frac{1}{2} \ln M \biggr). \label{bhback}
\end{eqnarray}
As $M \to 0$ this approaches the linear dilation
vacuum~(\ref{lindbg}).  To first order in $M$ it is just the
perturbation
\begin{equation}
\delta G_{00}\ \propto M e^{2\sqrt 2 X^1}. \label{grav}
\end{equation}

We should note that, unlike the case of four dimensions where the
size of the black hole is proportional to $M$, here the size is
independent of $M$.  The mass appears in the background~(\ref{bhback})
only as an additive term in $\Phi$ and as an additive shift
of $X^1$.  The size is set by the only other scale in the problem,
the string scale.  The spacetime action~(\ref{gdact}) is only valid
at scales long compared to the string scale, so the derivation of
the solution~(\ref{bhback}) is not strictly valid.\footnote
{This comment does not apply to dilaton gravity as a {\it field}
theory, where the scales are set by hand.}   However, an exact CFT
corresponding to this solution, a variation of the WZW theory, is
known~\cite{Wbh}.  Also, because the scale of the geometry is of the
same size as the string itself, we cannot be sure that the string
sees it as a black hole.  For example, we cannot make a clean
causality argument in the information paradox, as we can do in four
dimensions.
But the exact solution does appear to resemble a black
hole---for example in the Bogoliubov transformation
between the asymptotic tachyon modes and those on the
horizon~\cite{DVVbh}.

Now let me state the result for the observable
Hilbert space at higher levels (I am following here
refs.~\cite{Wgr}; see these for earlier references).  Extra physical
states  (the {\it discrete states}) are found at momenta $(k^0,k^1) =
(i n \sqrt{2}, \pm i s \sqrt{2})$ for $2s$ and
$s-n$
integers such that $|n| \leq s-1$, as shown in fig.~27.
\begin{figure}
\begin{center}
\leavevmode
\epsfbox{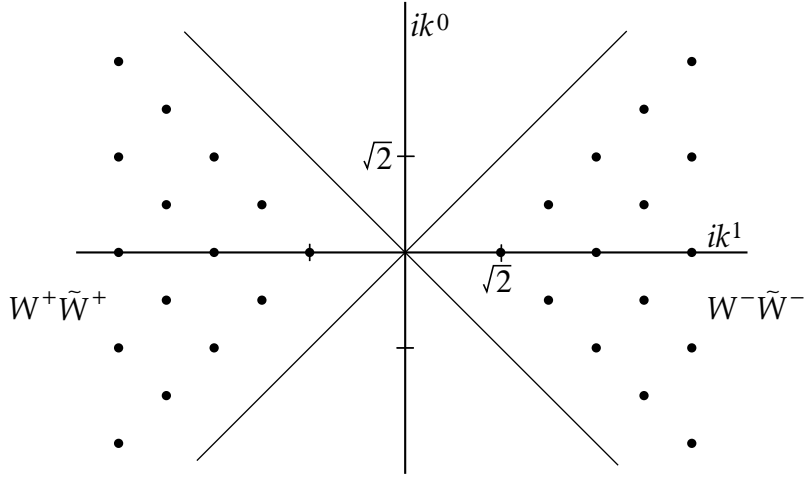}
\end{center}
\caption[]{Extra physical states in the $(i k^0,i k^1)$ plane.}
\end{figure}
The
corresponding operators are denoted $W_{s,n}^{\mp}(z) \tilde
W_{s,n}^{\pm}(\bar z)$.  The lines $i k^0 = \pm i k^1$ correspond
to (imaginary) tachyon momenta.  The
$L_0$, $\tilde L_0$ conditions imply that the state $(s, n)$ is at
level $s^2 - n^2$.

Note also the conserved currents $\partial X^0$ and $\bar\partial
X^0$, closely related to the operator $W_{1,0}^{+}\tilde
W_{1,0}^{+} = \partial X^0\bar\partial X^0$.  Actually, although
these currents are separately conserved, their difference couples to
the winding number in the $X^0$ direction.  For the non-compact
theory that we are concerned with this vanishes, so only one
symmetry is nontrivial.
At higher levels there are
additional symmetries $A_{s,n}$, in one-to-one
correspondence with the
$W_{s,n}^{+}\tilde W_{s,n}^{+}$ (recall the
result~(\ref{sginv}) that a spacetime symmetry corresponds to a BRST
invariant operator of appropriate ghost number).

So the $D=2$ string theory does have a few graviton-dilaton states,
and some interesting physics associated with them.  In addition it
has a large number of similar states at higher levels.
The state $W_{1,0}^{-}\tilde W_{1,0}^{-}$ corresponds to a black
hole background.  It
seems likely that the other $W_{s,n}^{-}\tilde W_{s,n}^{-}$ give
rise to some infinite-parameter generalization of the black hole,
but little is known about this; a formal argument for the
existence of these solutions is given in ref.~\cite{Sbhgen}.

The tachyon state was obvious in the matrix model.  So are the
symmetries $A_{s,n}$---they are just the $v_{mn}$ introduced above
for positive integer $m,n$~\cite{AJw}-\cite{MPW},\footnote
{Incidentally, the $v_{mn}$ are well-defined for general $m$, $n$
only if the sea is entirely in the quadrant $x < |p|$, but for
integer $m$, $n$ they are always well-defined.}
\begin{equation}
A_{s,n} \ \propto\ v_{s+n, s-n}.
\end{equation}
This identification is consistent with the $(k^0,k^1)$ values.
Also the linear term in the variation of the
tachyon field is known to be the same, as is the algebra
\begin{equation}
\{ v_{mn}, v_{m'n'} \} = 2 (m'n - n'm) v_{m+m'-1,n+n'-1}.
\end{equation}
This is the same as the $w_\infty$ commutator (\ref{winf}), with
$4 V^i_m \equiv v_{i+m+1,i-m+1}$.  Only the range of the indices
differs.

However, the remainder of the physical states are not so evident.
The $W_{s,n}^{+}\tilde W_{s,n}^{+}$ do mix with the tachyons
at the appropriate momenta and can be seen in matrix
model correlators~(reviewed in chapter~6 of ref.~\cite{Krev}).  The
$W_{s,n}^{-}\tilde W_{s,n}^{-}$ are, like the black hole, singular at
large $X^1$ and do not appear in the matrix model in any simple
way~\cite{Sliou}.  In the next section we will see that
much of the physics of the discrete states is not contained within
the matrix model by itself but also involves the mapping between the
matrix model and the $D=2$ string theory.

\subs{Spacetime Gravity in the Matrix Model}

The results of the previous section seem quite exciting.  The $D=2$
string theory has black hole solutions, and we have in the matrix
model an exact quantum solution to this theory.  But things are not
yet so rosy.  First, we do not know how to describe the black hole
in the matrix model language---what deformation of the matrix model
corresponds to nonzero Schwarzchild mass $M$?  Second, as we will
discuss in section~5.8, we do {\it not} yet have the exact solution
to the theory, only its perturbation expansion.

To begin to study this issue, we will first ask a much more basic
question~\cite{NP}.  How, in the matrix model, do we see the
gravitational interaction even between lumps of matter too diffuse to
form a black hole?  I will phrase this as an S-matrix question.
Imagine sending two successive tachyon pulses in from the asymptotic
region.  The first pulse carries energy and momentum, so the second
pulse must feel its gravitational field.  There will then be some
amplitude for the process shown in fig.~28a, in which part of the
second pulse backscatters from the gravitational field of the first
pulse. \begin{figure}
\begin{center}
\leavevmode
\epsfbox{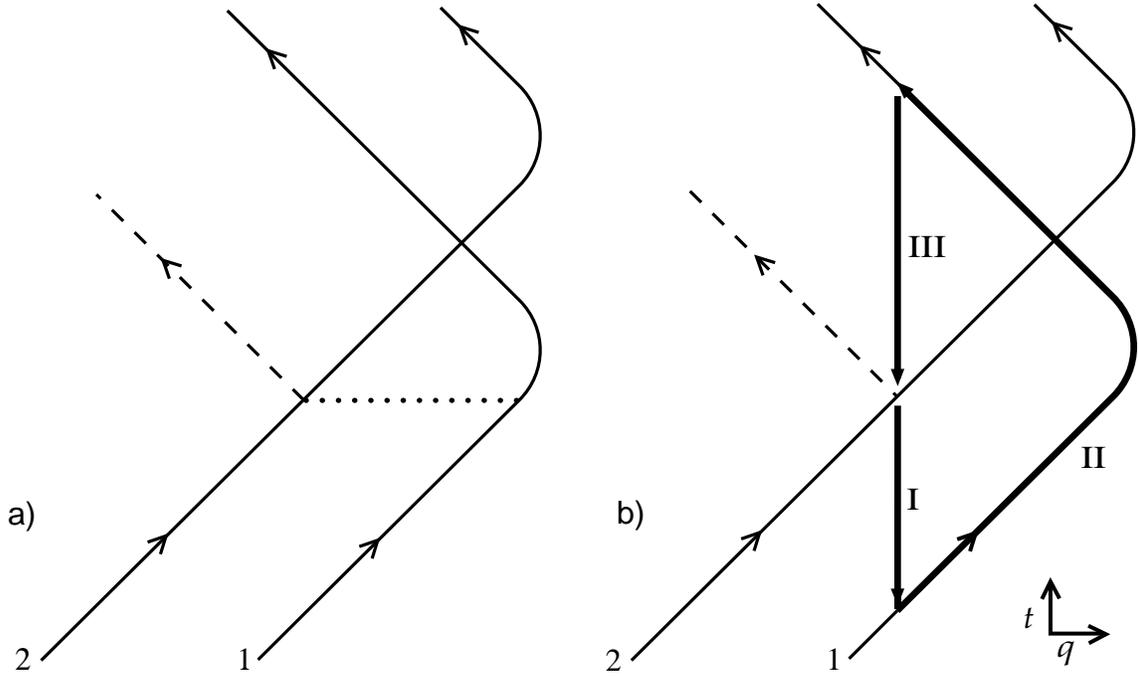}
\end{center}
\caption[]{a) Successive pulses moving in the $t-q$ plane.
Gravitational field (dotted) of pulse~1 will cause part of pulse~2
to backscatter, producing an outgoing wave (dashed) which precedes
the main reflection from the wall.
b) How the matrix model represents this process.  The
initial wavefunction renormalization (I) produces a tail on pulse~2
which overlaps pulse~1; the combined pulse reflects from the wall
(II); and the final renormalization (III) produces the outgoing wave.
}
\end{figure}
But this does not happen in the matrix model!  The pulses consist
of free fermions, which move independently in the inverted
oscillator potential.

This would seem to be a contradiction, because I have set this up
as an S-matrix question, and earlier I told you that the string and
matrix model S-matrices agree.  However, I also told you that there
is a renormalization of the states, the leg pole
factor~(\ref{tachren}), and this is
in fact essential to seeing all of the physics.  To see how the
renormalization can have this effect, let us write it in coordinate
space, in terms of the asymptotic incoming and outgoing tachyon
fields:
\begin{eqnarray}
{\rm (I)}:\quad &&
\overline {\cal S}_+(t - q) =
\int_{-\infty}^\infty d\tau\, K(\tau) {\cal S}_+(t - q - \tau)
\nonumber\\
{\rm (III)}:\quad &&
{\cal S}_-(t + q) = \int_{-\infty}^\infty d\tau\, K(\tau) \overline
{\cal S}_-(t + q - \tau).  \label{conv}
\end{eqnarray}
Here, a bar is again used to distinguish the matrix model quantity
from the corresponding string quantity.  The same kernel appears in
both transformations,
\begin{eqnarray}
K(\tau) &=& \int_{-\infty}^\infty \frac{d\omega}{2\pi}
e^{i \omega \tau}
\biggl( \frac{\pi}{2} \biggr)^{-i\omega/4}
\frac{\Gamma(-i \omega )}{\Gamma(i \omega )}
\nonumber\\
&=& -\frac{z}{2} J_1 (z), \qquad z = 2 (2/\pi)^{1/8} e^{\tau/2}.
\end{eqnarray}
This has asymptotic behaviors
\begin{eqnarray}
K(\tau) &\sim&
-\biggl( \frac{\pi}{2} \biggr)^{-1/4} e^{\tau}, \qquad \tau \to
-\infty \nonumber\\
&\sim& \biggl( \frac{\pi}{2} \biggr)^{-1/16}
\frac{e^{\tau/4}}{\sqrt{\pi}}
\cos( z + \pi/4 ), \qquad \tau \to \infty . \label{kasym}
\end{eqnarray}
To describe the scattering of an incoming ($+$) string tachyon
pulse, one must (I) transform to the matrix model tachyon field
via~(\ref{conv}), (II) evolve the
pulse as described in section~5.5, and (III) transform back.
The transformation~(\ref{conv}) is nonlocal, and it is the
early-time behavior that is important.  Although this falls
exponentially, so does the gravitational effect (\ref{grav}) that we
seek.  As long as we use narrow enough pulses, gaussians, the
exponential tail can be discerned~\cite{Nzm},~\cite{NP}.

The string amplitude I-II-III does
indeed display the gravitational scattering.  This works as shown in
fig.~28b.  The transformation I on the incoming pulse~2 produces an
early tail that overlaps pulse~1.\footnote
{The amplitude of $K$ grows at late times, but it oscillates so
rapidly that it falls effectively to zero, at least in all cases I
have yet encountered.}
These propagate together through
the turning point and interact locally, and the transformation III
produces the outgoing gravitationally scattered wave.

It will be useful for the later discussion to provide some
details, but it is simpler for this to look at the slightly
simpler process of $2 \to 1$ bulk tachyon scattering.
That is, as shown in fig.~29a, a pair of tachyons in one pulse can
interact and scatter into one outgoing tachyon before reaching the
wall.
\begin{figure}
\begin{center}
\leavevmode
\epsfbox{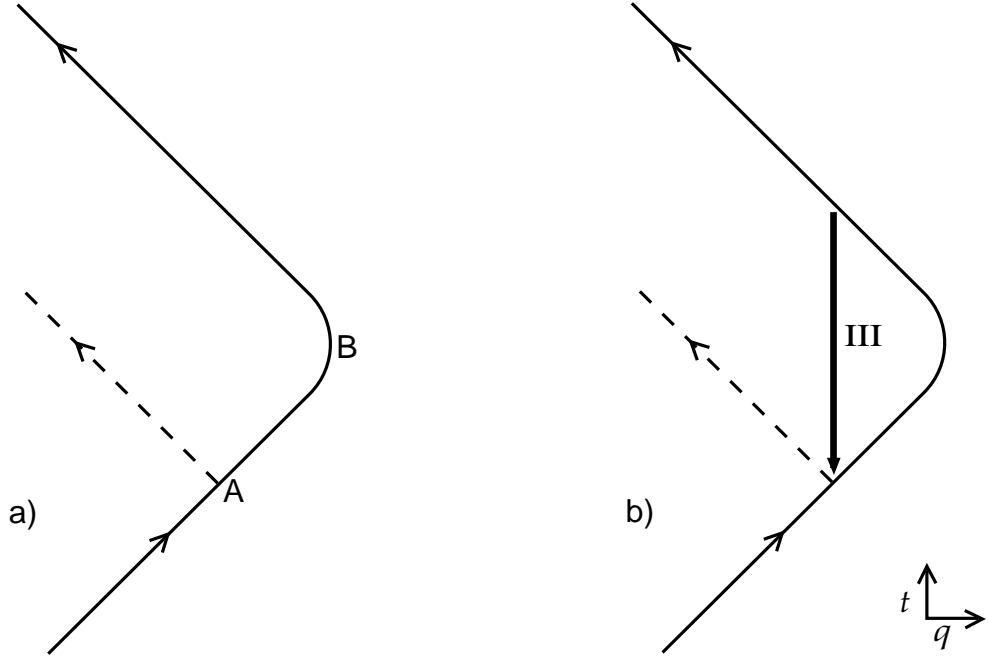}
\end{center}
\caption[]{a) Two tachyons in an incoming pulse can backscatter into
one (dashed line) at a given point~A.  In the matrix model all
tachyons pass through the turning region B. b) The tail from
transformation~III produces the outgoing wave.}
\end{figure}
This occurs in the $D=2$ string theory, from direct calculation of
the three string interaction, but it again does not occur in the
matrix model because the fermions are free---they travel on
hyperbolic phase space orbits and so reach the turning point of the
potential.  The leg pole transformation III is again responsible
for the difference, as shown in fig.~29b; the tail on
transformation~I is not important in this case.  Explicitly,
\begin{eqnarray}
\lim_{t + q \to -\infty} {\cal S}_{-}(t + q)
&=& -2^{1/4} \pi^{-1/4} e^t \int_{t}^\infty
dt'\, e^{- t'} \overline{\cal S}_{-}(t' + q) \nonumber\\
&=& -2^{1/4} \pi^{-1/4} e^t \int_{-\infty}^\infty
dt'\, e^{- t'} \overline{\cal S}_{-}(t' + q) \nonumber\\
&=& 2^{-3/4} \pi^{1/4} e^{t + q} v_{10}.
\end{eqnarray}
In the second line we have used the narrowness of the wavepacket to
extend the range of integration, and in the third we have noted that
the result is simply proportional to the conserved charge $v_{10}$.
Now expressing this in terms of the incoming field gives
\begin{eqnarray}
\lim_{t + q \to -\infty} {\cal S}_{-}(t + q)
&=& 2^{-3/4} \pi^{1/4} e^{t + q}
\int_{\infty}^\infty dt'\, e^{-t'+q}
\Bigl\{ (\partial_t \overline{\cal
S}_+(t-q))^2 + \overline\mu \partial_t\overline{\cal
S}_+(t-q) / \sqrt{\pi} \Bigr\} \nonumber\\
&=& 2^{-1/2} e^{t + q} \int_{\infty}^\infty dt'\,  e^{-t'+q}
({\cal S}_+(t'-q))^2 \ +\ O(\overline\mu)\ .
\label{2to1}
\end{eqnarray}
In the second line we have carried out the
renormalization~(\ref{conv}), leading to a simple result
after integration by parts.  The $O(\overline\mu)$ term
is from
$1 \to 1$ scattering on the tachyon background.

The result~(\ref{2to1}) is the correct bulk scattering, as found
from a string theory or effective field theory calculation~\cite{NP}.
Note in particular the exponential factor $e^{t - t' + 2q}$.  At
$t=t'$, which is the point A in figure~29 where the incoming and
outgoing rays meet, this is the trilinear string coupling $e^{2q}$.

In a sense, nothing here is new.  In momentum space, we have
simply found the graviton pole.  It is well-known that this pole
is not present in the matrix model amplitude but comes from the
pole factors for the external legs~\cite{dFK}-\cite{GKs1c1},
\cite{STlp}-\cite{Llp}.
However, this coordinate
space analysis makes a number of things much clearer, and dispels
some confusions in the literature.  One confusion is the frequent
assertion that the leg pole factor is not relevant for physics
in Minkowski space (real $\omega$),
because it is a pure phase.  We have seen that it is indeed
essential for important Minkowski space physics---this is possible
because the phase is a function of the momenta of the particles and
so can be seen in interference.  A second confusion is an
identification of the string tachyon field with the macroscopic
loop amplitude.  This is not the same as the
identification~(\ref{conv}), and does not have the correct
spacetime physics.

The coordinate space calculation also provides a physical
interpretation for the discrete states in the cohomology.  That is,
they correspond to long-range forces not associated with propagating
degrees of freedom.\footnote{Although they fall exponentially, I
use the term `long-range' because they extend far enough to
allow them to be distinguished from the essential gaussian
nonlocality of the string.}
The imaginary values of the momenta
for the discrete states correspond to exponential falloff of the
forces in spacetime.

Let me emphasize that it is
only the full string S-matrix I-II-III that is physical.
That this factorizes into three separate parts, the first and third
being linear in terms of the bosons, and the second being linear
in terms of the fermions, is an accident of the $D=2$ kinematics
and one should not give particular significance to the individual
factors.  Note also that the matrix model, in spite of its
qualitative resemblance to the string spacetime,
misrepresents the causal structure of the physics.  The $2 \to 1$
bulk process actually occurs at the point A in fig.~29a, before
the pulse reaches the Liouville wall, but in the matrix model
calculation the information in the incoming pulse always travels
through the turning region~B.

The scattering calculations here become more tedious at
higher orders in the gravitational field and at higher
levels.  One would like to carry out the redefinition~(\ref{conv})
directly in the Hamiltonian, converting the known matrix model
Hamiltonian into a string Hamiltonian, with propagating tachyons
plus long-ranged fields.  The latter would presumably correspond
to some gauge-fixing of string field theory.  But
one must add to the linear
transformation~(\ref{conv}) appropriate nonlinear terms to bring
the Hamiltonian to this form, and this is does not seem to be
simple.

\subs{Strong Nonlinearities}

Now that we have detected weak field gravity, we would like to
find the black hole solution.  The first thing that comes to
mind is to keep the field while omitting the source pulse~1,
producing a source-free solution.  But it is not evident how to
do this, since the construction in fig.~28a depends in an
essential way on the direct interaction between the two pulses.
Various groups have proposed
matrix model representations of the black hole
background~\cite{mmbh}.  They take points of view different from my
own, so it is hard to make a direct comparison.  Several do emphasize
the important point that the string theory is defined not by the
matrix model alone  (step II) but also by the mapping between the
matrix model and string theory (steps I and III).
But it is not clear that any of the proposal pass the test that
they give the correct scattering in the weak field, weak coupling
region studied in the previous section.

Since we have a complete description of the scattering of
incoming pulses, we might try to {\it make} a black hole, by
sending in a large enough pulse of matter.  The
perturbation~(\ref{grav}) of the metric is $M/x^4$ in matrix model
variables, so in order to form a black hole we need at least for
this to be of order~1. Consider now a pulse on the Fermi sea, of
width ${\mit \Delta} x$ and height ${\mit \Delta} p$.  The total
energy $M$ in the pulse is of order the number of fermions times
the height above the Fermi level, or ${\mit \Delta} x\,{\mit
\Delta} p \, {\mit \Delta} (p^2)$.  Now, ${\mit \Delta} x
\,\roughly{<}\, x$ and ${\mit \Delta} (p^2) \sim x {\mit \Delta} p$, so
we need
\begin{equation}
({\mit \Delta} p)^2\ \roughly{>}\ x^2
\end{equation}
to produce a gravitational effect of order~1.  That is, the
height of the pulse is comparable to the height of the whole sea.
The tachyon self-interaction is then also of order~1, since the
dispersion of the fermions in the pulse is large.  In
contrast to four dimensions, we cannot get into a situation
where the gravitational interaction dominates.\footnote
{The $D=2$ string also differs from dilaton gravity, where the
matter is introduced by hand and one may take it to be free and
conformally invariant.}
It has a long
enough range to be detected, but not long enough to dominate.
So we cannot be sure a priori that it is possible to make a
black hole.  We will have to be experimentalists, throwing in
pulses with various properties and seeing if the result has any
of the characteristic signatures of a black hole.

In fact, two interesting qualitative behaviors
are found with large pulses~\cite{Pclass}.  The first is depicted in
fig.~30a. \begin{figure}
\begin{center}
\leavevmode
\epsfbox{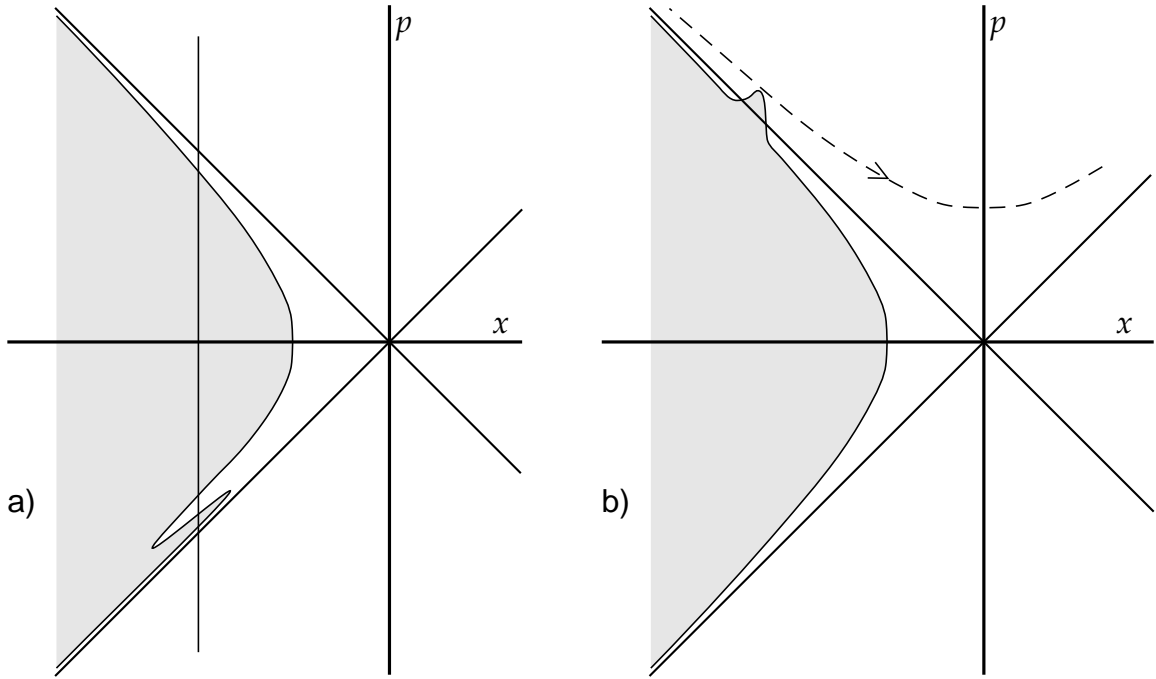}
\end{center}
\caption[]{a) Outgoing pulse in which dispersion has produced a
fold.  At the indicated $x$, the Fermi sea has two filled bands
separated by an empty band.  b) Incoming pulse extending above
the line $p = -x$.  The fermions above the line pass over the
potential barrier to $x = \infty$.}
\end{figure}
For a sufficiently tall or steep pulse dispersion will cause it
to broaden past the vertical, producing an empty band in the
Fermi sea and a double-valued Fermi surface.  What does this
correspond to in bosonic language?  There is a one-to-one
correspondence between single-valued Fermi surfaces and
classical states of the bosonic field.  By a classical state
I mean one for which $\langle A^2 \rangle = \langle A \rangle^2$
for any observable $A$, in the classical limit where the fermions
become a continuous fluid.  The double-valued surface does not,
in this classical limit, correspond to a classical bosonic
state. One can check this explicitly by calculating in such a sea
the one-point and two-point functions of the boson modes
$\overline\alpha(\omega)$, which are just the fermion currents.
Even without calculation, a little thought will show that by moving
fermions down into the empty band one can reduce the energy
$\frac{1}{2} \int \{ (\partial_t \overline{\cal S})^2 +
(\partial_q \overline{\cal S})^2 \Bigr\}$ while keeping the
expectation values of $\partial_t \overline{\cal S}$ and
$\partial_q
\overline{\cal S}$ fixed.

The double-valued surface is thus a bosonic state
with a large amount of energy in addition to that in the classical
field---that is, a large amount of radiation.  There is a
superficial resemblance here to formation of a black hole.  In
both cases, for incoming matter below a threshold (to form a
horizon or a fold respectively), the incoming energy comes out
again in the classical field.  Above the threshold a
non-negligible fraction of the incoming classical energy returns
as radiation, Hawking radiation in the black hole and fold
radiation in the matrix model.  But here the resemblance ends.
The fold radiation comes out too promptly, and it is too
hot---typical energies of the quanta scale as $\hbar^{1/2}$
rather than $\hbar^1$.

It is not obvious how we would see the fold
radiation if we had only the stringy, bosonic, description and not
the fermionic one.  In bosonic language the fold corresponds to
an intersection of characteristics (null lines), which occurs
because the velocity is a function of
the tachyon field.  In most situations where such a thing
occurs, there are higher-derivative terms in the equation of
motion which prevent double-valuedness and lead to formation of
a shock wave.  But that is not the case here---it is evident in
the free fermionic language that pulse turns into a fold.
So instead of a shock wave we get radiation, but I don't know
how to see this purely from the bosonic Hamiltonian.
One rather different possibility is that the fold is a coordinate
singularity, and that we need to unfold it by an appropriate
coordinate transformation in phase space, but I do not know how
to make sense of this.

The second interesting behavior is shown in fig.~30b.  This is a
pulse which extends above the line $p = -x$, so that some of the
fermions pass over the barrier to the asymptotic region on the
right.  This certainly sounds like a black hole---we throw
matter in and it doesn't all come back out!  But now we have to
face a problem that we have deferred, the interpretation the other
side of the potential barrier.  In fact there is an infinite number
of ways to define the matrix model, all of which give a unitary
S-matrix and all of which have the same perturbation
expansion~\cite{MPR} (and presumably the same as that of the string).
One class of
theories (type~I in the terminology of ref.~\cite{MPR}) eliminates
the second asymptotic region by modifying the potential.  For
example, a sharp infinite barrier, $V(x) = \infty$ for $x > A$,
$A \geq 0$,
leaves the perturbation theory with fixed numbers of incoming strings
unaffected for any $A \geq 0$.
So does any other modification such that $V(x)$ is
$-\frac{1}{2} x^2$ for
$x < 0$ and rises to infinity as $x \to \infty$.  All incoming fermions
eventually return ot $x = -\infty$, so these are unitary
quantum theories within the Hilbert space of incoming and outgoing
fermions in the left asymptotic region (or the bosonized
equivalent).  The type~II theory, on the other hand, leaves the
potential unmodified and and fills both sides of the barrier to
the same level.  It is a unitary quantum theory but with {\it two}
asymptotic Hilbert spaces.  So there appears to be an infinite number
of consistent nonperturbative definitions of the matrix model.
Note that the ambiguity appears not only in the nonperturbative
single-fermion tunneling amplitudes, but also in the large-field
classical behavior.  This is because a large incoming pulse can
propagate into the strongly coupled region.

Our study of gravitational and bulk scattering now pays an
unexpected bonus, for we have a new consistency condition which
must be satisfied.  Suppose we modify the potential by an
infinite barrier at $x = A \geq 0$.  A fermion reaching this
barrier will jump suddenly from phase space point $(x,p) =
(A,p)$ to $(A, -p)$.  The quantities $v = (-x-p) e^{-t}$ and $w
= (-x+p) e^t$ are no longer conserved, and neither will be the
$v_{mn}$.  Eq.~(\ref{2to1}) no longer holds and we get
the {\it wrong} bulk amplitude.  This
violates causality---the bulk scattering occurs at the point~A
of fig.~29a, where string perturbation theory is valid and the
incoming wave weak; the nonlinearity becomes large only later,
at point~B\@.  A further acausality will be a
nonconservation of gravitational mass in the process of
fig.~28.
This same argument applies to all modifications of type~I,
even if
the potential rises more smoothly.  Since
$(-x-p) e^{-t}$ is negative for incoming fermions with $p > - x$
but positive on all outgoing trajectories on the left side, it can
never be conserved.  So all theories of type~I are inconsistent.
One can also see this another way---the $v_{mn}$
correspond to unbroken spacetime gauge symmetries and one would expect
these to be conserved exactly.

The type~II theory with unmodified potential conserves the
$v_{mn}$, but it still gives rise to an inconsistent string
theory, at least with the natural interpretation that the two
asymptotic regions of the matrix model correspond to two
asymptotic regions of spacetime.  The point is that part of the
conserved $v_{mn}$ passes over the barrier, so the first equality
in eq.~(\ref{2to1}), relating $v_{10}$ to the outgoing field on
the left only, no longer holds.

So I am saying that in spite of the substantial effort that went
into matrix models {\it no} consistent nonperturbative
construction of the $D=2$ string has yet been given.  I think that
this is much better than the previous situation of an infinite
number of theories.  The study of the bulk and gravitational
scattering in coordinate space has not only dispels some
confusions and clarifies the nature of the discrete states, but
also provides a criterion for selecting among the possible
nonperturbative definitions of the theory.  The next step is to
search for a solution to the consistency conditions, considering
modification both of the matrix model itself and of the mapping
between the matrix model and string theory.  This is an important
problem, bearing directly on the $e^{-C/g_{\rm c}}$ behavior, and I
am optimistic that progress can be made.

\subs{Conclusion}

In conclusion, let me say again that I think it likely that string
theory will involve rich new dynamics that play an essential
role in determining the ground state.  Sometimes I feel that the
tools I have described, such as the matrix model and the various
high energy limits, are crude and unsophisticated.
This is why I found the parallels with Wilson's lecture so
encouraging---both the feeling of clutching at straws, and the
resemblance between his models and the matrix model.  So perhaps
the same method of attack that was so successful in quantum field
theory will be useful again.

\subsection*{Acknowledgments}

I would like to thank Mark Bowick, Shyamoli Chaudhuri, Jacques
Distler, Jeff Harvey, Igor Klebanov,
David Lowe, Makoto Natsuume, Phil Nelson,
Andy Strominger, Lenny Susskind, and Lars Thorlacius for
discussions of various points.  I would also like to thank the
students, organizers, and fellow lecturers of the school for their
questions and comments, for an enjoyable experience, and for foosball
lessons.  This work was supported by NSF grants PHY-89-04035,
PHY-91-16964, and PHY-94-07194.

\vfill

\pagebreak

\end{document}